\documentclass[useAMS,usenatbib,usegraphicx]{mn2e}
\usepackage{times}
\usepackage{graphicx}
\usepackage{epsfig}
\usepackage{amssymb}
\usepackage{natbib}
\usepackage{array}
\usepackage{hyperref}
\usepackage{multirow}
\newcommand{\low}{\textit{standard-res}}
\newcommand{\med}{\textit{high-res}}
\newcommand{\high}{\textit{very high-res}}
\newcommand{\slow}{\textit{slow-rotator}}
\newcommand{\slows}{\textit{slow-rotators}}
\newcommand{\fast}{\textit{fast-rotator}}
\newcommand{\fasts}{\textit{fast-rotators}}

\newcommand{\atlas}{\texttt{ATLAS$^{\mbox{3D}}$}}

\title[Formation of slowly rotating early-type galaxies]{Formation of slowly rotating early-type galaxies via major mergers:\\ a Resolution Study}
\author[M. Bois et al.]{M. Bois$^{1,2}$\thanks{E-mail:mbois@eso.org}, F. Bournaud$^{3}$, E. Emsellem$^{1,2}$, K. Alatalo$^{4}$, L. Blitz$^{4}$, M. Bureau$^{5}$, \newauthor M. Cappellari$^{5}$, R. L. Davies$^{5}$, T. A. Davis$^{5}$, P. T. de Zeeuw$^{2,6}$, P.-A. Duc$^{3}$, \newauthor S. Khochfar$^{7}$, D. Krajnovi\'c$^{2}$, H. Kuntschner$^{8}$, P.-Y. Lablanche$^{1}$, R. M. McDermid$^{9}$,\newauthor  R. Morganti$^{10}$, T. Naab$^{11,12}$, T. Oosterloo$^{10}$, M. Sarzi$^{13}$, N. Scott$^{5}$, P. Serra$^{10}$, \newauthor A. Weijmans$^{14}$, and L. M. Young$^{15}$ \\ \\
$^{1}$  Universit\'e Lyon 1, Observatoire de Lyon, Centre de Recherche Astrophysique de Lyon and Ecole Nationale Sup\'erieure de Lyon, 9 avenue Charles Andr\'e, \\ F-69230 Saint-Genis Laval, France \\
$^{2}$  European Southern Observatory, Karl-Schwarzschild-strasse 2, 85748 Garching, Germany \\
$^{3}$  CEA, IRFU, SAp et Laboratoire AIM, CEA Saclay -- CNRS -- Universit\'e Paris Diderot, 91191 Gif-sur-Yvette, France  \\
$^{4}$  Department of Astronomy and Radio Astronomy Laboratory, University of California, Berkeley, CA 94720, USA  \\
$^{5}$  Denys Wilkinson Building, University of Oxford, Keble Road, Oxford OX1 3RH, UK \\
$^{6}$  Sterrewacht Leiden, Leiden University, Postbus 9513, 2300 RA Leiden, the Netherlands \\
$^{7}$  Max-Planck-Institute for Extraterrestrial Physics, Giessenbachstrae, 85748 Garching, Germany \\
$^{8}$  Space Telescope European Coordinating Facility, European Southern Observatory, Karl-Schwarzschild-Str 2, 85748 Garching, Germany \\
$^{9}$  Gemini Observatory, Northern Operations Centre, 670 N. A'ohoku Place, Hilo, Hawaii 96720, USA \\
$^{10}$ Netherlands Foundation for Research in Astronomy (ASTRON), Postbus 2, 7990 AA Dwingeloo, The Netherlands \\
$^{11}$ Universit\"ats-Sternwarte M\"unchen, Scheinerstr. 1, D-81679 M\"unchen, Germany \\
$^{12}$ Max-Planck-Institute for Astrophysics, Karl-Schwarzschild-strasse 1, 85741 Garching, Germany \\
$^{13}$ Centre for Astrophysics Research, University of Hertfordshire, Hatfield, Herts AL1 09AB, UK \\
$^{14}$ Dunlap Institute for Astronomy \& Astrophysics, 50 St. George Street, Toronto, Canada \\
$^{15}$ Department of Physics, New Mexico Institute of Mining and Technology, Socorro, NM 87801, USA \\
}

\begin{document}

\date{Accepted 2010 April 20.  Received 2010 February 5; in original form 2009 July 23}

\pagerange{\pageref{firstpage}--\pageref{lastpage}} \pubyear{??}

\maketitle

\label{firstpage}

\begin{abstract}
We study resolution effects in numerical simulations of gas-rich and gas-poor major mergers, and show that the formation of slowly-rotating elliptical galaxies often requires a resolution that is beyond the present-day standards to be properly modelled. Our sample of equal-mass merger models encompasses various masses and spatial resolutions, ranging from about 200~pc and $10^5$ particles per component (stars, gas and dark matter), i.e. a gas mass resolution of $\sim 10^5$~M$_{\sun}$, typical of some recently published major merger simulations, to up to 32~pc and $\sim 10^3$~M$_{\sun}$ in simulations using $2.4 \times 10^7$ collisionless particles and $1.2 \times 10^7$ gas particles, among the highest resolutions reached so far for gas-rich major merger of massive disc galaxies. We find that the formation of fast-rotating early-type galaxies, that are flattened by a significant residual rotation, is overall correctly reproduced at all such resolutions. However, the formation of slow-rotating early-type galaxies, which have a low residual angular momentum and are supported mostly by anisotropic velocity dispersions, is strongly resolution-dependent. The evacuation of angular momentum from the main stellar body is largely missed at standard resolution, and systems that should be slow rotators are then found to be fast rotators. The effect is most important for gas-rich mergers, but is also witnessed in mergers with an absent or modest gas component (0--10\%\ in mass). The effect is robust with respect to our initial conditions and interaction orbits, and originates in the physical treatment of the relaxation process during the coalescence of the galaxies. Our findings show that a high-enough resolution is required to accurately model the global properties of merger remnants and the evolution of their angular momentum. The role of gas-rich mergers of spiral galaxies in the formation of slow-rotating ellipticals may therefore have been underestimated. Moreover, the effect of gas in a galaxy merger is not limited to helping the survival/rebuilding of rotating disc components: at high resolution, gas actively participates in the relaxation process and the formation of slowly-rotating stellar systems.
\end{abstract}


\section{Introduction}
Numerical simulations have been intensively used for more than two decades to study the properties of the remnants of galaxy mergers and the role of hierarchical merging in the formation of elliptical-like early-type galaxies \citep{HB91,Barnes92, mihos95}. With the increasing resolution and large statistical samples \citep[e.g.,][]{naab03, bour05, dimatteo1,dimatteo2}, modern work tends to quantify in details the properties of major and minor merger remnants, and accurate comparisons with observed properties of early-type galaxies can now be envisioned \citep[e.g.,][]{burkert-sauron}.

A general concern, though, is that the impact of the spatial and the mass resolutions on the detailed properties of the systems under scrutiny remains largely overlooked, and whether or not simulations of mergers have converged with today's typical resolution remains unexplored. Obviously, increasing resolution enables simulations to directly resolve cold gas clouds and clustered star formation \citep[e.g.][]{B08,kim09}, but whether these additional small-scale ingredients can significantly impact the global, large-scale morphology and kinematics of merger remnants has not been studied in detail. In cosmological simulations, an increase in resolution (i.e., an increase in the number of particles and/or decrease of the softening length) can affect the baryonic density and circular velocity profiles of individual galaxies in a halo (\citealt{naabresol}). \citet{navresol} also studied numerical convergence via a suite of $\Lambda$CDM simulations and confirmed that the halo mass distributions were better described by Einasto profiles that are not, stricly speaking, universal.

While many resolution studies have been made in cosmological simulations, few have focused on galaxy merger simulations. \citet{coxresol}, \citet{hopkinsresol} and e.g., \citet{dimatteo2} included some checks of the effect of resolution on the star-formation activity of on-going mergers. But a resolution study aimed at examining the detailed morphology and kinematics of relaxed merger remnants (i.e., galaxies which tend to be roughly S0- or elliptical-like) has not yet been conducted.

Models of galaxy mergers have reached particularly high resolution with the work of \citet{wetzstein07} (70~pc softening with $4\times 10^6$ particles in total -- but only 45,000 for the gas component), \citet{li04} (10 to 100~pc and $5 \times 10^5$ gas particles per galaxy), \citet{naabresol} ($8\times 10^6$ particles with a 125~pc resolution in a cosmological re-simulation of an individual galaxy halo). The highest resolution for gas-rich mergers have been achieved recently by \citet{B08} for mergers of bright spiral galaxies, with a total of $3.6 \times 10^7$ particles including more than $10^7$ gas particle, and a 32~pc softening size, and \citet{kim09} with a spatial resolution of 3.8~pc and a mass resolution of $2\times 10^3$~M$_{\sun}$ (for dwarf or low-mass spirals, though). But such high-resolution studies have focused on small-scale gas physics and structure formation, without studying the impact of high resolution on the global properties of the elliptical-like galaxies formed in major mergers.

Large samples of simulations of idealized galaxy mergers remain typically limited to softening lengths of about 100--300~pc, and $\sim 10^5$ particles per galaxy \citep[see samples in][]{naab03,NJB06,bour04,bour07,coxal06,coxal08,dimatteo1}. Whether or not the relatively limited numerical resolution used in such studies affects the global properties of merger remnants is still a largely open question: for instance, the detailed comparison of major merger remnants with the observed anisotropy-flattening relation by \citet{burkert-sauron} relies on simulations with $2 \times 10^4$ gas particles per galaxy, a gas particle mass $\sim 3 \times 10^5$~M$_{\sun}$, and a spatial resolution (softening) of about 200~pc.

Within the context of the \atlas{} project (\href{http://purl.org/atlas3d}{http://purl.org/atlas3d}), an extensive set of numerical simulations is being conducted to support the multi-wavelength survey of a complete sample of early-type galaxies within the local (40Mpc) volume, in terms of various formation mechanisms of early-type galaxies: binary mergers, multiple mergers, disc instabilities, etc. An ambitious series of simulations of mergers are being specifically performed and analysed for this purpose (Bois et al. in preparation). To properly interpret the results from these simulation efforts, as well as to understand the robustness of existing and past studies of galaxy mergers, we first probe the effect of spatial and mass resolutions on the global structure of binary disc merger remnants.

In this paper, we study the effect of numerical resolution on the global morphology and the kinematics of the simulated remnants of binary, equal-mass major mergers. We wish to examine resolutions ranging from the typical resolutions used in recent, large simulations samples, to some of the highest merger simulations ever performed. We study both dry (collisionless) and wet (gas-rich) mergers of disc galaxies. The modeled interaction orbits lead to the formation of both fast rotators, i.e., early-type galaxies flattened by  significant rotational support, and slow rotators, i.e., early-type galaxies with low residual rotation, supported (and flattened) by (anisotropic) velocity dispersions, following the classification detailed in \citet[][see also Section~\ref{sec:nbody}]{ems07}. We find that the formation of fast rotators is overall correctly reproduced with numerical simulations at modest resolutions. In contrast, the formation of slow-rotating systems is correctly reproduced only at high resolution (Section~\ref{sec:results}), above the resolution of most of the recently published merger simulations. The influence of gas on the structure of merger remnants, compared to dry mergers, also differs at high resolution, and is not limited to easing the survival and/or rebuilding of rotating disc components. In Section~\ref{sec:origins}, we further examine the origin of this observed resolution effect in the formation of slow-rotating systems. We show that it is not an artefact from different initial conditions or interaction orbits, but that the physical treatment of the merging process is actually biased when the resolution is too low. The effect of the resolution has been tested on other simulations producing slow rotators and we find that it is a systematic one (Section~\ref{sec:systematic}). We summarize our results, discuss the required resolution for accurate studies and the general implications for the formation of elliptical galaxies in Section~\ref{sec:conclusions}.

\section{Simulations and analysis}
\label{sec:nbody}

   \subsection{Method \label{method}}
      \subsubsection{Code \label{sec:code}}
We use the particle-mesh code described in \citet[][]{B08} and references therein.

This code uses a Cartesian grid on which the particles are meshed with a "Cloud-In-Cell" interpolation. The gravitational potential is computed with an FFT-based Poisson solver and particle motions are integrated with a leap-frog algorithm, and a time-step of 0.5~Myr.

Interstellar gas dynamics is modelled with the sticky-particle scheme with elasticity parameters $\beta_t=\beta_r= 0.6$. This scheme neglects the temperature and thermal pressure of the gas, assuming it is dominated by its turbulent pressure, which is the case for the star-forming interstellar medium at the scales that are studied here \citep{ES04,burkert06}. The velocity dispersion of the particles model the turbulence, and their mutual collisions are inelastic to ensure that the turbulence dissipates over about a vertical crossing time \citep{ml99}.

The star formation rate is computed using a Schmidt-Kennicutt law: it is then proportional to the gas density in each cell to the exponent 1.5. Gas particles are converted to star particles with a corresponding rate in each cell. Energy feedback from supernovae is accounted for with the scheme proposed by \citet{MH94}. Each stellar particle formed has a number of supernovae computed from the fraction of stars above  8 M$_{\sun}$ in a Miller-Scalo IMF. A fraction $\epsilon$ of the $10^{51}$~erg energy of each supernova is released in the form of radial velocity kicks applied to gas particles within the closest cells. We use $\epsilon = 2 \times 10^{-4}$ , as \citet{MH94} suggest that realistic values lie around $10^{-4}$ and less than $10^{-3}$.

      \subsubsection{Set-up for initial disc galaxies \label{sec:init}}
The baryonic mass of our model galaxies is $10^{11}$~M$_{\sun}$. In dry merger simulations, this mass is purely stellar. In wet merger simulations, 80\% of this mass is stellar, and 20\% is gaseous. The initial gas and stellar discs are Toomre discs, with a scale-length of 4~kpc and a truncation radius of 10~kpc for the stars, resp. 8~kpc and 20~kpc for the gas. 20\% of the stars are in a spherical bulge, modeled with a \citet{hernquist-profile} profile with a 700~pc scale-length.
The dark matter halo is modelled with a Burkert profile \citep{burkert-profile}, a 7~kpc scale-length and a truncation radius of 70~kpc, inside which the dark matter mass is $3\times 10^{11}$~M$_{\sun}$.

The two "progenitor" disc galaxies in each simulation are identical, the total mass of the remnant will be $2 \times 10^{11}$~M$_{\sun}$ which is consistent with the \slow{} mass range observed in the \atlas{} sample \citep{ems07}.

      \subsubsection{Orbits \label{sec:orbit}}
We have used two interacting orbits, for each kind of merger (dry and wet) and each resolution level. None corresponds to a very specific and unlikely configuration like coplanar discs, or polar orbits.

The first orbit is called {\em "fast"} because it forms fast-rotating early-type galaxies. The velocity at an infinite distance is 170~km~s$^{-1}$ and the pericentre distance is 30~kpc. This orbit is prograde with respect to the first progenitor disc, with an inclination of the orbital plane w.r.t. the disc plane of 25\degr. The orbit is retrograde w.r.t. the other progenitor disc, with an inclination of 45\degr.

The second orbit is called {\em "slow"} because it forms slow-rotating early-type galaxies (at least at high-enough resolution). The velocity at an infinite distance is 140~km~s$^{-1}$ and the pericentre distance is 25~kpc. This orbit is prograde with respect to the first progenitor disc, with an inclination of the orbital plane w.r.t. the disc plane of 45\degr. The orbit is retrograde w.r.t. the other progenitor disc, with an inclination of 25\degr.

These orbits as well as those used in additional tests (Section~5) have a total energy $E > 0$ or $E \simeq 0$, corresponding to initially unbound galaxy pairs. Such orbits are representative of the most common mergers in $\Lambda$-CDM cosmology \citep{khochfar06}.

   \subsubsection{Standard-, high- and very high-resolutions \label{sec:resols}}
Wet and dry mergers have been simulated for each orbit at three resolution levels. The detail for these resolutions are indicated in Table~\ref{tab:prop}. The {\em very high} resolution arguably corresponds to the highest resolution simulation of a wet major merger performed so far \citep[see][]{B08}.

\begin{table}
\begin{tabular}{|c|c|c|c|}
\hline
Label & Softening Length & Particles / component & Total particles \\
\hline 
{\em standard}  & 180 pc & $10^5$          & $6 \times 10^5$    \\
\hline 
{\em high}      & 80 pc  & $10^6$          & $6 \times 10^6$    \\
\hline 
{\em very high} & 32 pc  & $6 \times 10^6$ &  $3.6 \times 10^7$ \\
\hline 
\end{tabular}
\caption{Label for the resolution, softening length, number of particles per component (stars, gas and dark matter) and total number of particles in the simulation for the three resolutions.} 
\label{tab:prop}
\end{table}

\begin{figure}
  \centering
  \epsfig{file=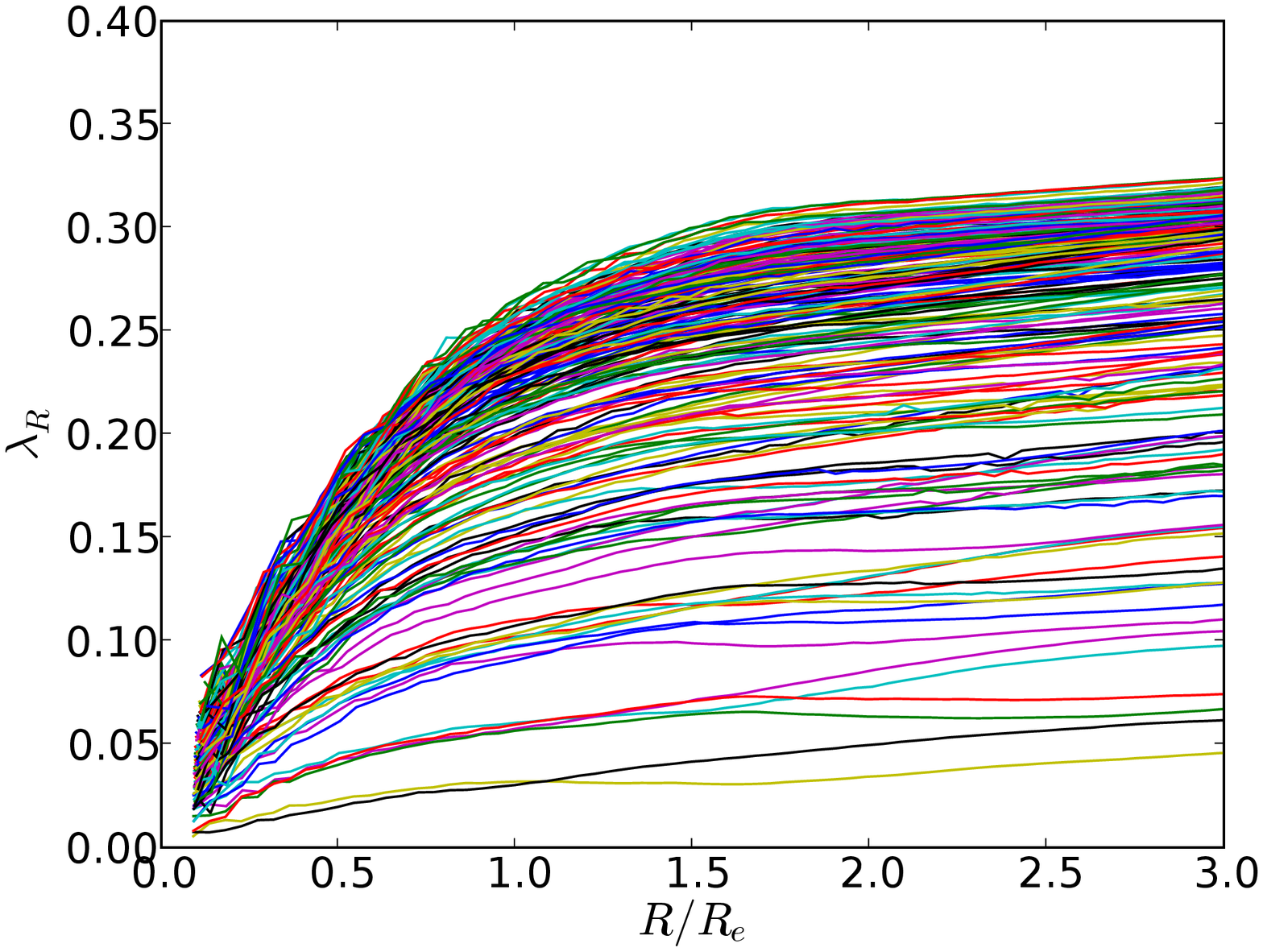, width=\columnwidth}\\
  \epsfig{file=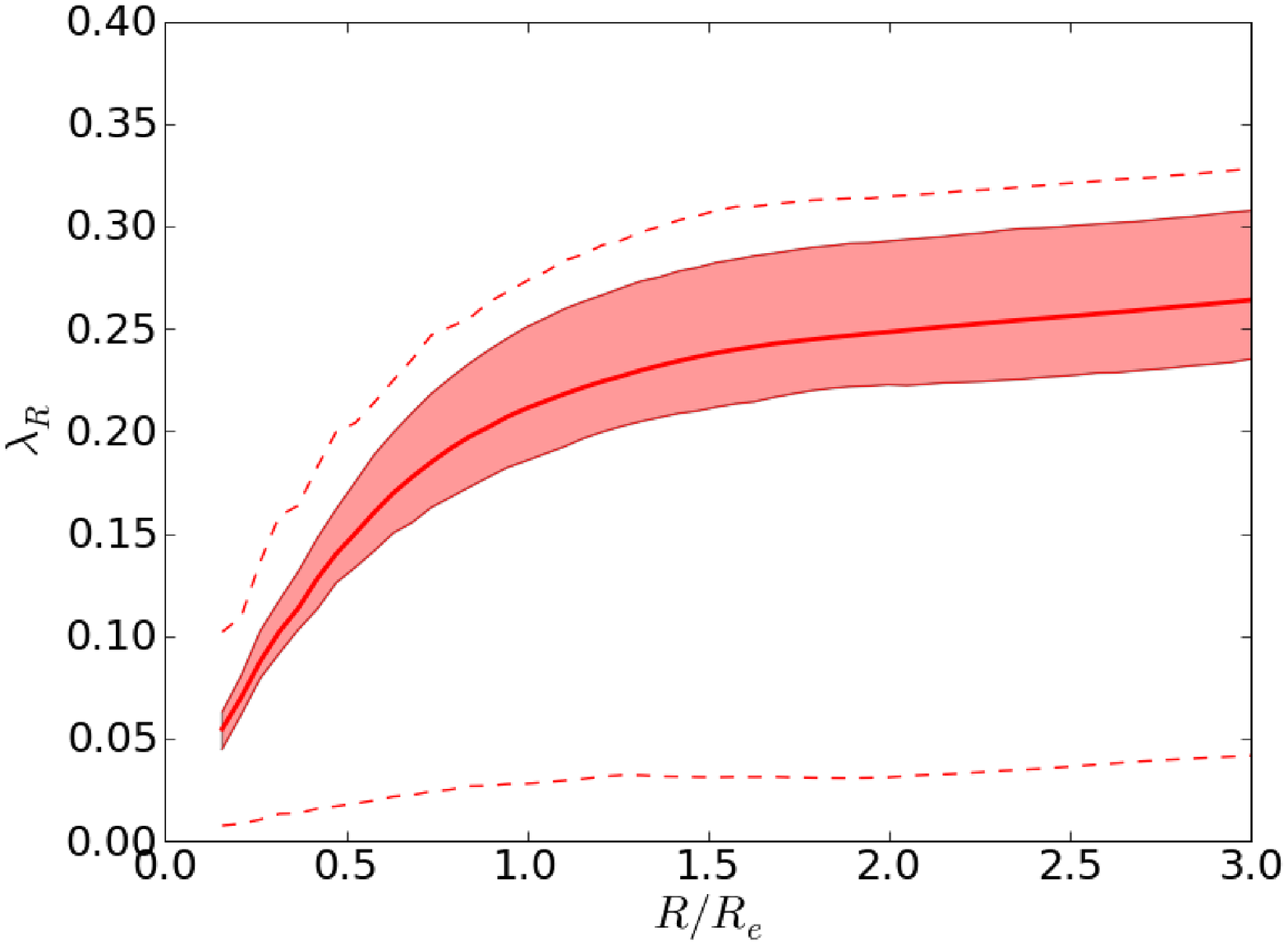, width=\columnwidth}
  \caption{{\bf Top panel} $\lambda_R$ profiles of all 200 projections for the {\em Dry-Fast-very High} simulation. The profiles are plotted as a function of $R/R_e$ (radius normalised by the effective radius for each projection). Each line here corresponds to a given projection. {\bf Bottom panel} Corresponding median (thick solid line), quartiles (thin solid line) and maximal and minimal values (dashed lines) at each radius. This representation of the results illustrates the fact that all values are between the dashed lines, and 50\% of the projections are in the filled area. Each line plotted in this panel does not correspond to one specific projection: the median or quartiles are derived for different projections at each radius.}
  \label{fig:projlr}
\end{figure}

\medskip
We will label each simulation with the following nomenclature:
\begin{itemize}
\item the first item indicates a {\em Wet} or {\em Dry} merger, i.e. gas-rich or gas-free progenitors.
\item the second item specifies the chosen orbit: the one producing \fasts{} or \slows{} (at least at high-enough resolution).
\item the last item indicates the resolution level: {\em standard}, {\em high} or {\em very high}.
\end{itemize}
For instance, the {\em wet-fast-high} simulation refers to the high-resolution models of a wet merger on the orbit producing a fast-rotating early-type galaxy.

   \subsection{Analysis of the relaxed merger remnants \label{sec:analysis}}
We analyse the merger remnants after 1.2~Gyr in the simulation, which is 800-900~Myr after the first pericentre passage, and 600-700~Myr after the central coalescence. The remnants are thus relaxed when the analysis is performed. Tidal debris can still be orbiting around the merger remnant, but the bulk of the stellar mass in the central body does not show significant evolution. Analysis performed at earlier and later instants did not show significant variations, so spurious effects related to time evolution should not affect the comparison of the three different resolution levels.

      \subsubsection{Projected maps \label{sec:projected}}

Intrinsic and apparent properties of the merger remnant (e.g the apparent ellipticity) are directly linked with its orbital structure \citep{JNB05}. To probe the relaxed merger remnants, we have therefore built projected maps of the stellar mass density, line-of-sight velocity and velocity dispersion fields. Two-dimensional maps are useful to reveal the wealth of photometric or kinematic structures associated with a galaxy merger, e.g., globular clusters or kinematic misalignments (see \citealt{BB00}; \citealt{JNPB07}).

The projected maps cover a $16\times16$~kpc$^2$ field of view around the density peak of each system: our analysis is conducted up to a limit of three effective radii $R_e$, which encloses most of the baryonic mass of early-type galaxies, and the typical effective radius of our merger remnants is 2.5~kpc. Each projection was computed on a $100\times100$ pixel grid. The pixel size is 160$\times$160~pc$^2$, which approximatively corresponds to the size of the softening length of our \low{} simulations, and is kept fixed for all resolutions.

To obtain statistically significant results, we have built such maps and performed the subsequent analysis with 200 isotropically distributed viewing angles (i.e., 200 different line-of-sights). In this way, we do not characterize and compare the merger remnant under a particular projection, but their global, statistical properties. As an example, Figure~\ref{fig:projlr} shows the effect of the projections on the radial $\lambda_R$ profiles for one simulation. Among these 200 profiles, the lowest (near zero) and highest values correspond respectively to the merger remnant seen nearly face-on (i.e., the lowest apparent ellipticity) or nearly edge-on (i.e., the highest apparent ellipticity). Our choice of 200 projections ensures that neighbouring projections are separated only by 10 degrees in any direction, so that intermediate viewing angles would not show significant differences.

      \subsubsection{Physical parameters \label{sec:parameters}}
Our analysis is based on a few simple morphological and kinematic parameters --  a choice mainly motivated by the fact that these parameters are often being used as standards in studies of nearby elliptical galaxies.

The morphological parameters pertains to the photometry : we measure the ellipticity $\epsilon$ (defined as $1 - b/a$, where $a$ and $b$ are the semi major- and minor-axes, respectively) and $a_4/a$ which is the fourth (cosine) Fourier coefficient of the deviation of isophotes from a perfect ellipse ($a_4/a > 0$ and $a_4/a < 0$ correspond to discy, and boxy isophotes, respectively). These two parameters are computed using the Kinemetry software tool\footnote{\href{http://www-astro.physics.ox.ac.uk/$~$dxk/idl/}{http://www-astro.physics.ox.ac.uk/$~$dxk/idl/}} which can be used to perform standard ellipse-fitting of galaxy images, as well as to study galaxy kinematics \citep{kinem}.  For the kinematic analysis, apart from the first two velocity moments (velocity and velocity dispersion), we use the $\lambda_R$ parameter, a robust proxy for the baryonic projected angular momentum, as defined in \citet[][hereafter E07]{ems07} :
\[
\lambda_R \equiv \frac{\langle R \, |V| \rangle}{\langle R \, \sqrt{V^2 + \sigma^2} \rangle}
\]

In E07, $\lambda_R$ was used to reveal two families of early-type galaxies, the \slows{} with $\lambda_R \leq 0.1$ and the \fasts{} with $\lambda_R > 0.1$ at one effective radius $R_e$. In a recent study, \citet{jesseitlr} have simulated binary disc mergers to investigate the $\lambda_R$ parameter: tests on their merger remnants reveal that $\lambda_R$ is a good indicator of the true angular momentum content in early-type galaxies. As emphasised in E07, \citet{cap07} and \citet{kra08}, fast and slow rotators exhibit qualitatively and quantitatively different stellar kinematics. $\lambda_R$ is thus an interesting parameter to probe, and should indicate whether or not the kinematics of the merger remnants are equally resolved at different resolutions.

For each above-mentioned parameter, we have computed the minimum, maximum, mean values, as well as the 1$^{st}$ and 3$^{rd}$ quartiles over all the projections at individual radii, to quantify the statistical distribution of these parameters in a simple way. An example is shown in Figure~\ref{fig:projlr}. Note that with this choice, the projection which minimises or maximises a parameter varies with radius.

\section{Effect of resolution on the formation of slow rotators}
\label{sec:results}
\begin{figure}
  \centering
  \epsfig{file=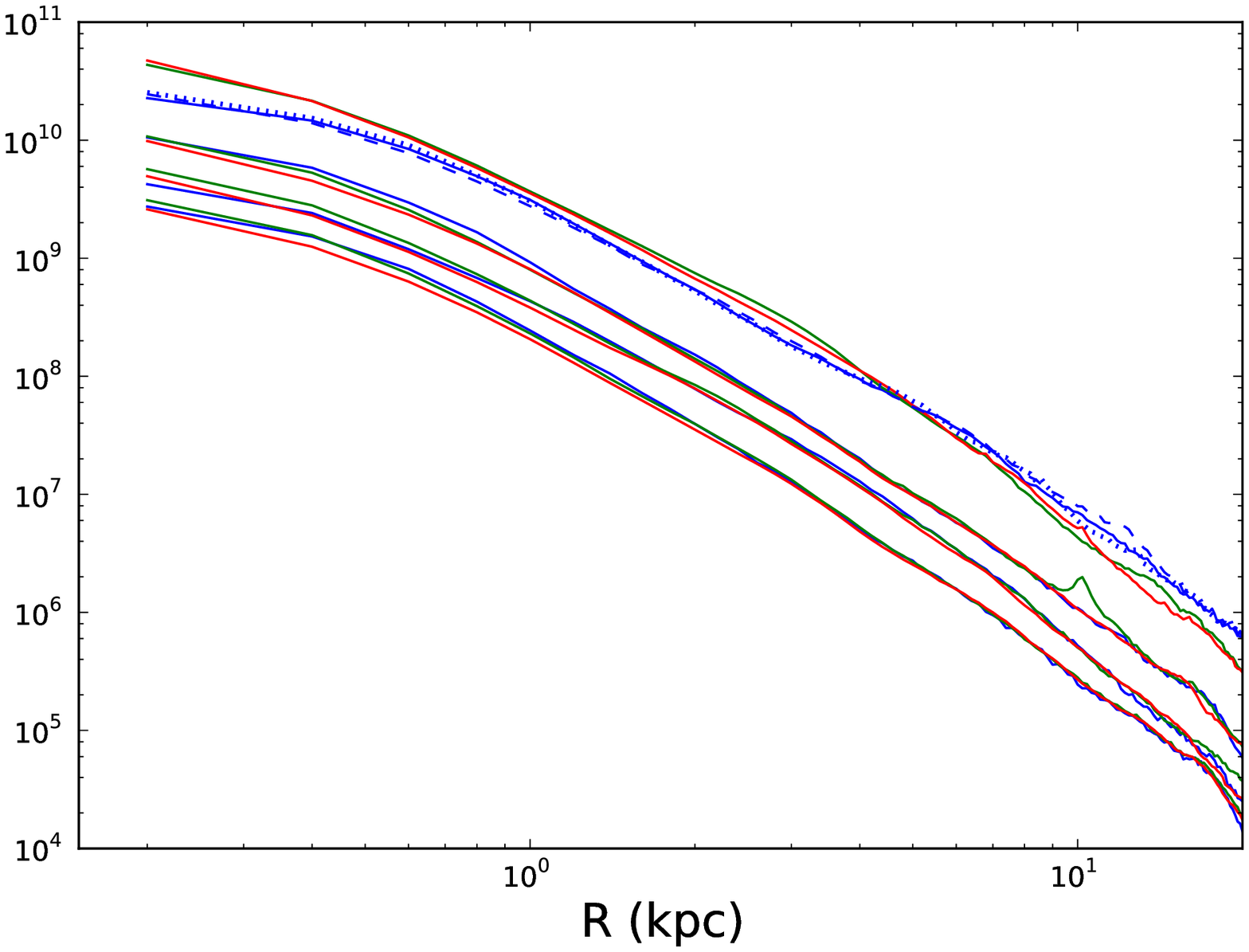, width=\columnwidth} \\
  \caption{Stellar density profiles for the Wet-Slow, Wet-Fast, Dry-Slow and Dry-Fast simulations (from top to bottom, respectively). The density of the Wet-Fast, Dry-Slow and Dry-Fast cases has been divided by a factor of five, ten and twenty respectively to improve the readability of the plot. Red lines correspond to very high-resolution models, green lines to high-resolution cases, and blue lines to standard-resolution models. The two alternative realisations of the Wet-Slow-Standard simulation (see Sect.~\ref{sec:crash}) are shown in dashed and dotted lines.}
  \label{fig:densprof}
\end{figure}

In this Section, we briefly describe the properties of the simulated mergers with the three different resolutions. The complete set of analysis results can be found in Appendix~A. We then focus the analysis on the simulations that show important differences, namely the cases producing slow rotators at high resolution.

\begin{figure*}
\begin{center}
  \begin{tabular}{rcccl}
 & \begin{Large}Dry-Fast \end{Large} & \vline & \begin{Large}Wet-Fast \end{Large} & \\
 & $\epsilon_{min}$ \hspace{1.3cm} $\epsilon_{max}$ \hspace{1.3cm} $\epsilon_{mean}$ & \vline & $\epsilon_{min}$ \hspace{1.3cm} $\epsilon_{max}$ \hspace{1.3cm} $\epsilon_{mean}$ &   \\
\low & \epsfig{file=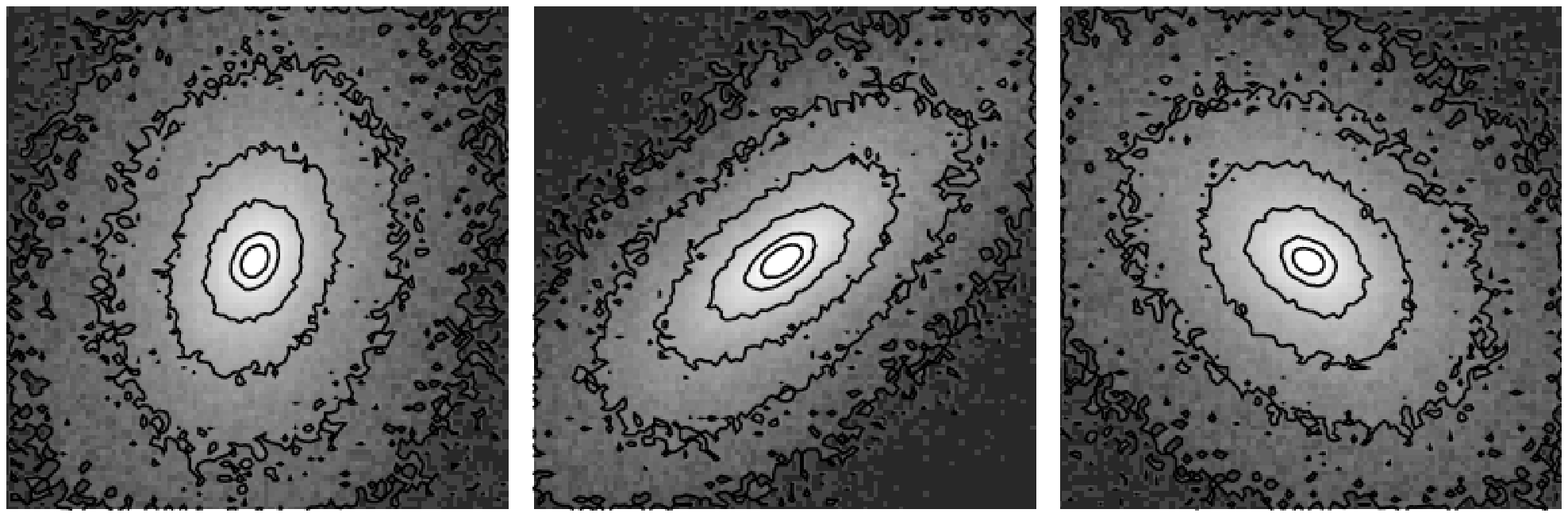, width=0.8\columnwidth} & \vline & \epsfig{file=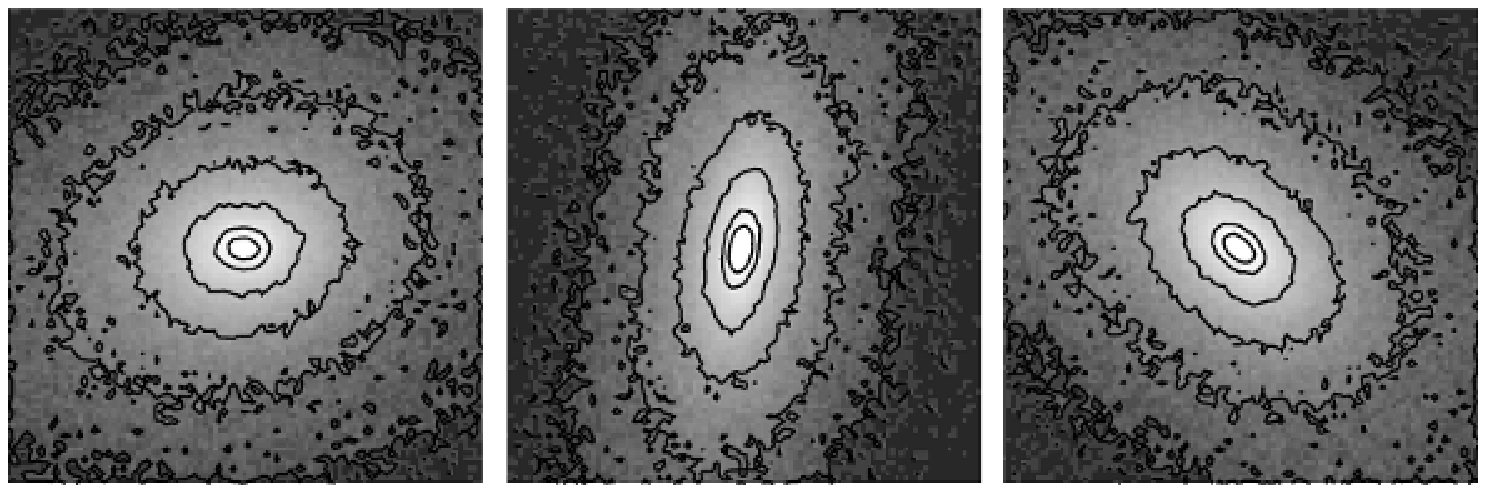, width=0.8\columnwidth} & \low \\
\med & \epsfig{file=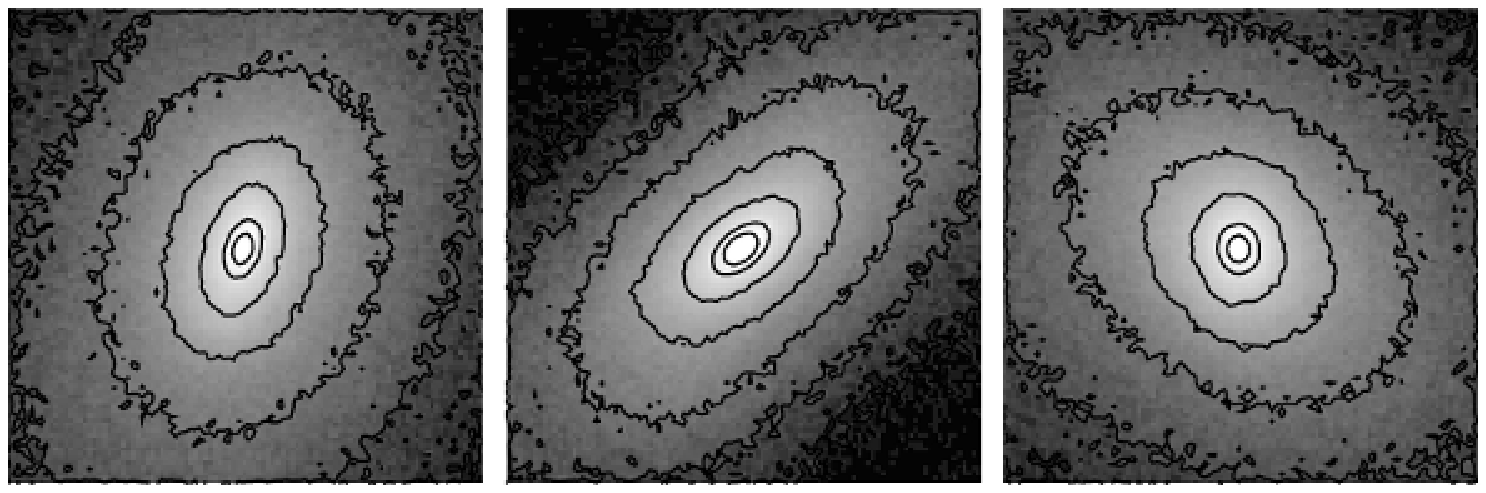, width=0.8\columnwidth} & \vline & \epsfig{file=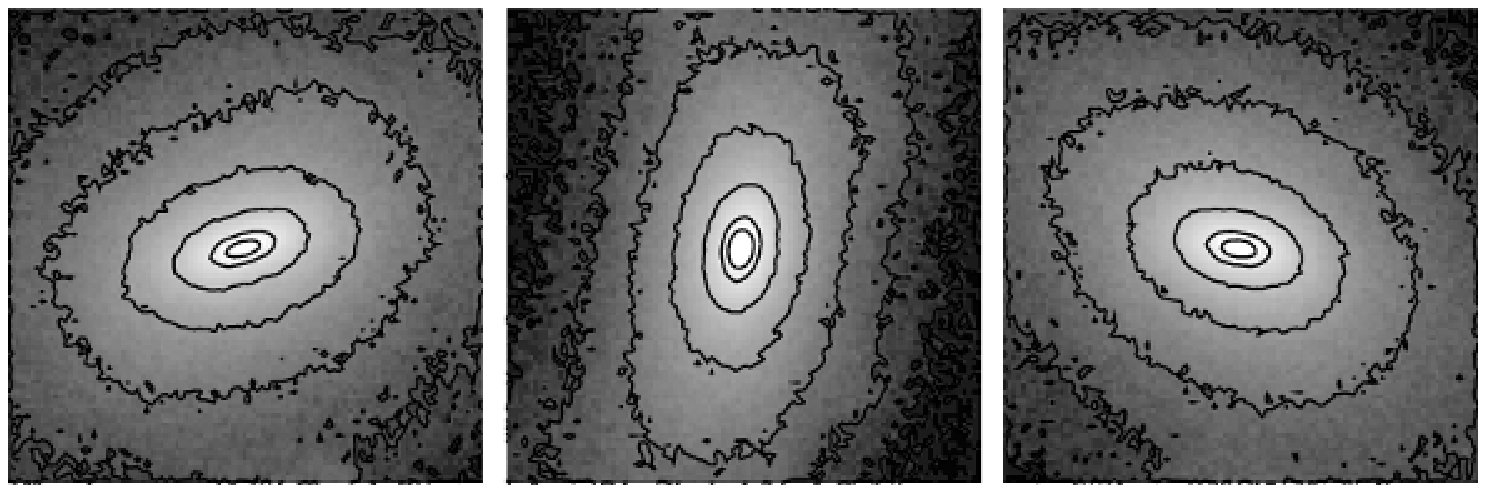, width=0.8\columnwidth} & \med \\
\high & \epsfig{file=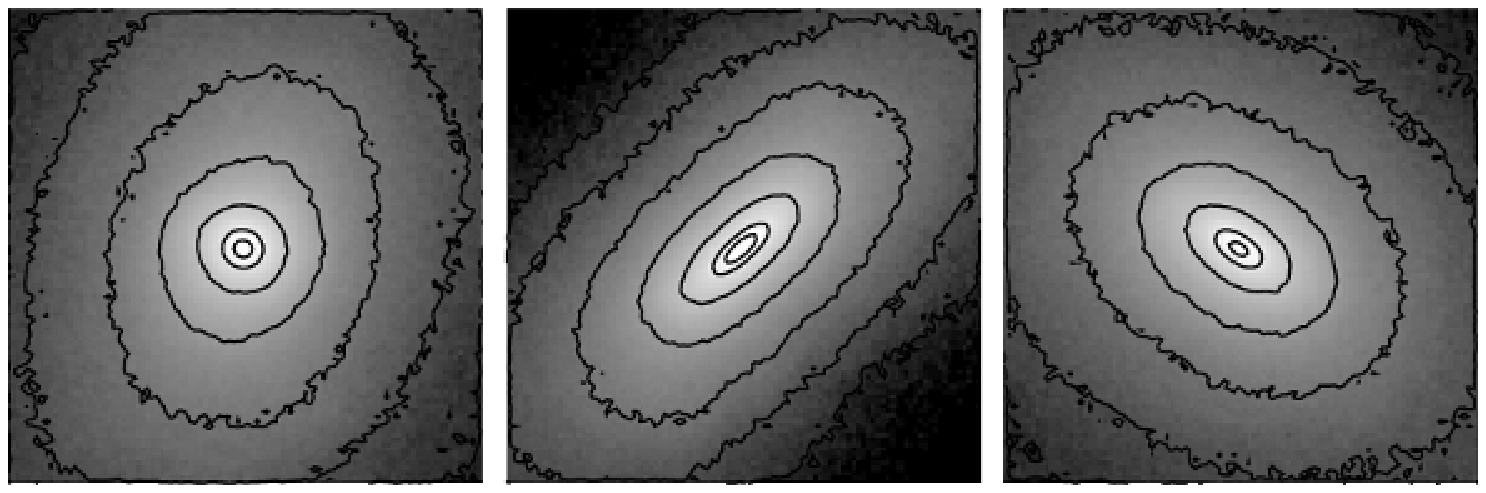, width=0.8\columnwidth} & \vline & \epsfig{file=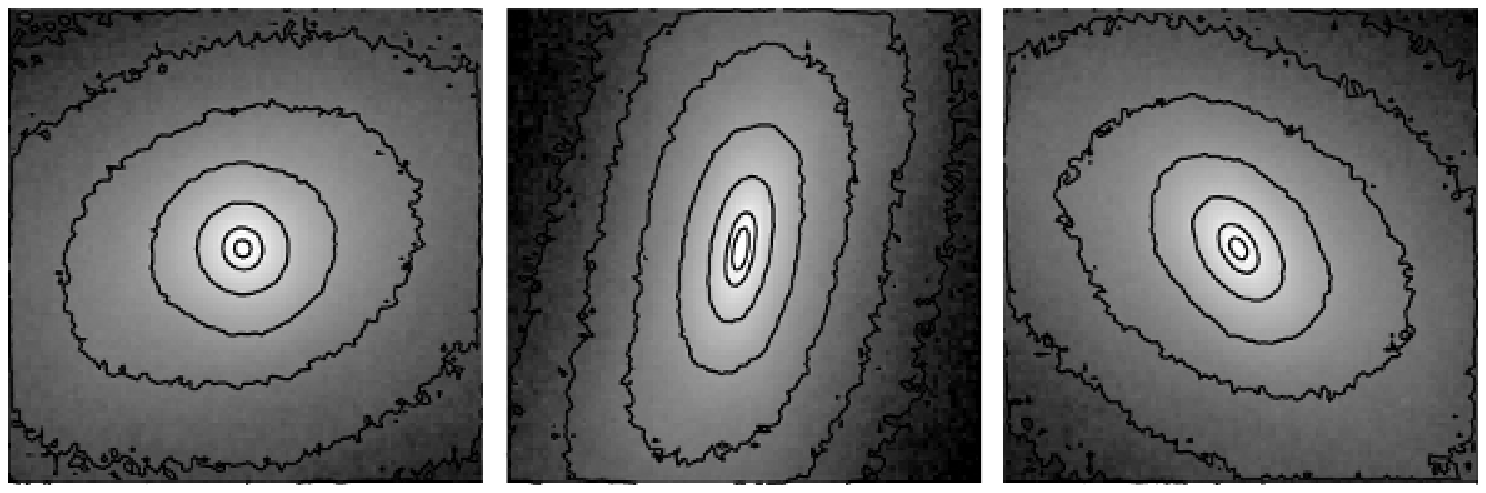, width=0.8\columnwidth} & \high \\
 & & \\
\hline
 & & \\
 & \begin{Large}Dry-Slow \end{Large} & \vline & \begin{Large}Wet-Slow \end{Large} & \\
 & $\epsilon_{min}$ \hspace{1.3cm} $\epsilon_{max}$ \hspace{1.3cm} $\epsilon_{mean}$ & \vline & $\epsilon_{min}$ \hspace{1.3cm} $\epsilon_{max}$ \hspace{1.3cm} $\epsilon_{mean}$  &  \\
 \low & \epsfig{file=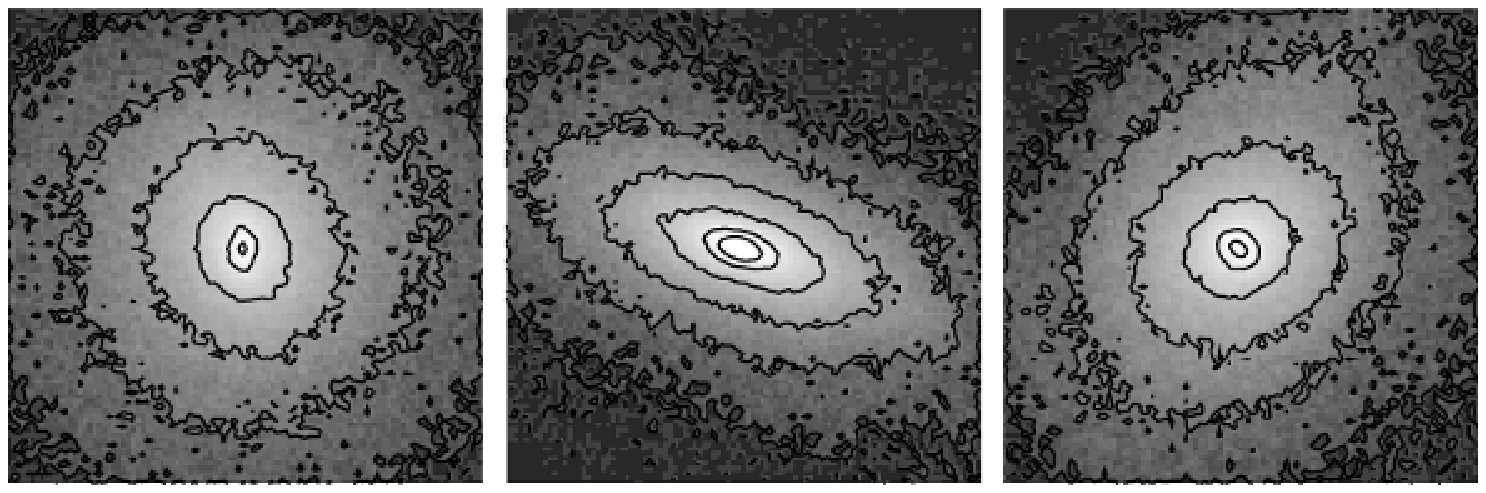, width=0.8\columnwidth} & \vline & \epsfig{file=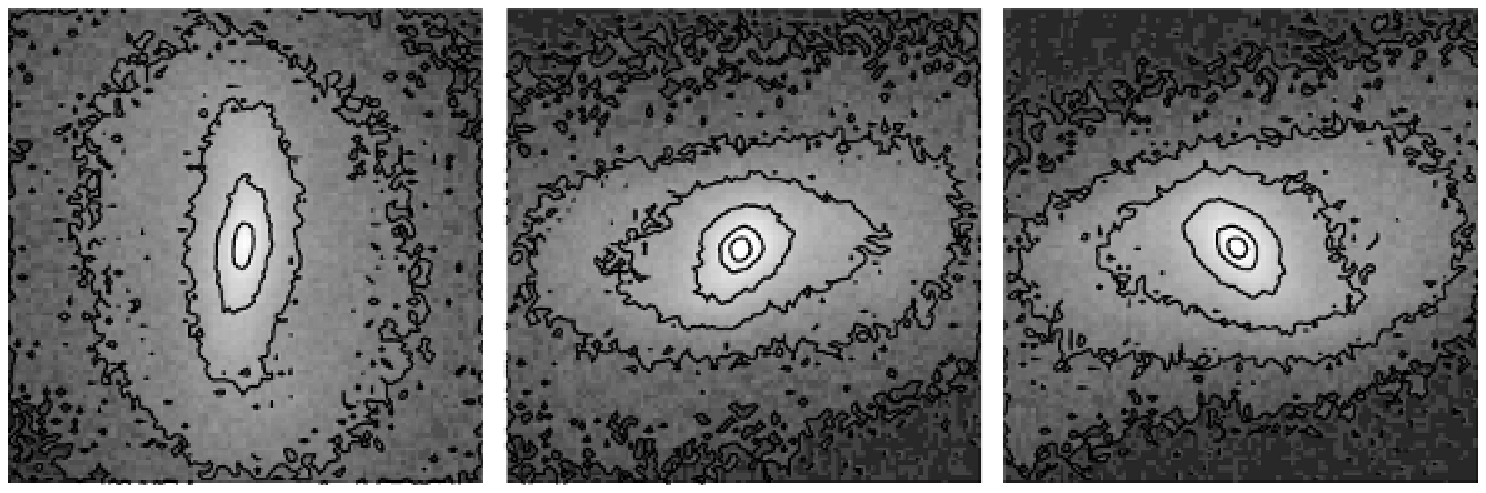, width=0.8\columnwidth} & \low \\
\med & \epsfig{file=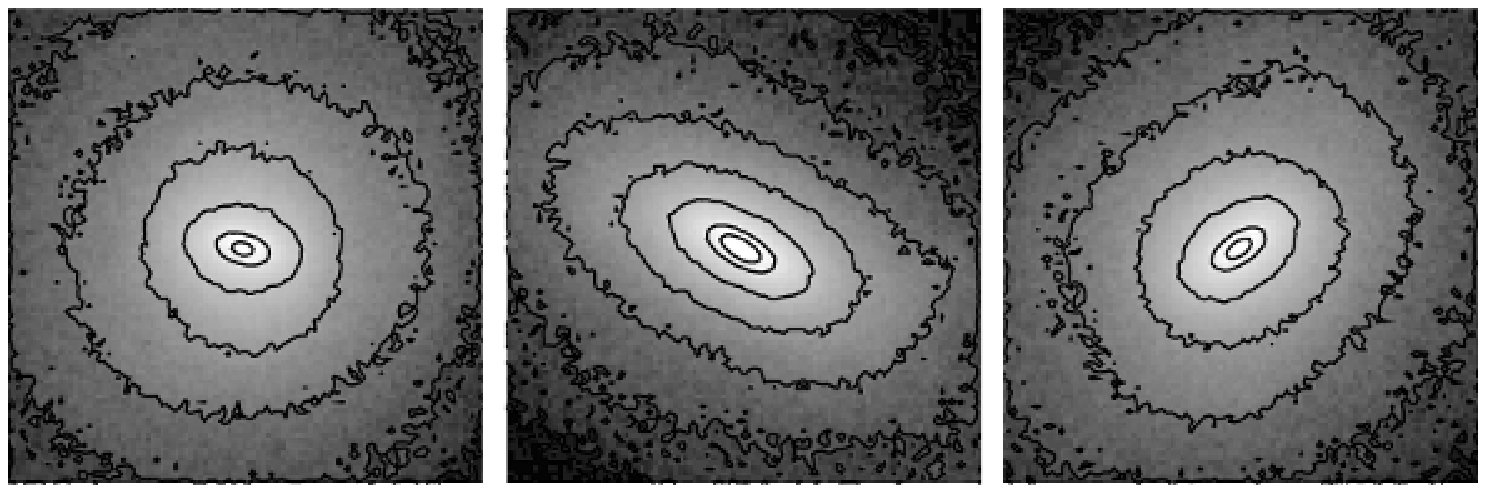, width=0.8\columnwidth} & \vline & \epsfig{file=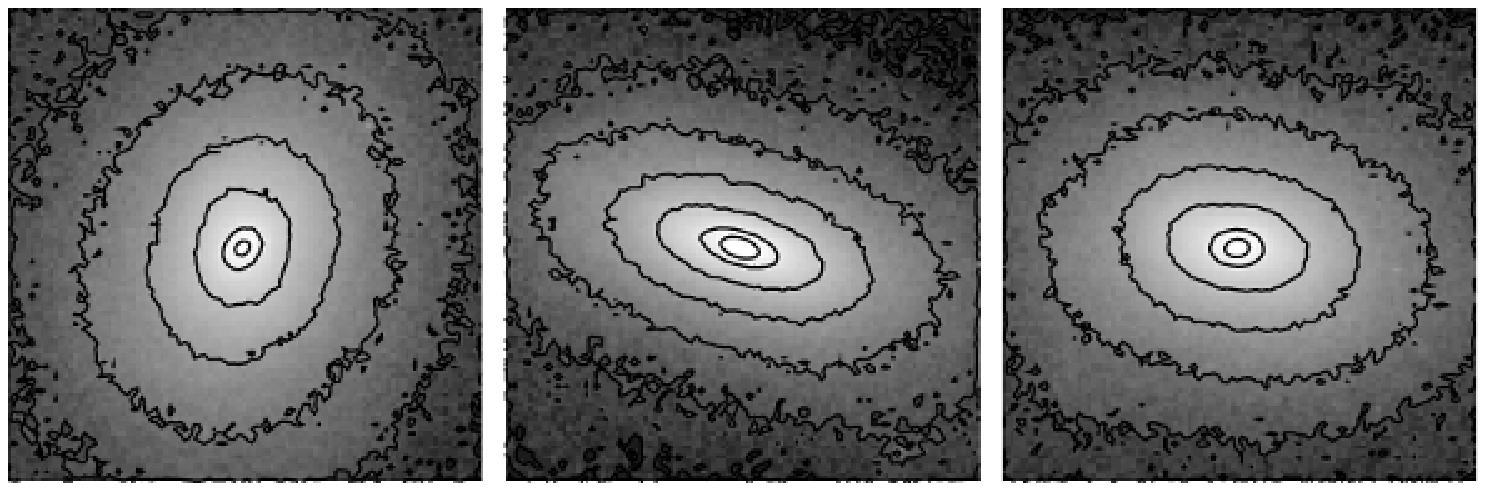, width=0.8\columnwidth} & \med \\
\high & \epsfig{file=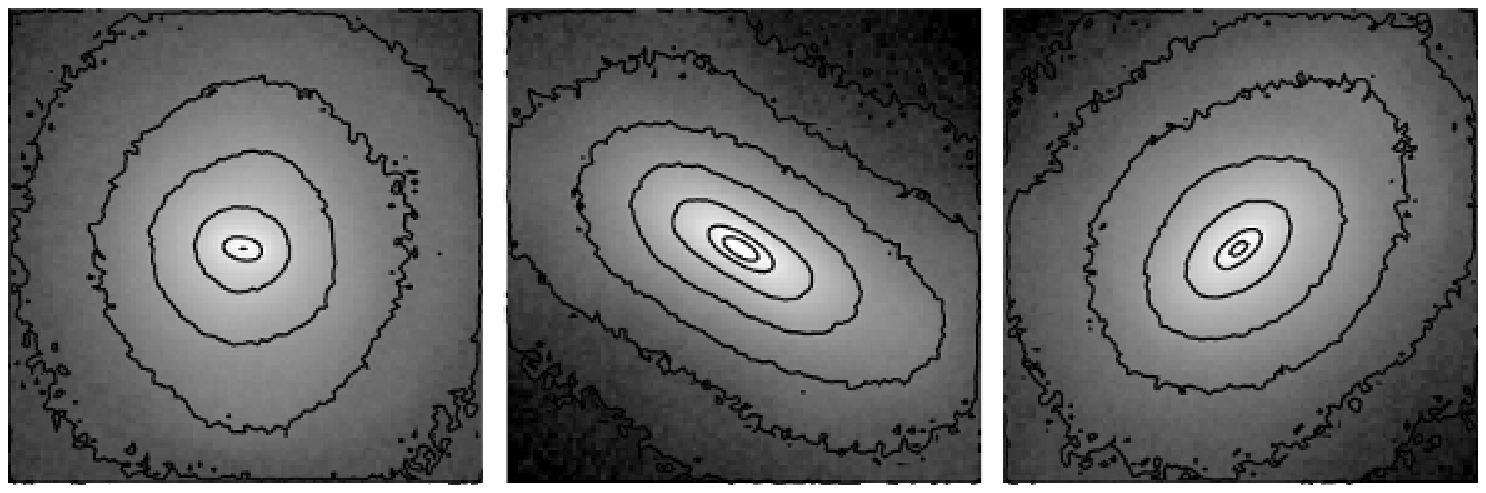, width=0.8\columnwidth} & \vline & \epsfig{file=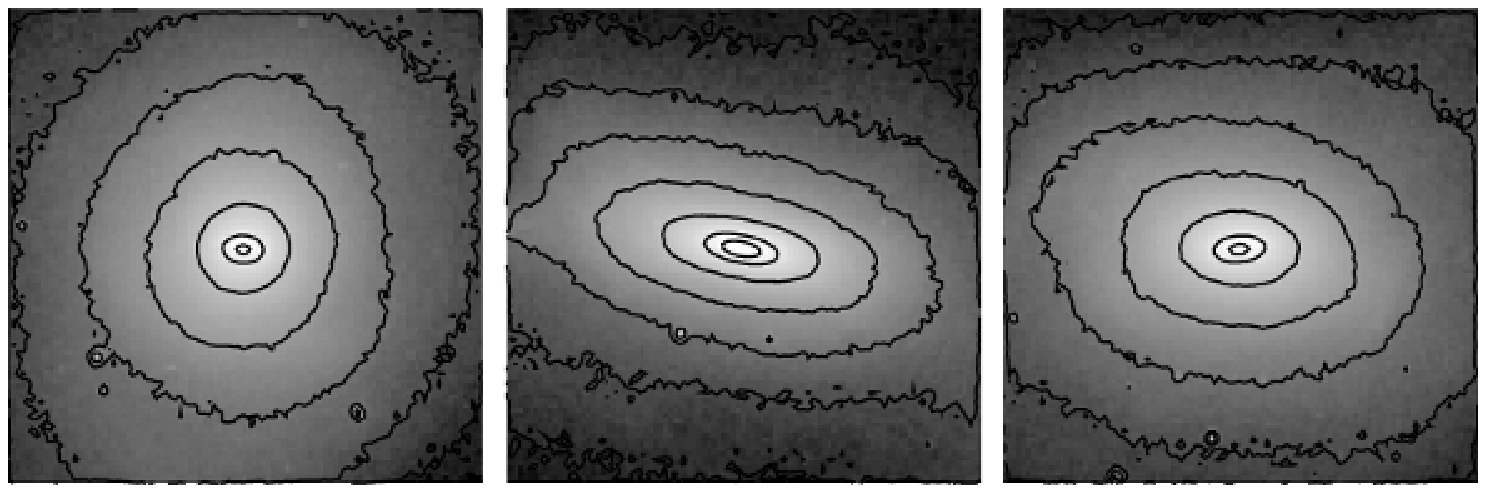, width=0.8\columnwidth}  & \high \\
  \end{tabular}
  \caption{The twelve normalized projected surface density maps (in log), for the four sets of simulations at three different resolutions (labelled accordingly). The field of view is $16\times16$~kpc$^2$. For each simulation, the projections corresponding to the minimum, maximum and mean ellipticities are shown. The viewing angle of these projections are defined at \high{} and re-applied for the \low{} and \med{} simulations: projections are thus established along the same line-of-sights for all resolutions. Luminosity contours are the same for all simulations and drawn with a spacing of 0.5 magnitude (except for the 2 inner contours with a step of 0.3). The effective radius is about at the edge of the 4$^{th}$ isophote for all simulations.}
  \label{fig:phot}
\end{center}
\end{figure*}

\begin{figure*}
\begin{center}
  \begin{tabular}{rcccl}
 & \begin{Large}Dry-Fast \end{Large} & & \begin{Large}Wet-Fast \end{Large} & \\
 & $\epsilon_{min}$ \hspace{1.2cm} $\epsilon_{max}$ \hspace{1.2cm} $\epsilon_{mean}$ & & $\epsilon_{min}$ \hspace{1.2cm} $\epsilon_{max}$ \hspace{1.2cm} $\epsilon_{mean}$ &   \\
 & & \multirow{4}{*}{\epsfig{file=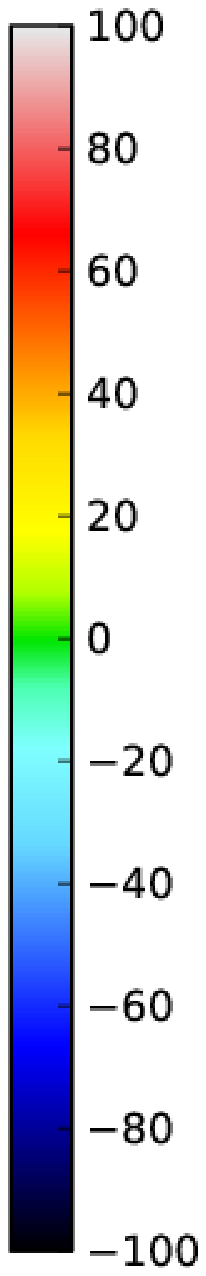, width=0.2\columnwidth}} & & \\
\low & \epsfig{file=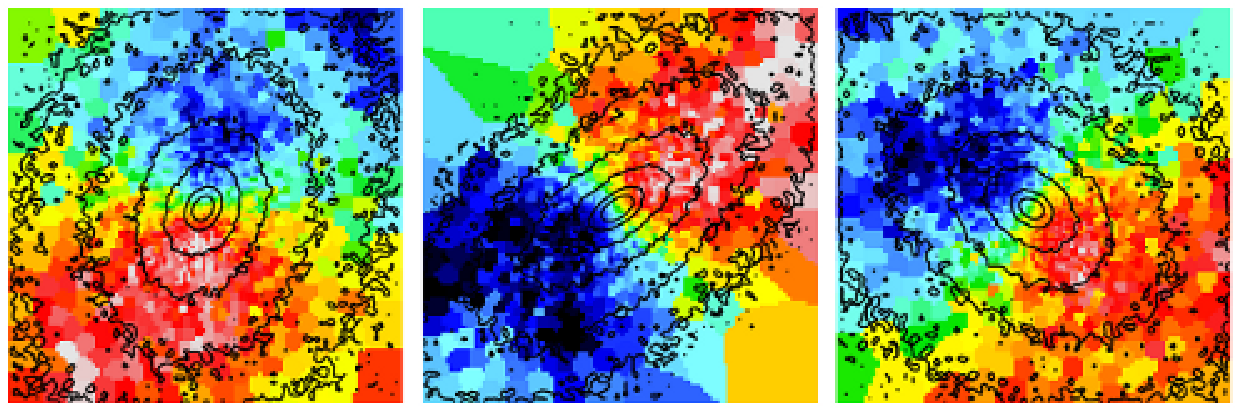, width=0.7\columnwidth} &  & \epsfig{file=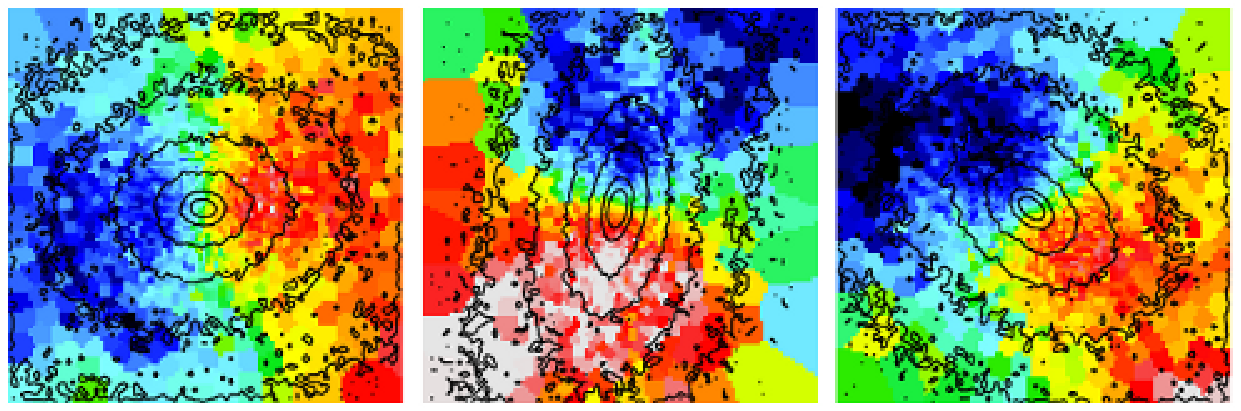, width=0.7\columnwidth} & \low \\
\med & \epsfig{file=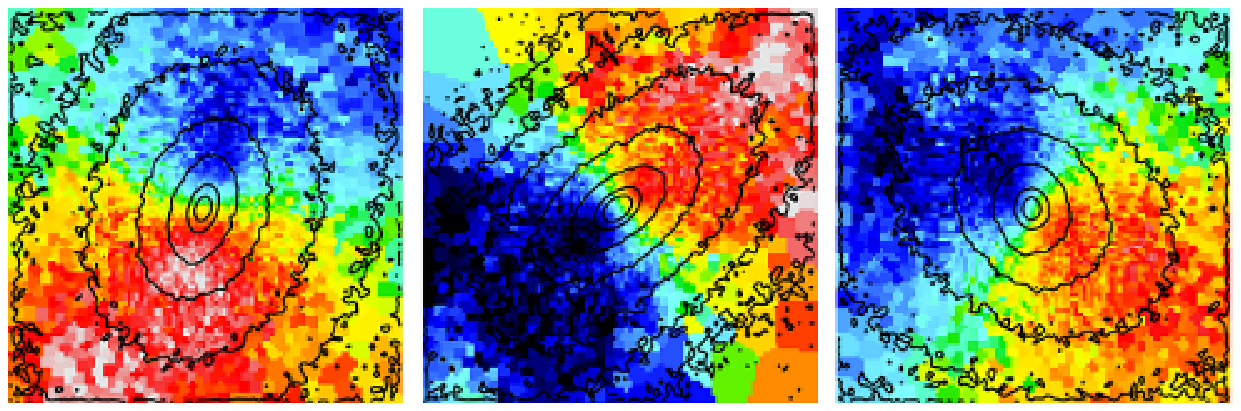, width=0.7\columnwidth} &  & \epsfig{file=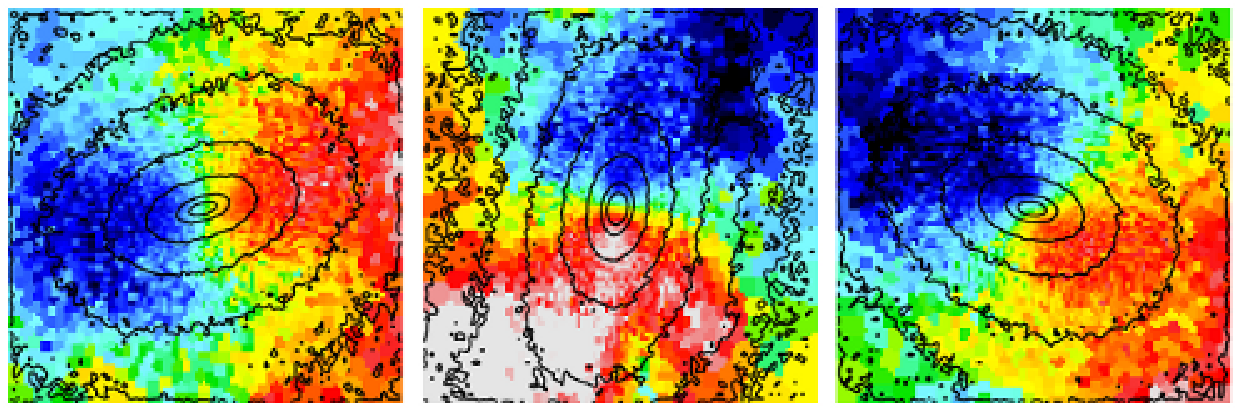, width=0.7\columnwidth} & \med \\
\high & \epsfig{file=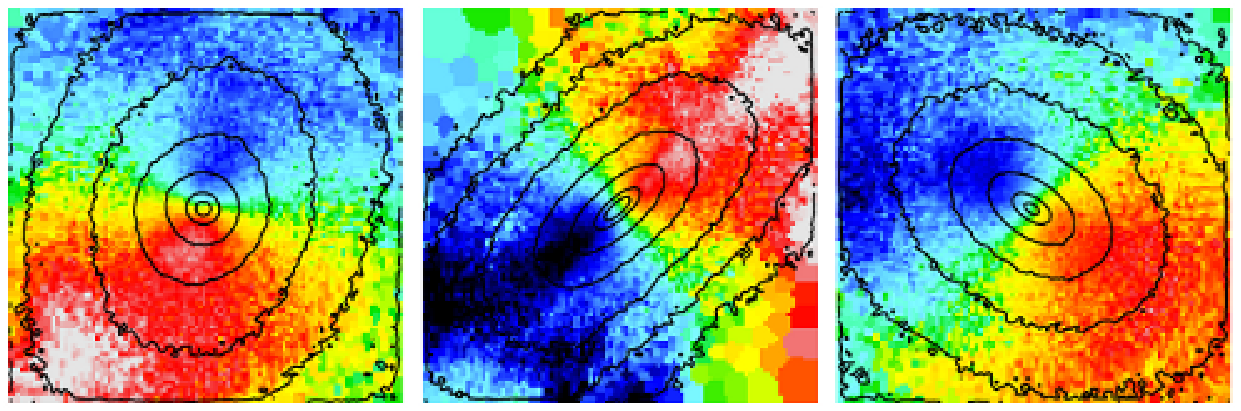, width=0.7\columnwidth} &  & \epsfig{file=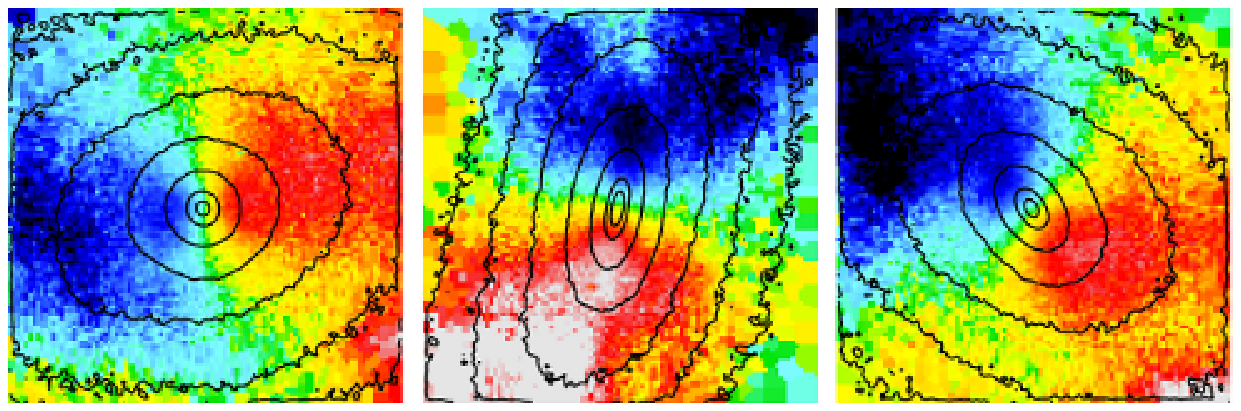, width=0.7\columnwidth} & \high \\
 & & \\
\hline
 & & \\
 & \begin{Large}Dry-Slow \end{Large} & & \begin{Large}Wet-Slow \end{Large} & \\
 & $\epsilon_{min}$ \hspace{1.2cm} $\epsilon_{max}$ \hspace{1.2cm} $\epsilon_{mean}$ & & $\epsilon_{min}$ \hspace{1.2cm} $\epsilon_{max}$ \hspace{1.2cm} $\epsilon_{mean}$  &  \\
 & & \multirow{4}{*}{\epsfig{file=colorbar2.eps, width=0.2\columnwidth}} & & \\
 \low & \epsfig{file=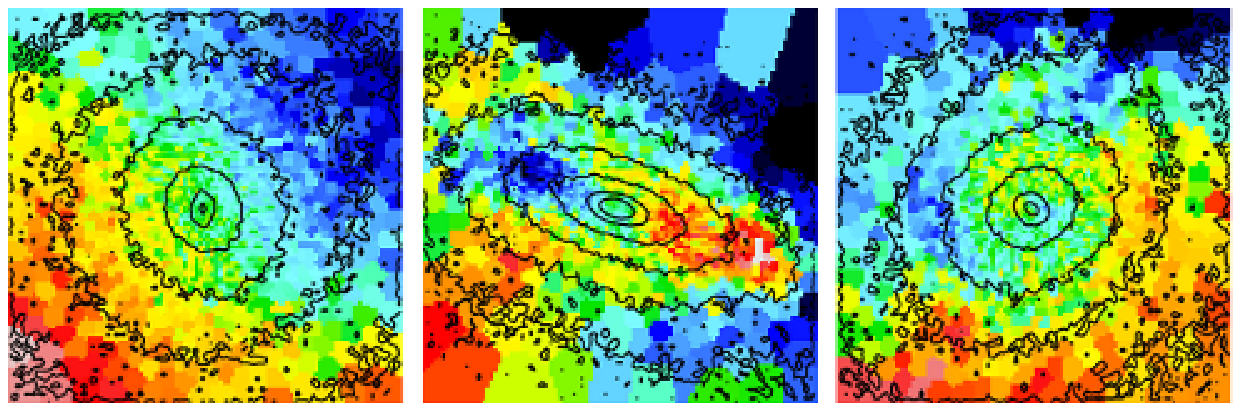, width=0.7\columnwidth} &  & \epsfig{file=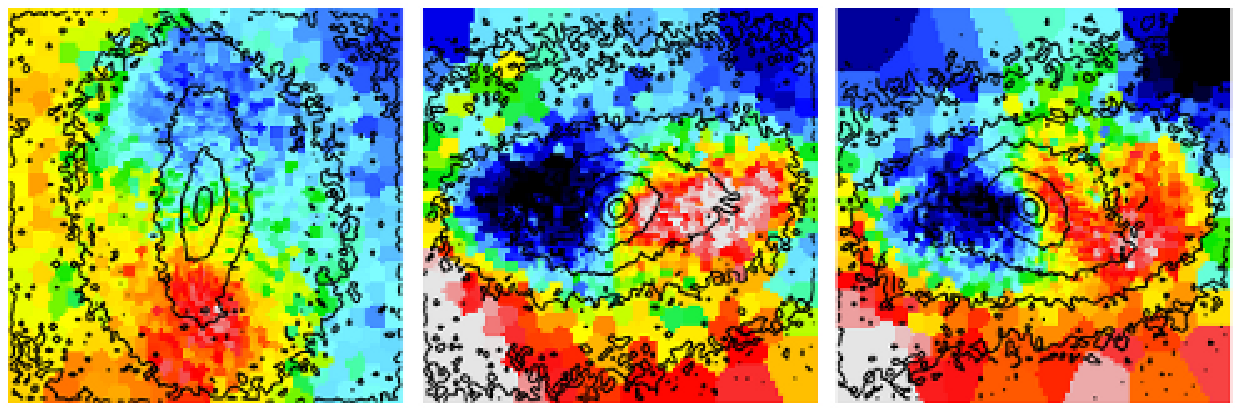, width=0.7\columnwidth} & \low \\
\med & \epsfig{file=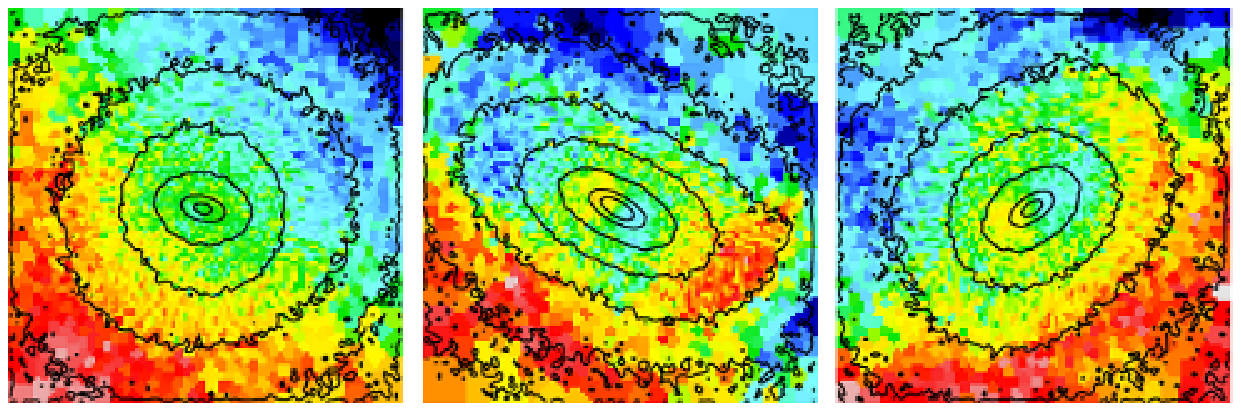, width=0.7\columnwidth} &  & \epsfig{file=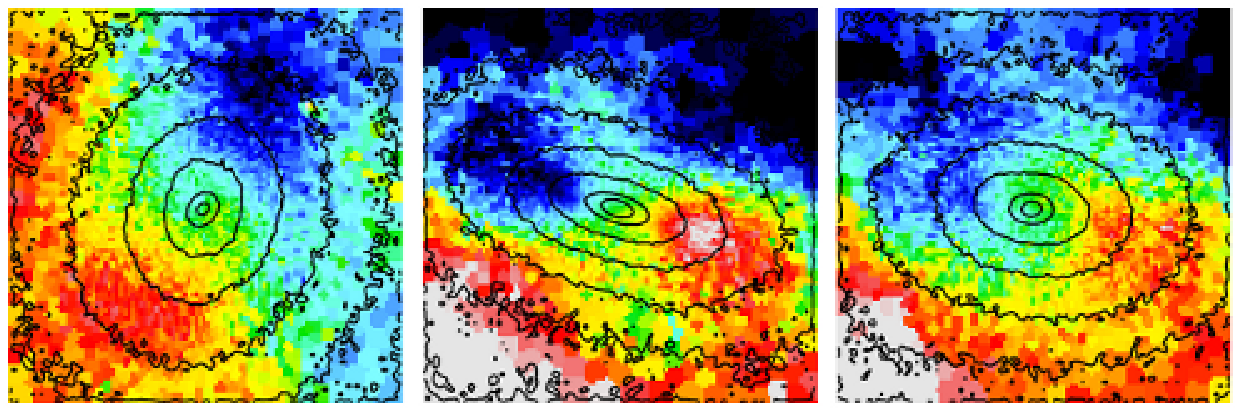, width=0.7\columnwidth} & \med \\
\high & \epsfig{file=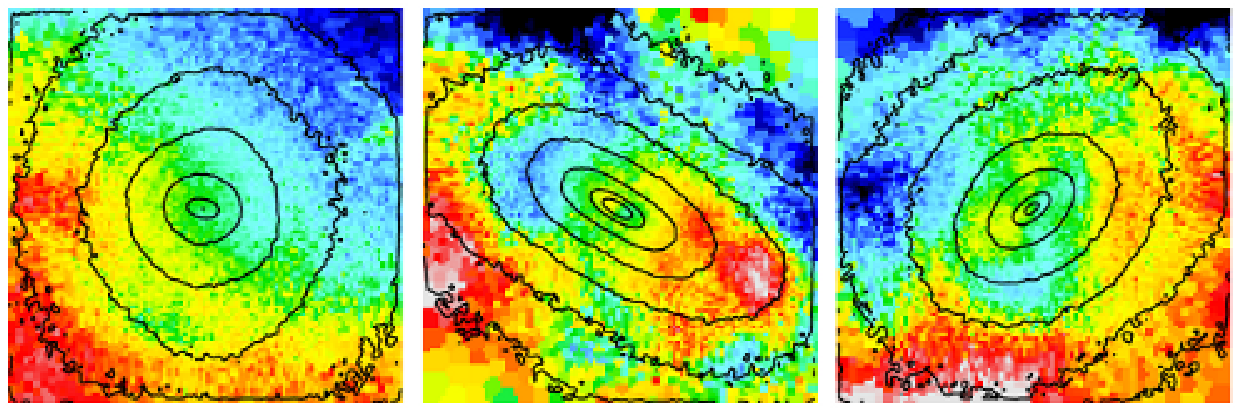, width=0.7\columnwidth} &  &  \epsfig{file=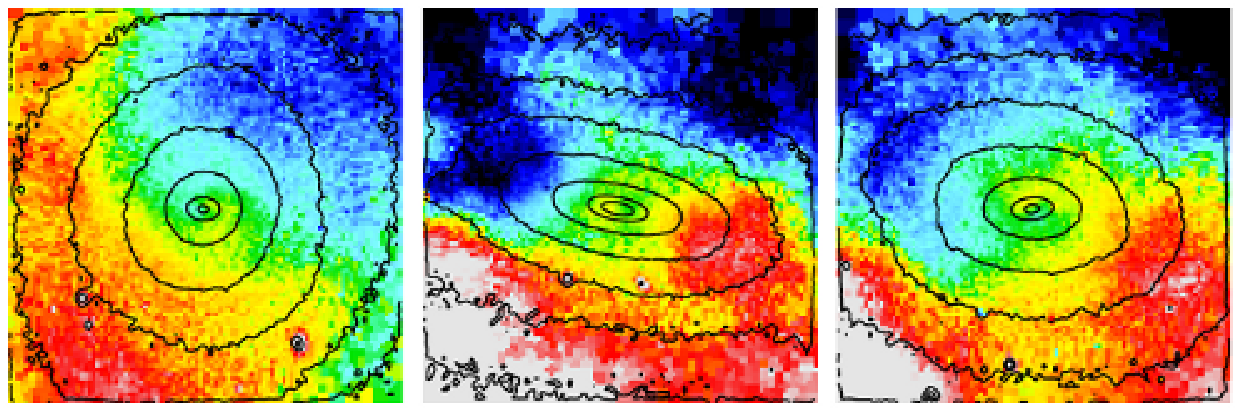, width=0.7\columnwidth}  & \high \\
  \end{tabular}
  \caption{The twelve projected stellar velocity fields. The field of view is $16\times16$~kpc$^2$, projections and contours are the same as in Figure~\ref{fig:phot}.}
  \label{fig:vel}
\end{center}
\end{figure*}

   \subsection{Morphology and kinematics\label{sec:IVmaps}}

Radial stellar density profiles are shown in Figure~\ref{fig:densprof}. We then show in Figure~\ref{fig:phot} the projected stellar density maps, of the relaxed merger remnants in all simulations, choosing the flattest and roundest projections as well as a projection representative of the mean ellipticity in each case. The corresponding line-of-sight stellar velocity fields are presented in Figure~\ref{fig:vel} for the same projections, the maps have been Voronoi binned \citep{CC03} to the same level of 15 particles minimum per bin. Further morphological or kinematics parameters are presented in Appendix~A.

This analysis reveals various similarities or differences, depending on which merger is considered. The most noticeable results are as follows:
\begin{itemize}
   \item {\em Mergers that produce \fasts{} at the highest resolution also result in fast rotating systems at the lower, standard resolution}. Overall, the apparent morphology for any projection of the Dry-Fast and Wet-Fast models is unaffected by the resolution (Figure~\ref{fig:phot}). The velocity fields are also quite similar (Figure~\ref{fig:vel}), with only minor misalignments between the apparent kinematic and photometric axes. Ellipticity and $\lambda_R$ profiles, provided in Appendix~A (see Figure~\ref{figA:ella4} and \ref{figA:vslr}), confirm these similarities and that all these mergers remnants are fast rotators, with a rotational support that is largely independent from the numerical resolution.
\item {\em Strong kinematic misalignments and kinematically decoupled cores (KDCs) are found only in slow-rotators, but really appear only at high resolution}. The Dry-Slow model has a KDC at standard resolution, but its amplitude is significantly lower than the one observed in the high and highest resolution models. The Wet-Slow model has a KDC only at high/very high resolution. Overall, kinematic misalignments increase at high resolution, as illustrated for instance by the flattest projections of the Wet-Slow case.
\item {\em Morphological and kinematic differences are most important for mergers that produce \slows{} at high resolution.} Striking morphological differences are seen in particular for the Wet-Slow case (Figures~\ref{fig:densprof} and \ref{fig:phot}) and both the amplitude and the shape of the velocity field change with resolution for the Wet-Slow and Dry-Slow cases (Figure~\ref{fig:vel}). For instance a rapidly rotating core is seen in the Wet-Slow merger remnant at standard-resolution, instead of a slow-rotating KDC at high- and very high-resolutions. The Dry-Slow remnant also shows up as a discy rotating system at standard resolution, in contrast with the observed remnant at higher resolutions. We also note on Figure~\ref{fig:densprof} that the stellar density profile is resolution-dependent in particular for the Wet-Slow case, with a much less concentrated merger remnant in the standard-resolution case (the mass within 5~kpc is about 25\% lower than at high or very high resolutions).
\end{itemize}

   \subsection{Formation of \slows{} at high resolution \label{sec:wetslow}}

We now focus on the detailed properties of the mergers for which the most important differences have been noticed, namely those producing slow rotators at the highest resolutions.

      \subsubsection{Morphology and Kinematics \label{sec:componants}}
\begin{figure}
  \hspace{1.0cm} Old stars \hspace{1.3cm} Young stars \hspace{1.7cm} Gas    \\
  \epsfig{file=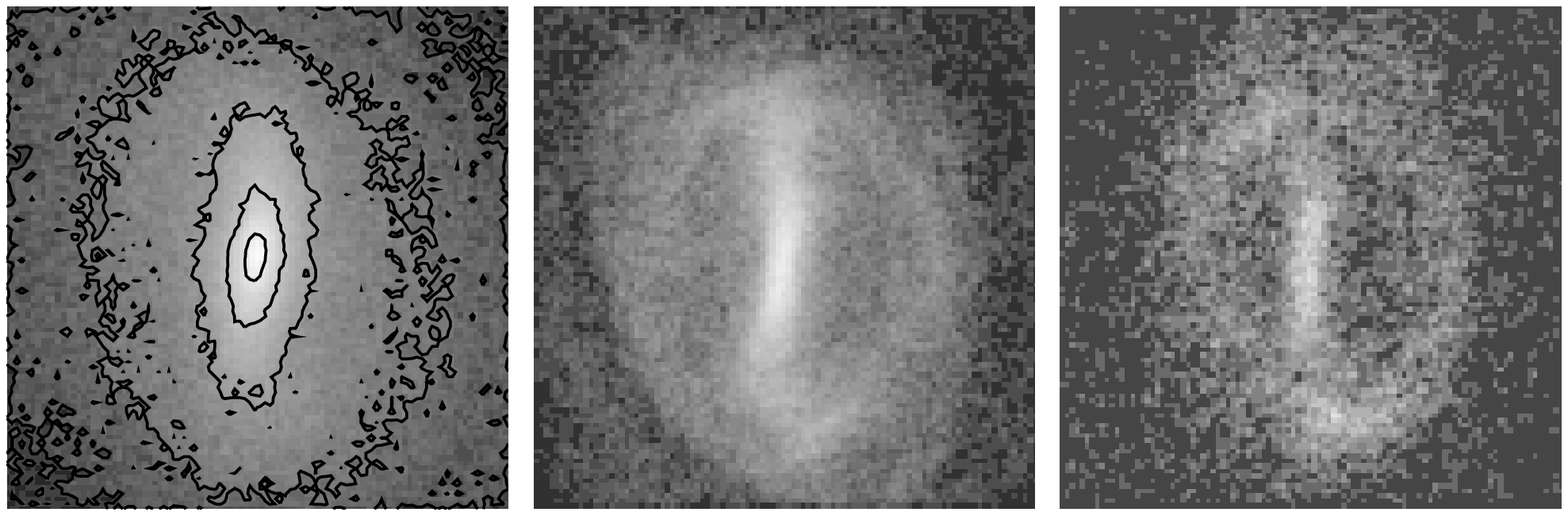, width=\columnwidth}   \\
  \epsfig{file=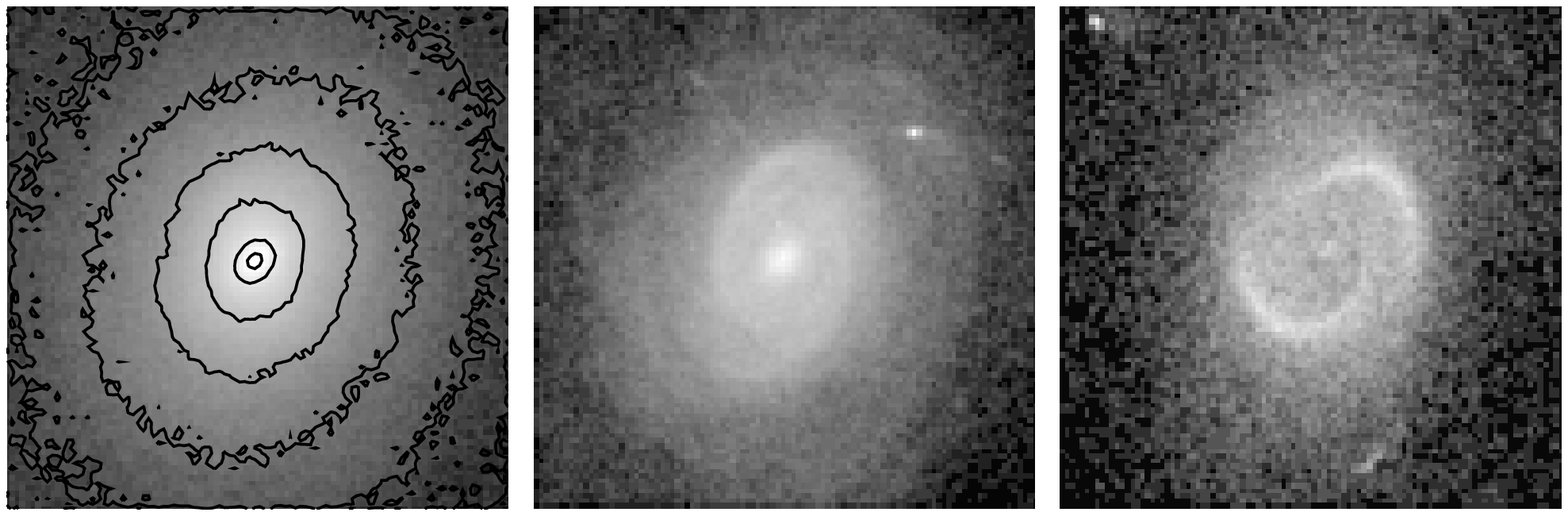, width=\columnwidth}   \\
  \epsfig{file=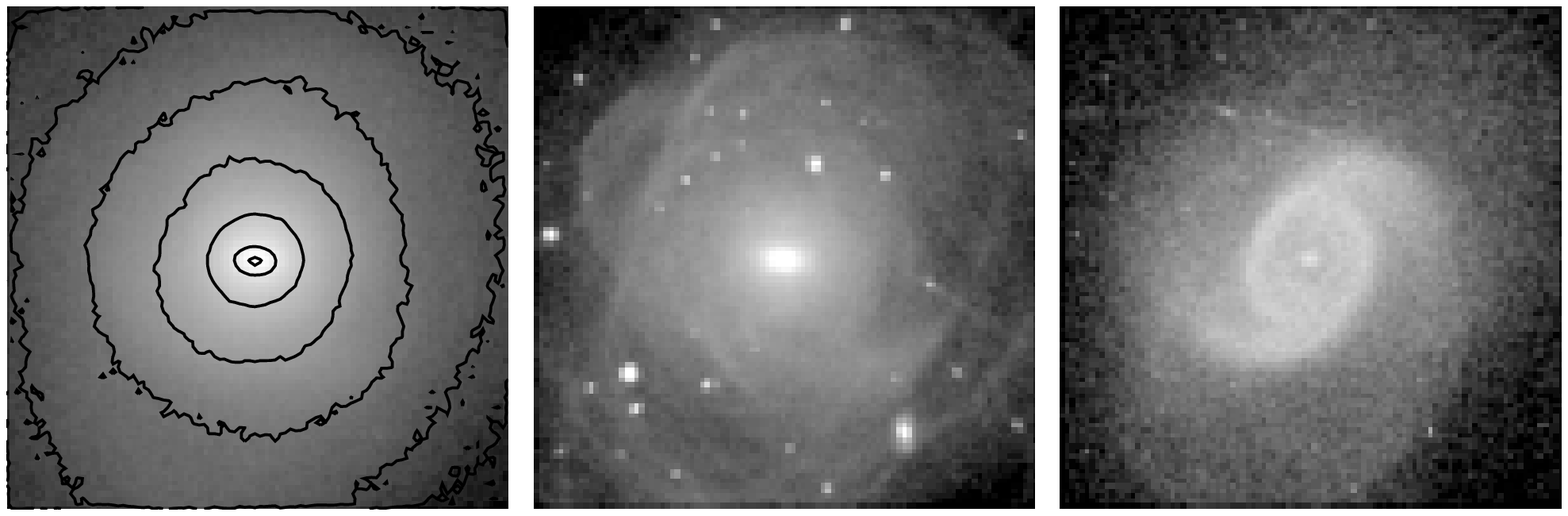, width=\columnwidth}   \\
  \caption{Projected density maps of the old stars, young stars, and gas in the wet-slow remnants; from top to bottom : {\em standard-}, {\em high-} and \high{} models, for the projection which minimises the ellipticity ($\epsilon_{min}$) as in Figure~\ref{fig:phot}. Old stars are those formed before the merger, young stars are formed during/after the merger. The field of view is $16\times16$~kpc$^2$, and the isocontours correspond to the projected old stellar component.}
  \label{fig:component}
\end{figure}

\begin{figure}
  \epsfig{file=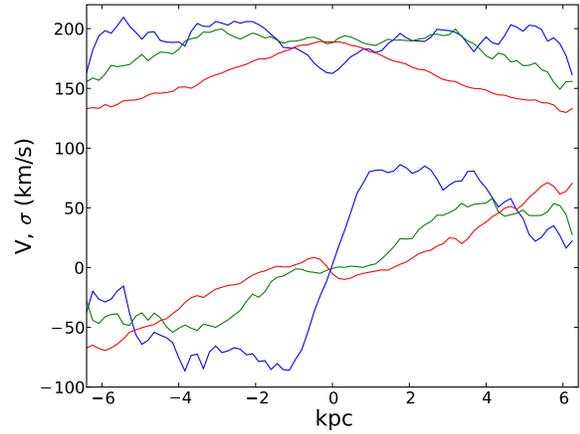, width=\columnwidth} \\
  \epsfig{file=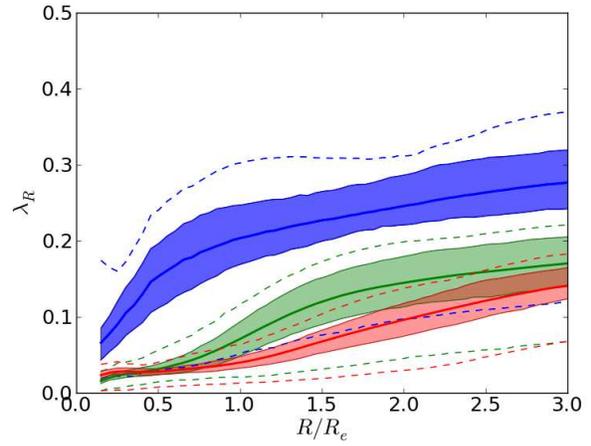, width=\columnwidth}
  \caption{{\bf Top panel} Radial line-of-sight velocity and velocity dispersion profiles (respectively bottom and top lines of the plot) for the mean ellipticity projection along the global kinematic position angle of the Wet-Slow simulation (right panels of Figure\ref{fig:vel}). {\bf Bottom panel} $\lambda_R$ profiles as a function of $R/R_e$, the minimum, median, maximum and quartiles values are presented as in Figure~\ref{fig:projlr}. In both panels, the \low{} is represented in blue, the \med{} in green, the \high{} in red.}
  \label{fig:wetslowvslr}
\end{figure}

\begin{figure}
  \epsfig{file=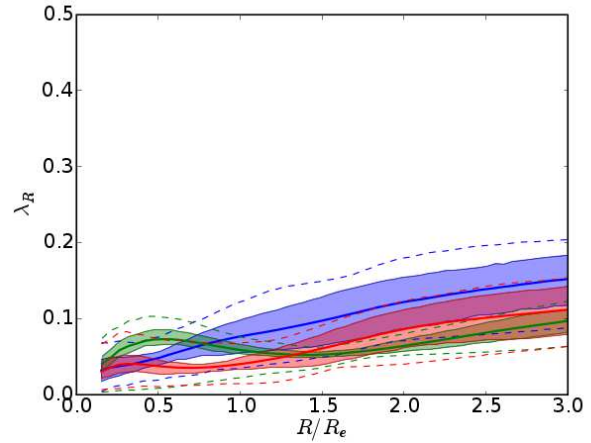, width=\columnwidth}
  \caption{$\lambda_R$ profiles as a function of $R/R_e$ for the Dry-Slow simulation. The \low{} is represented in blue, the \med{} in green, the \high{} in red.}
  \label{fig:dryslowlr}
\end{figure}

To better understand the differences seen in the morphology of the Wet-Slow simulations, we have examined the three included baryonic components of the merger remnants separately, namely the ``old'' stars formed before the beginning of the merger event, the ``young'' stars formed during/after the merger event, and the gas left over after the merger (see Figure~\ref{fig:component}). Within the central 10~kpc, the \low{} remnant exhibits a prominent bar, the inner distribution of the gas and young stars being driven by this tumbling structure with e.g., a ring-like structure at a radius of $\sim$ 6~kpc. In the \med{} and \high, the gaseous component and the young stars have a smoother distribution more closely following the overall old stellar distribution. In addition, many young star clusters are visible in the maps from the \high, a few in the \med, and none in the \low. High spatial resolution of course allows to resolve the formation of stellar clusters \citep[see also][]{B08} but there is also a larger number of other young stellar substructures at increasing resolution, like filaments, tidal streams, and a compact nucleus (Figure ~\ref{fig:component}).

The kinematic discrepancies discussed above in the velocity fields are quantified globally in the radial velocity and $\lambda_R$ profiles (Figure ~\ref{fig:wetslowvslr}). The \low{} displays significant rotation inside 3-4~kpc (up to $\sim85$~km\ s$^{-1}$) and a decreasing rotation velocity at larger radii. There is a drop in the velocity dispersion in the central 2~kpc, and no sign of a KDC. This is in stark contrast with both the \med{} and \high{} which overall show much lower rotational velocity support (below $\sim 50$~km\ s$^{-1}$ and particularly low in the central 2~kpc), and a KDC in the central 1~kpc. Overall the \med{} and \high{} have similar velocity rotation curves, apart from a more pronounced KDC signature in the \high{} (partly due to the KDC having a slightly different position angle in these two remnants).

The general discrepancies of the \low{} versus \med{} and \high{} realisations are confirmed by the $\lambda_R$ profiles (Figure~\ref{fig:wetslowvslr}). The merger remnant made at \low{} is clearly a \fast{}. The \med{} and \high{} are both classified as \slows{} with respectively a maximum value of $\lambda_R$ of 0.1 and 0.06 at one effective radius. The $\lambda_R$ profile goes up somewhat more rapidly with radius in the \med{} case than in the \high{}, but the difference remains of the order of the scatter between different projections of each case. The presence of a bar in the stellar component of the \low{} is likely a result of the significantly higher rotational support (see also Sect.\ref{sec:systematic}).

Beyond one $R_e$, the $\lambda_R$ profiles of the \med{} and \high{} are rising: there is less angular momentum in the center, which has been expelled outwards (see also E07). However, even at these large radii, the \slows{} have less angular momentum than \fasts{} (see Figure~\ref{fig:wetslowvslr}). Observations conducted up to two or three R$_e$ \citep{Weijmans09, Coccato09} would bring additional constraints on the formation scenario of slow-rotating early-type galaxies.
\medskip

\subsubsection{Role of gas on the properties of merger remnants \label{sec:rolegas}}

The Dry-Slow simulations show smaller differences in the stellar density maps and velocity fields. They also exhibit smaller differences in their $\lambda_R$ profiles (Figure~\ref{fig:dryslowlr}). Nevertheless, the \low{} simulation is again a faster-rotator than the \med{} and \high{} cases at 1, 2 or 3 effective radii. A KDC is also found only in the \med{} and \high{} cases, associated to a peak of $\lambda_R$ inside one effective radius. 

A lower specific angular momentum in the main stellar body at higher resolution is not only found in Wet-Slow mergers, but also in Dry-slow mergers, the differences being still much more pronounced in the Wet case.

Gas plays an important role in shaping merger remnants \citep[][]{NJB06,rob06,H09} and it is interesting to compare the Wet-Slow and Dry-Slow merger remnants at fixed resolution, to better understand its specific impact (Figures~\ref{fig:wetslowvslr} and \ref{fig:dryslowlr}):
\begin{itemize}
\item at \low{}, the wet merger remnant has a much higher rotational support than the dry case. This is consistent with the usually known effect of gas helping the survival of rotating stellar discs during major mergers, and/or re-building of discs after mergers (Robertson et al. 2006, Hopkins 2009b).
\item at \med{} and \high{}, the rotational support of the merger remnant is not increased when gas is present. The angular momentum, traced by $\lambda_R$, is actually lower by about 20\% inside one effective radius in the \high{} wet case, compared to the corresponding dry merger.
\end{itemize}

It thus seems that the impact of gas on the global properties of major merger remnants is more complex than originally thought, and can even be weakened at high resolution. This suggests that the global dynamics of gas during the major merger or in a young merger remnant can be significantly affected by resolution. As seen in Figure~\ref{fig:component}, gas at \low{} largely lies in smooth structures and the formation of new stars during the merger proceeds in a relatively smooth way. At increased resolutions, thinner gas structures are resolved during the merger, which can result in clustered star formation and the formation of numerous young stellar structures, as observed in the final merger remnant in Figure~\ref{fig:component}.

\subsection{Summary of the resolution tests \label{se:summary}}

The resolution does not seem to significantly affect the morphology and kinematics of the mergers remnants that are fast rotators at high resolution: they are still fast rotators at lower resolution, with very similar morphological and kinematic properties. This contrasts with the fact that resolution has a major effect on the formation of slow-rotating systems. The systems that are slow rotators at high resolution rotate more rapidly when the resolution decreases, and can be observed as true fast rotators at \low{}. The effect is small in dry mergers, but is dramatic in our wet merger model. KDCs in these slowly-rotating systems are also significantly better resolved at high resolution. The role of gas in shaping merger remnants is found to vary with resolution: at low resolution, gas rebuilds rotating disc components, increasing the overall disciness and rotational support. At higher resolution, the effect cancels out: a merger that forms a slowly-rotating system in a dry case still forms an equally slow-, or even a bit slower-rotator in the corresponding wet case.

The next section focusses on interpreting the origin of the resolution effect in the formation of slowly-rotating ellipticals. We in particular show that it is not an artifact caused by different initial conditions or a bias in the simulated orbits, but a real effect related to the way the violent relaxation during the merger itself is treated.

\section{Origin of the resolution effect}
\label{sec:origins}

We here show that the above-mentioned discrepancies observed in the simulations that produce slow-rotating ellipticals at high resolution, are really attributable to the physical modeling of the merging process. They are not artefacts related to initial conditions of the progenitor galaxies and/or interaction orbits that would vary with the resolution.

   \subsection{The progenitor galaxies \label{sec:progenitors}}
\begin{figure}
  \centering
  \epsfig{file=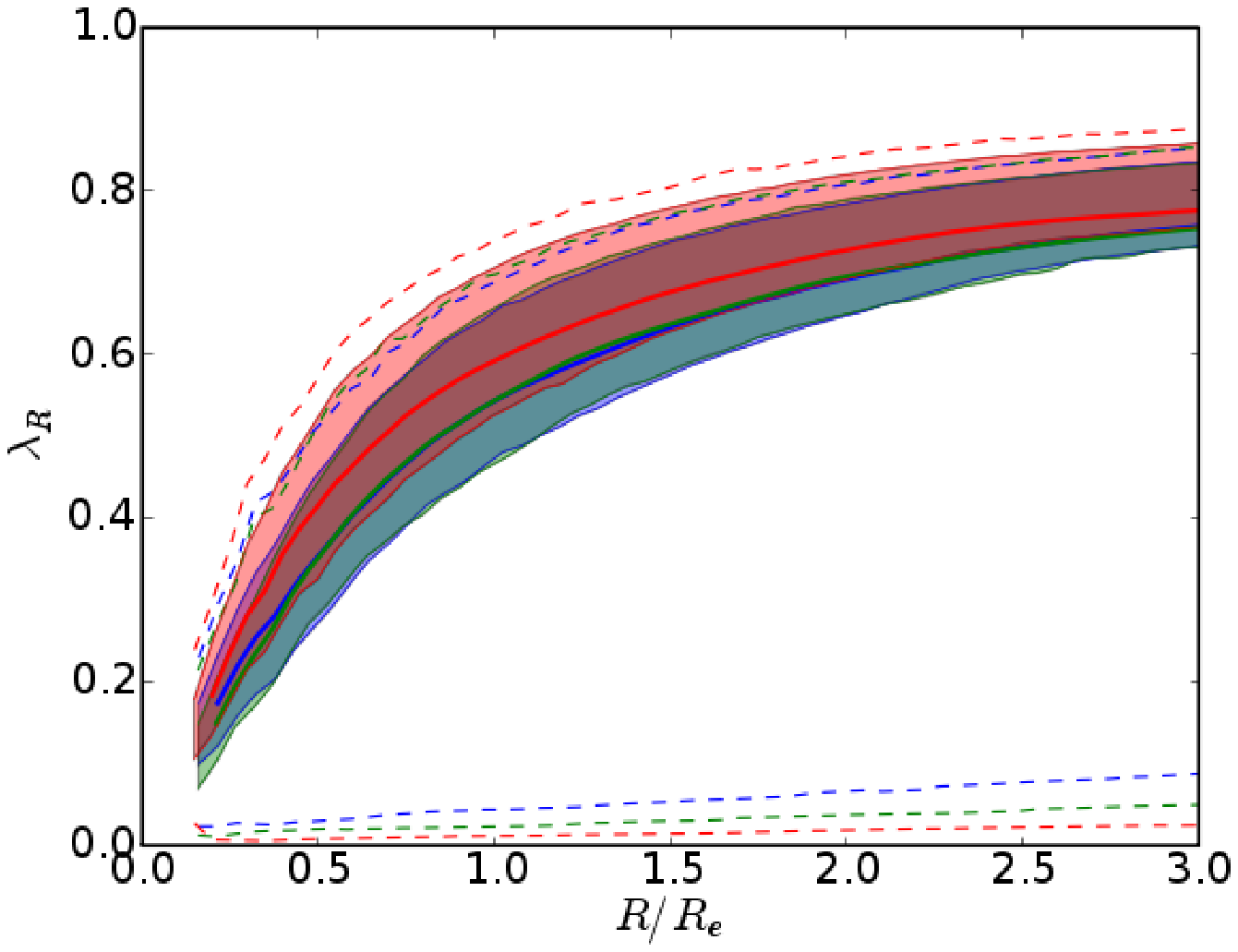, width=\columnwidth} \\
  \epsfig{file=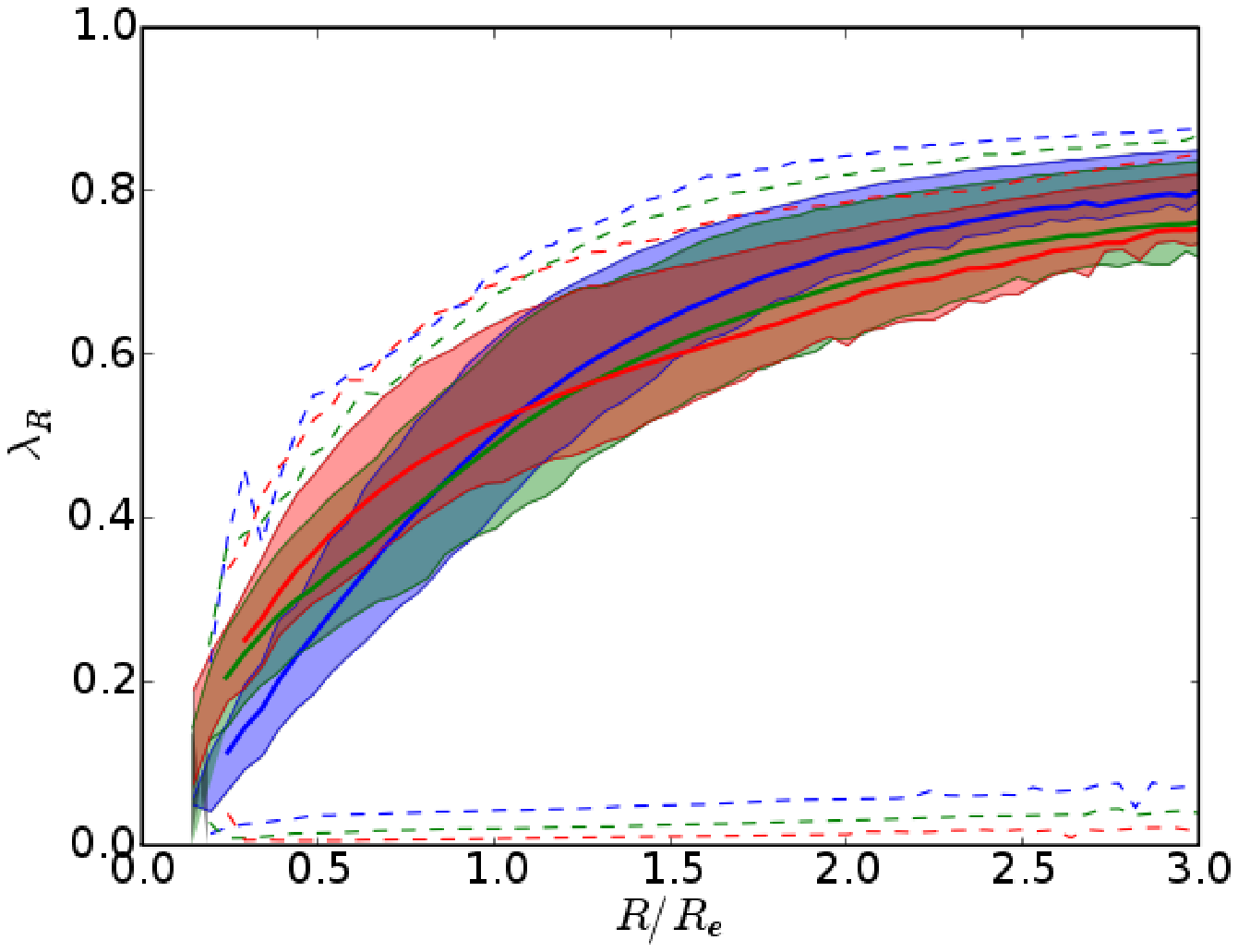, width=\columnwidth}
  \caption{$\lambda_R$ profiles of the two Wet progenitors as a function of $R/R_e$. Colours, as in previous Figures, with the \low{} in blue, the \med{} in green, the \high{} in red.}
  \label{fig:lrprog}
\end{figure}

We first check that the progenitors galaxies are similar at any resolution. To this aim, we analyze their kinematic properties, in particular the $\lambda_R$ profiles --  $\epsilon$ and $a_4/a$ parameters are less relevant for disc-dominated galaxies. Since the merger simulations were performed after an isolated relaxation of each progenitor galaxy (see Sect.~\ref{sec:nbody}), we analyzed the progenitors from a snapshot right after this relaxation period, so that the results (Figure~\ref{fig:lrprog}) are representative of the conditions under which the mergers occur.

The two progenitor galaxies have quite similar angular momentum profiles (Figure~\ref{fig:lrprog}). There are some fluctuations, but they are not systematically corresponding to an increase or decrease of $\lambda_R$ with resolution. They are also weaker than the discrepancies found in the final merger remnants. Actually, they result for a large part from the effective radius changing slightly with the resolution, and profiles of $\lambda_R$ as a function of the absolute radius (in kpc) show smaller differences than the profiles in units of the effective radius. These fluctuations cannot therefore be the main cause for the observed resolution effects in the merger remnants.

   \subsection{Interaction orbits \label{sec:galorbits}}
\begin{figure}
  \centering
  \epsfig{file=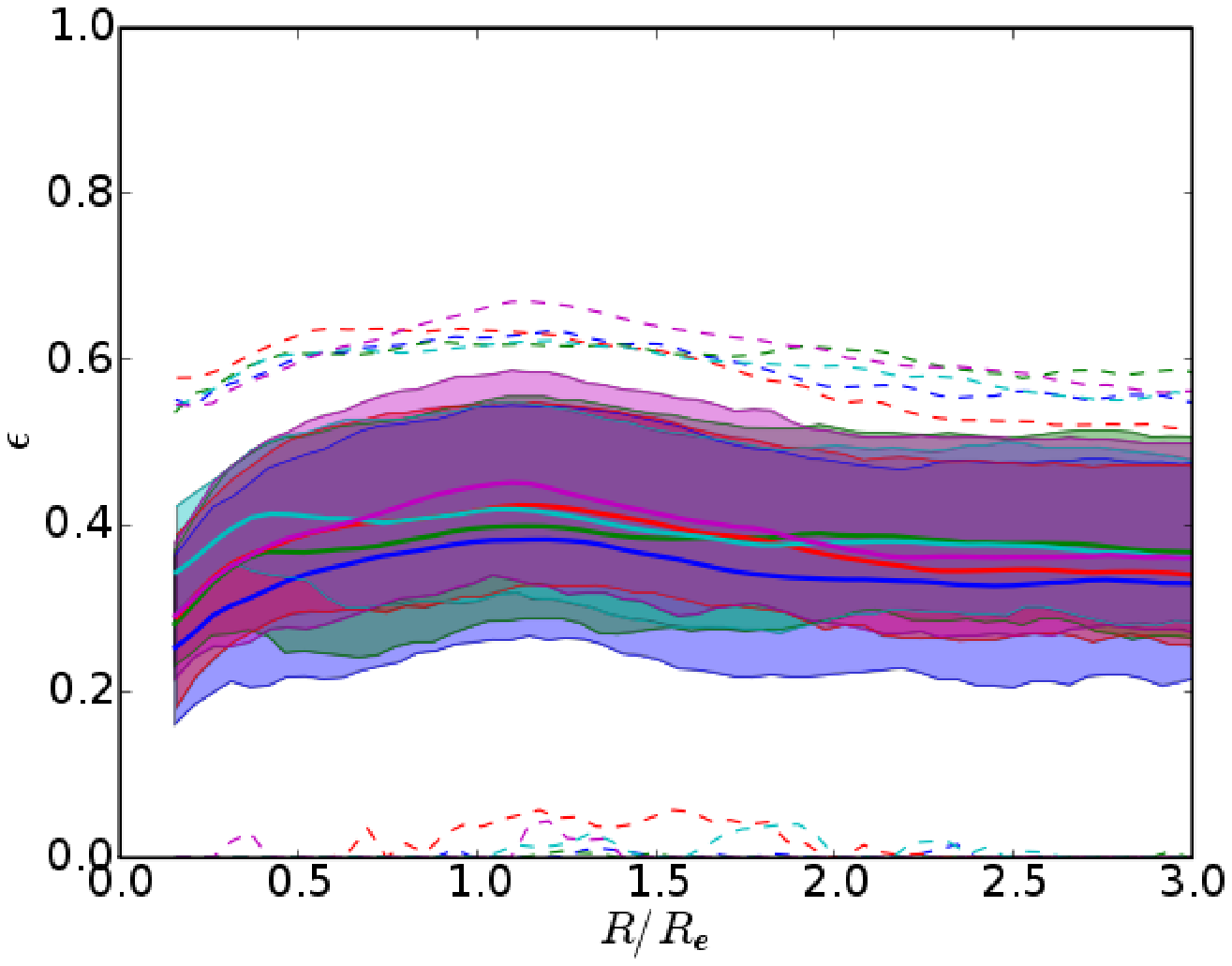, width=\columnwidth} \\
  \epsfig{file=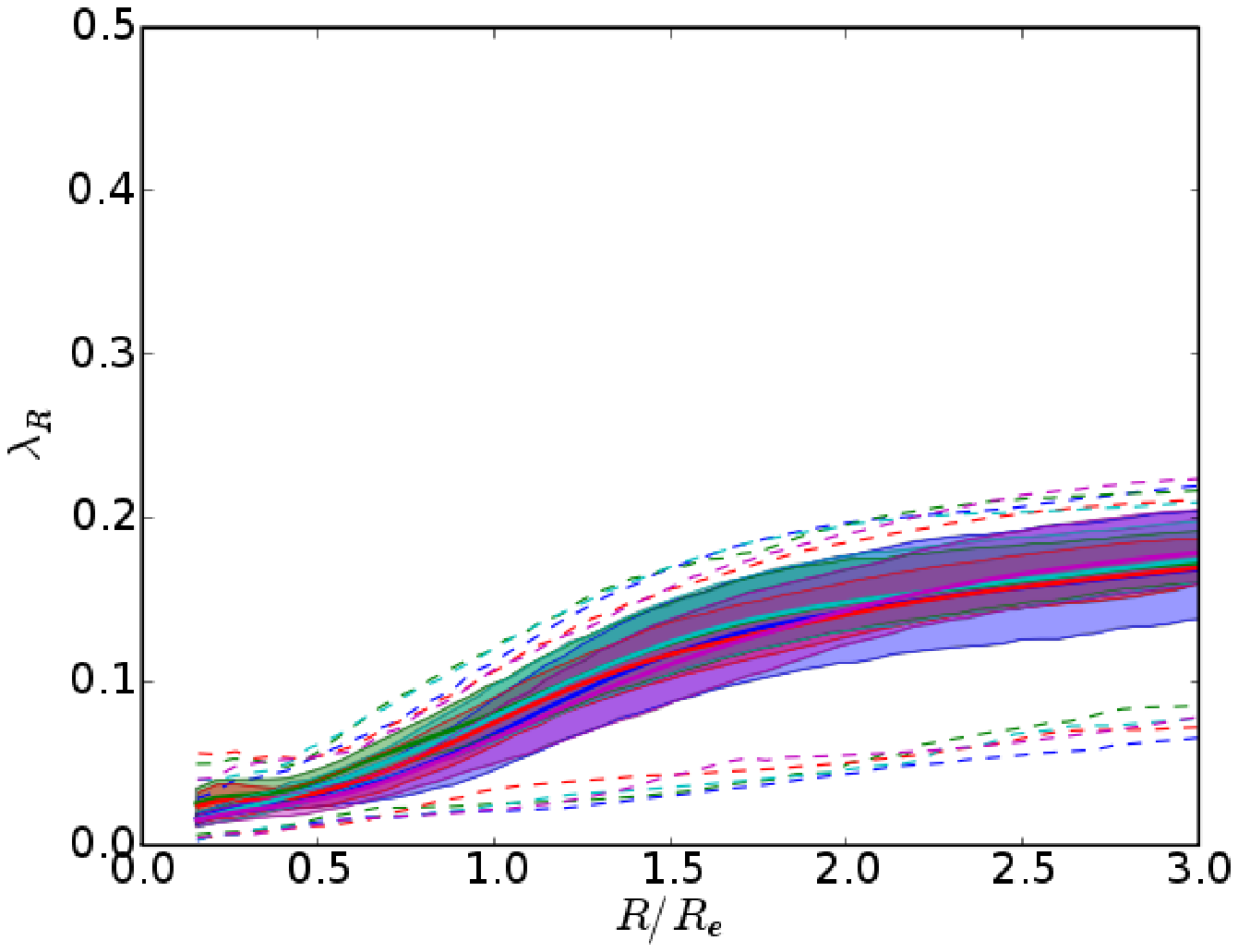, width=\columnwidth}
  \caption{$\epsilon$ (top) and $\lambda_R$ (bottom) profiles in function of $R/R_e$ for 5 slightly different orbits for the Wet-Slow-High simulation.}
  \label{fig:orbites}
\end{figure}

Simulations at the three resolutions are started with the same relative position, velocity and inclination for the two interacting progenitors. However, varying the resolution may result in slight differences in dynamical friction and angular momentum exchanges, if these processes are resolved differently, and the interaction orbits might diverge before the merger actually takes place. If this were the case, our results would be attributable to different orbits rather than different treatments of the merging process itself.

We found that the positions at the first pericentric passage vary by 2.1~kpc on average, and the velocities by 9~km~s$^{-1}$. Although these differences seem small and no systematic variation with resolution appeared, we further investigated their potential effect. To this aim we performed four new realizations for the Wet-Slow model at \med{}, with variations of the position or the velocity twice larger than the average values above (i.e., $\pm$3.6~kpc and $\pm$18~km~s$^{-1}$, respectively). 
The results are shown in Figure~\ref{fig:orbites} for the morphological and kinematic profiles of $\epsilon$ and $\lambda_R$. Changes are minor and differences arising in the interaction orbits cannot explain the variation of the results with resolution.

   \subsection{Robustness of the Wet-Slow-Standard simulation \label{sec:crash}}

   As the resolution effect found in the formation of slow rotators, in particular in the Wet-Slow model, cannot be attributed to a change in the initial conditions and interaction orbit, it likely relates to the physical treatment of the merging process itself. Nevertheless we wanted to check whether or not this could still be attributed to the presence of particle noise, which is higher in the \low{} cases. 

   The Wet-Slow model at the \low{} shows a strong stellar bar, contrary to the \med{} and \high{} cases. We wanted to make sure that this bar is a robust consequence of the high rotational support of the standard-resolution case, and is not a misinterpreted effect that arised from a particular realization of the particle noise.

\begin{figure}
  \centering
  \epsfig{file=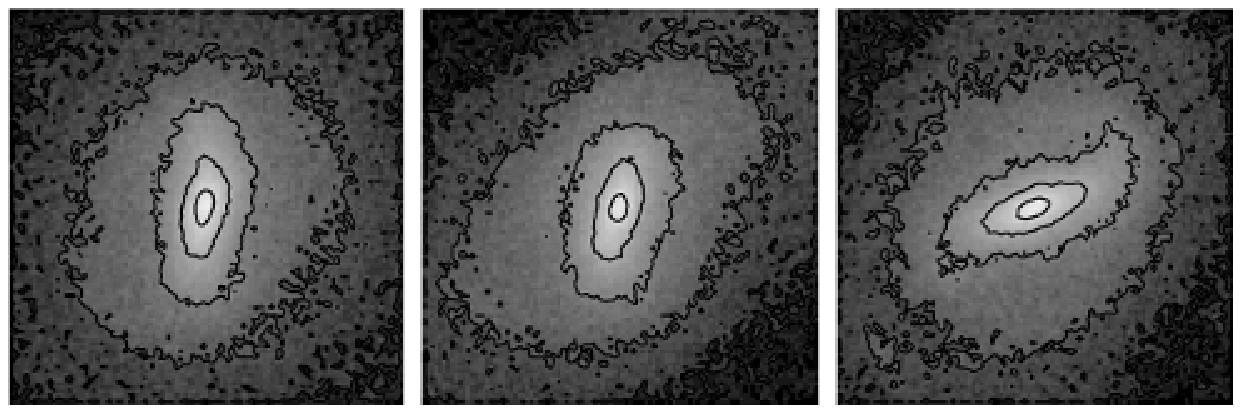, width=\columnwidth} \\
  \epsfig{file=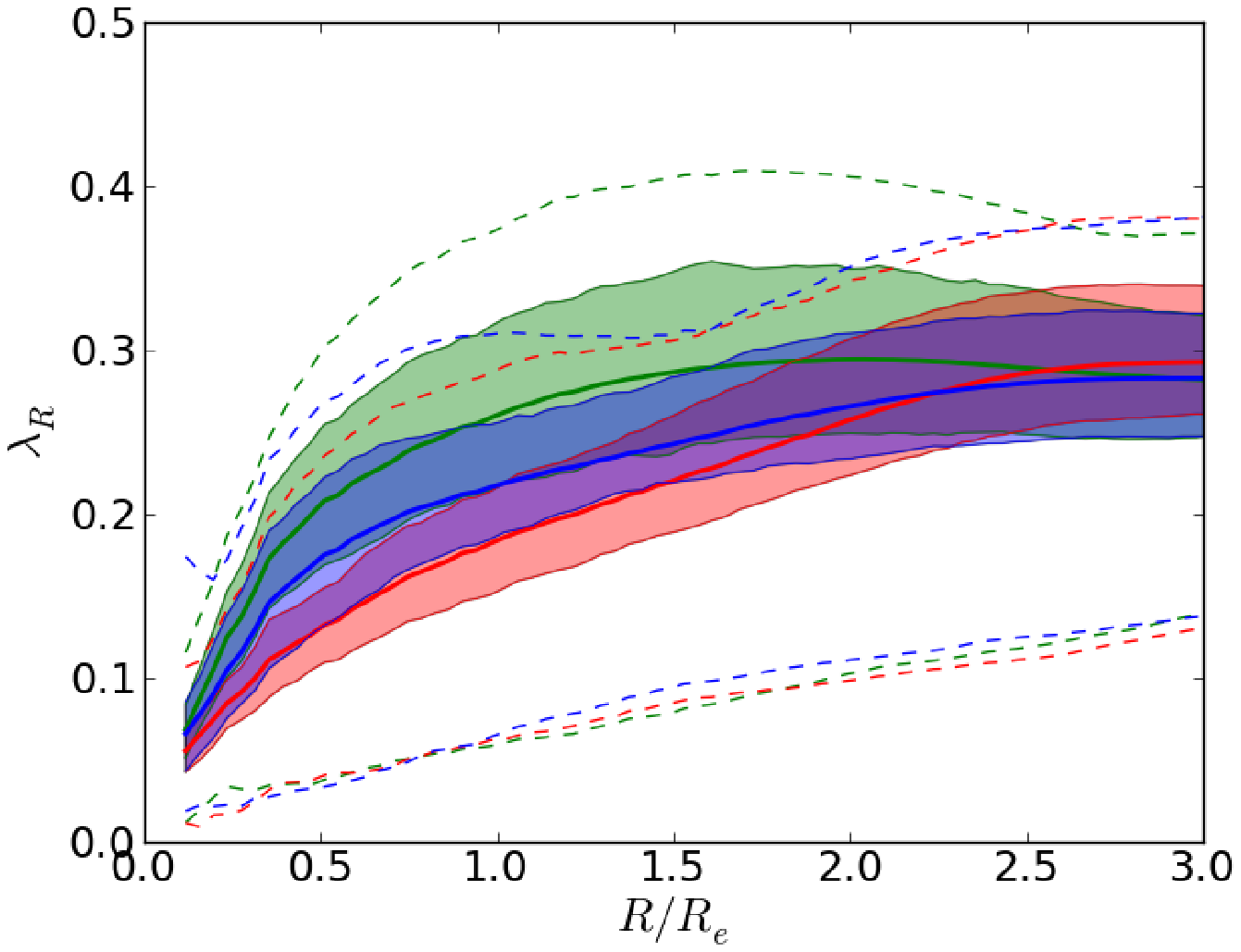, width=\columnwidth}
  \caption{{\bf Top panel} Stellar (old plus young) intensity maps for three different realisations of the Wet-Slow-Standard simulation. The field of view is $16\times16$~kpc$^2$. The simulation used in the study is in the left panel. {\bf Bottom panel} $\lambda_R$ profiles as a function of $R/R_e$ of the three above simulations.}
  \label{fig:crash}
\end{figure}

To this aim, we performed two other Wet-Slow-Standard simulations with the same initial conditions but different, random realizations of the particle noise. The final stellar distribution, shown in Figure~\ref{fig:crash} all show a similar bar, and the $\lambda_R$ profiles are also relatively similar to the original Wet-Slow-Standard model -- there are some variations, but the $\lambda_R$ distributions of the three realizations overlap with each other, and the three systems are equally fast rotators. These two new realisations are also shown in dashed and dotted lines on Figure~\ref{fig:densprof} and \ref{fig:ongoingdphi} and again share common properties with the initial Wet-Slow-Standard model, and hence the same differences compared to the higher-resolution cases.

Thus, the role of bars and  spiral patterns in redistributing the mass and angular momentum in the standard-resolution Wet-Slow model is robust, independent of a particular realization of the particle noise. We also find (see next subsection) that the time variations of the gravitational potential during the interaction and merger are similar for the three \low{} realizations.

   \subsection{The on-going merger \label{sec:ongoing}}

\begin{figure}
  \epsfig{file=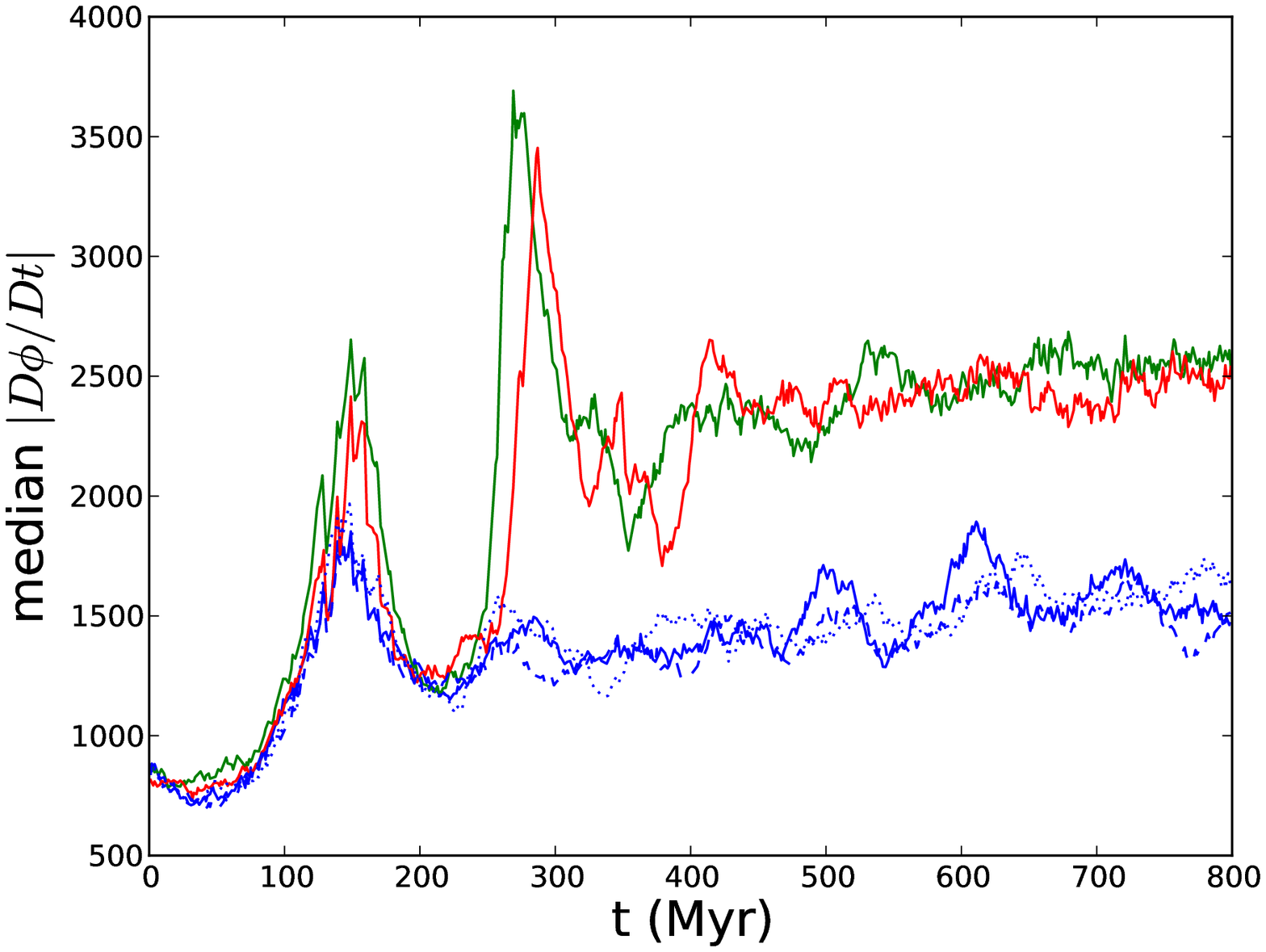, width=\columnwidth} \\
  \epsfig{file=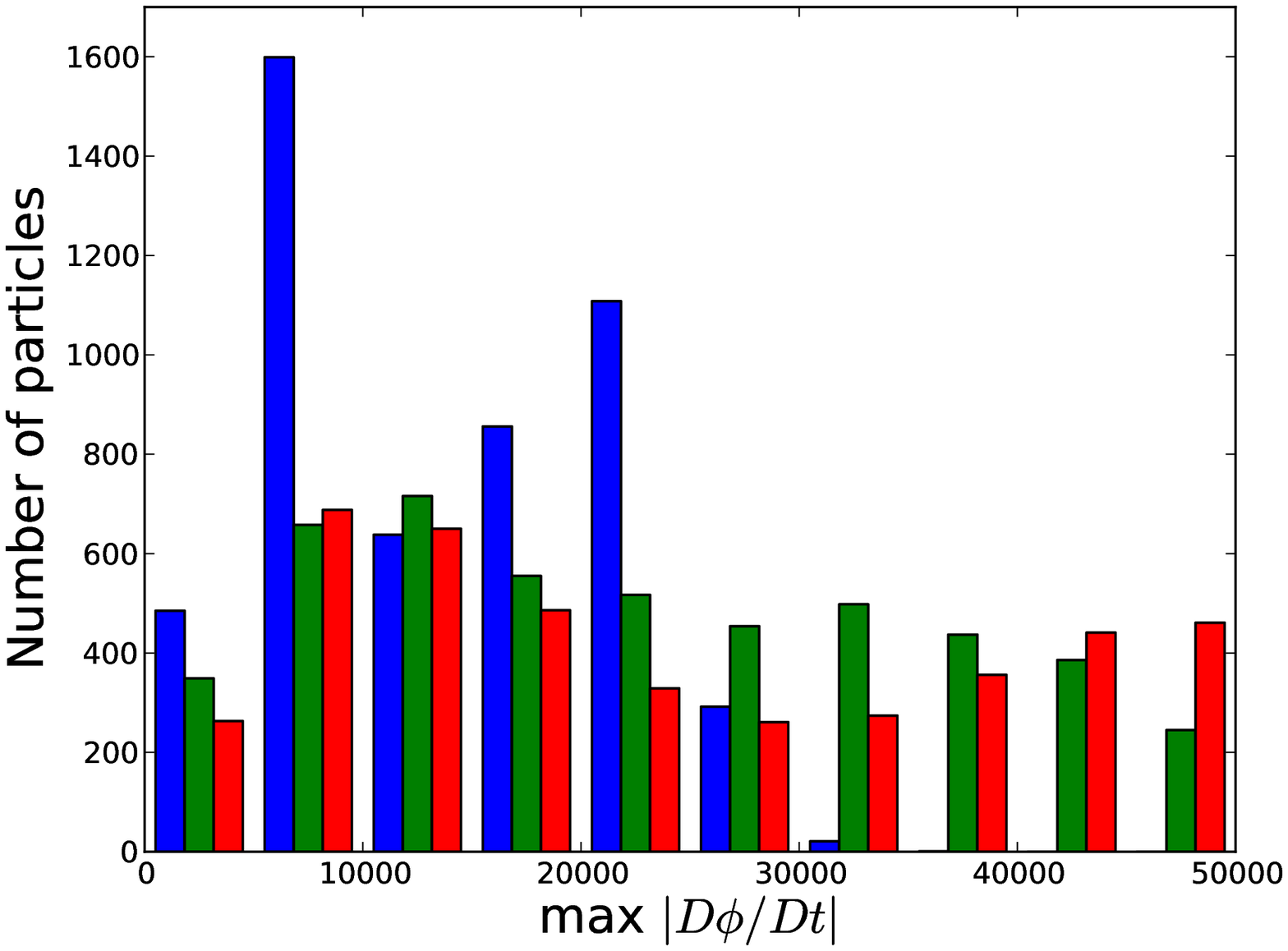, width=\columnwidth}
  \caption{{\bf Top panel} Median variations of the gravitational potential (in arbitrary units) over 5000 test particles as a function of time. The two other realisations of the Wet-Slow-Standard simulation are shown in dashed line. {\bf Top panel} Histogram of the maximum variation of the gravitational potential (in arbitrary units) of each particle over the simulation.}
  \label{fig:ongoingdphi}
\end{figure}

At this point, we have established that the differences observed in the merger remnants do not result from variations in the initial conditions, interaction orbits, or particle noise. The differences should then arise in the physical treatment of the merging process, which would mean that they are ``robust'' effects, potentially affecting any simulation with any numerical code. Varying the spatial and mass resolution could affect the detailed evolution of the dissipative component (including star forming structures), and this could in turn modify the overall orbital structure of the merger remnant \citep[see][]{BH96,NJB06,coxal06}. However we have seen that the resolution effect does not completely disappear in dry mergers. A more general effect can be the treatment of the violent relaxation, i.e. the rapid changes of gravitational potential that are responsible for the evacuation of energy and angular momentum from the main body of the merger remnant -- these quantities being carried away by a low fraction of the mass expulsed at large radii. This process of course plays a more important role in the formation of slow rotators than in the formation of fast rotators. The resolution effects are much more important for slow rotators than fast ones (Sect.~\ref{sec:results}), which suggests that they do actually relate to the violent relaxation process.

To quantify the importance of violent relaxation in our merger simulations, we followed, in the Wet-Slow models, the variations of the gravitational potential $D\phi / Dt$ of 5000 randomly chosen ``test'' particles, all of which are stellar particles existing at $t=0$, all along the simulations. The derivate is Lagrangian, since it follows the motion of each particle. In an isolated galaxy, $D\phi / Dt$ relates to the variation of potential along the orbit of each particle, in particular their radial excursion in the potential well of the galaxy. During the interaction and mergers, peaks of $D \phi /Dt$ should trace the importance of scattering by local density fluctuations through the violent relaxation process.

The top panel of Figure~\ref{fig:ongoingdphi} shows the median value of $| D\phi / Dt |$ as a function of time -- we take the absolute value for each particle, as a particle moving inward or outward can be considered with the same behaviour. Before the merging ({\em i.e} before $\sim$150~Myr) the three resolutions are identical, meaning that there is no difference in the progenitors during the approach phase, modest values of $| D\phi / Dt |$ simply correspond to modest radial excursions of particle in the progenitor disc galaxies. 

After the merger, each simulation shows a relatively constant $| D\phi / Dt |$ in a relaxed system, but the value is higher at \med{} and \high{}, indicating larger radial excursions of stellar particles compared to the \low{} case. More radial orbits are indeed expected for slow rotators compared to the \low{} fast rotator. We note again that the different orbital structure does not only affect the gas and the young stars formed during the mergers, but also the old stars present before the merger itself (see also Figure~\ref{fig:component} and Sect.~\ref{sec:results}). 

During the merging process, a first peak in the median $| D\phi / Dt |$ occurs at the first pericenter passage, after about 150~Myr, but is more pronounced at high(est)-resolution. Another peak is found at the \med{} and \high{} during the final coalescence at $t \sim$ 280~Myr , but is much weaker in the \low{} case. The final coalescence does take place at the same moment for the three resolutions, but is a smooth process in the \low{} case, while it is accompanied by rapid variations of the potential undergone by stellar particles at high resolution. The bottom panel of Figure~\ref{fig:ongoingdphi} shows the maximum variation of $|D\phi / Dt|$ for each particle over all the simulation. The distribution at \low{} is very different from the distribution at \med{} and \high. This confirms that the particles at \low{} undergo less rapid variation of the potential, i.e. lower peaks of gravitational forces.

This overall demonstrates that the relaxation process, during the merging of galaxies, is smoother at low resolution than at high resolution. We have shown previously that the \med{} and \high{} simulations resolve much more dense substructures, like gas filaments, stellar clusters, compact cores, etc. Our interpretation is then that these local density peaks are accompanied by rapid variations of the gravitational potential, which scatter the stellar orbits, evacuate the angular momentum, and form, for favorable orbits, slowly-rotating elliptical galaxies. At low resolution, these rapid and local fluctuations of the density and potential are largely missed, hence the merging process is smoother, and more angular momentum remains in the main stellar body of the merger remnant.

Density fluctuations are of course stronger in the dissipative component (gas) and the young stars formed therein, which likely explains why the resolution effect is stronger in wet mergers. Nevertheless, old stars are clearly affected as well, as was shown above.

This also explains why the effect of gas in a wet merger, compared to a dry merger at fixed resolution, is different for standard-resolution models and high-resolution ones (Sect.~\ref{sec:rolegas}). At \low{}, the gas remains relatively smooth, promotes the survival/rebuilding of a stellar disc component, thus increasing the rotational support in the final merger remnants. At higher resolution, the presence of gas forms many dense small-scale substructures of gas and young stars \citep[consistent with observations, see e.g.][]{B08}, these substructures increase the degree of relaxation during the merging process, not just for the gas and young stars but also for the old stars. Thus, while the presence of gas should still promote the survival/rebuilding of a disc component in the merger remnant (our high-resolution wet-slow remnant does have a low-mass disc componnent of gas and young star), it also promotes orbital scattering and evacuation of the angular momentum for the whole baryonic mass, but the latter effect is missed if the resolution is too low. This explains why, at high resolution, the merger remnant (in the Wet-Slow case) does not have a higher rotation support or a more prominent disc component than the corresponding Dry-Slow case, an in fact even has a somewhat lower $\lambda_R$ at one effective radius.

The high-resolution simulations, compared to the standard cases, resolve the formation of dense and relatively massive substructures (clusters, cores, filaments of $10^{5-7}$ solar masses) that scatter the stellar orbits and evacuate the angular momentum from the main body of the merger remnant. {\em Very high-res} simulations show a relatively reasonable convergence compared to the \med{} ones: they resolve the same massive substructures, plus lower mass ones ($\sim 10^{4-5}$ solar masses) that are more numerous but are much less efficient to scatter the orbits and affect the relaxation of the merger remnant, as the corresponding relaxation timescale is much longer. It is thus expected that results converge at a high-enough resolution. 

\subsection{Timestepping and code specificities \label{sec:timestep}}

Our results have been obtained with a given code and one can naturally wonder whether or not other codes would show the same resolution effect. We in fact expect no fundamental differences in the output from different codes, given that similar substructures are formed initially: this relaxation effect is mostly gravitational, and this should be treated rather similarly in grid-based and tree-codes. The main question remains then whether or not other codes would form substructures similar to those found in our simulations \citep[e.g., with a similar mass spectrum,][]{B08}: this a priori depends on the modeling of gas cooling and turbulence dissipation processes.

Another specificity of the code employed is its fixed timestep. A small timestep may better resolve the scattering of stellar orbits by dense substructures, in particular at high resolution. This would actually tend to increase the effect of resolution that we have found, which justifies studying the resolution effect at fixed time-step rather than decreasing the timestep at increasing spatial resolution. This way, the effects found can be robustly attributed to the spatial resolution. The time-step itself may have additional, separate effects, in our code or any other, that should be studied separately at fixed (high) resolution.

\section{A systematic effect in the formation of slow rotators}
\label{sec:systematic}

\begin{figure*}
\begin{center}
  \begin{tabular}{cccccccc}
\multicolumn{2}{c}{\begin{Large} Slow-1 \end{Large}} & \vline & \multicolumn{2}{c}{\begin{Large} Slow-2 \end{Large}} & \vline & \multicolumn{2}{c}{\begin{Large} Slow-3 \end{Large}} \\
standard-res & high-res & \vline & standard-res & high-res & \vline & standard-res & high-res \\
\epsfig{file=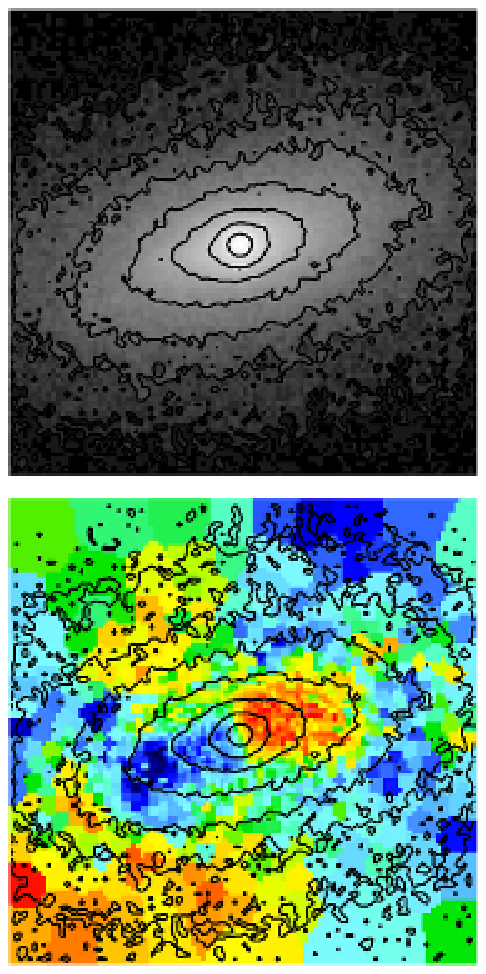, width = 0.3\columnwidth} & \epsfig{file=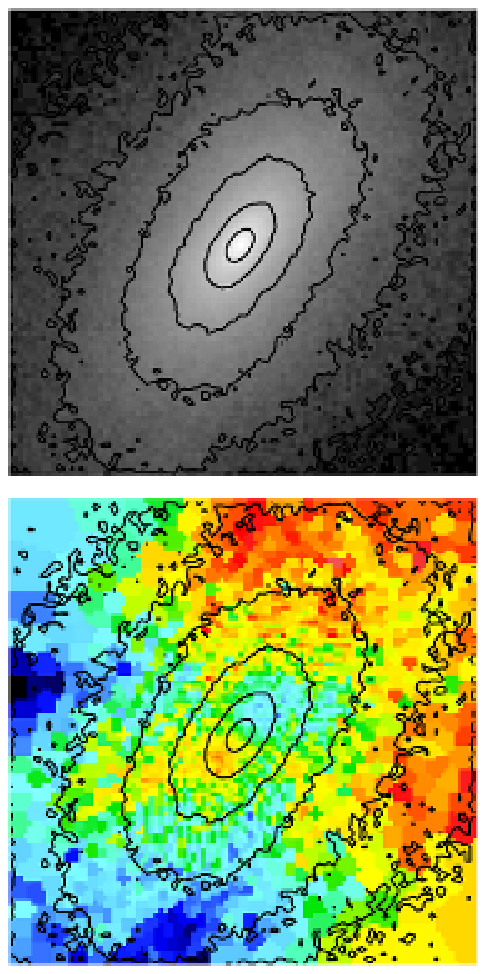, width = 0.3\columnwidth} & \vline & 
\epsfig{file=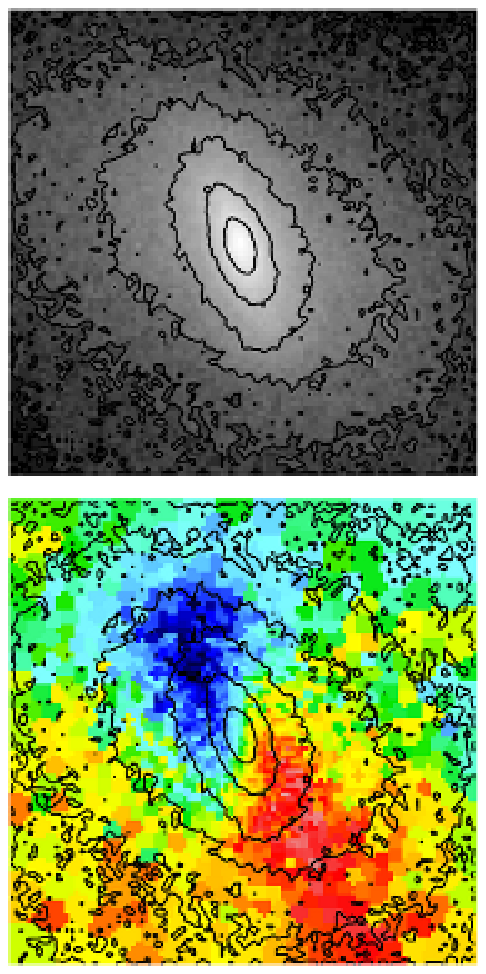, width = 0.3\columnwidth} & \epsfig{file=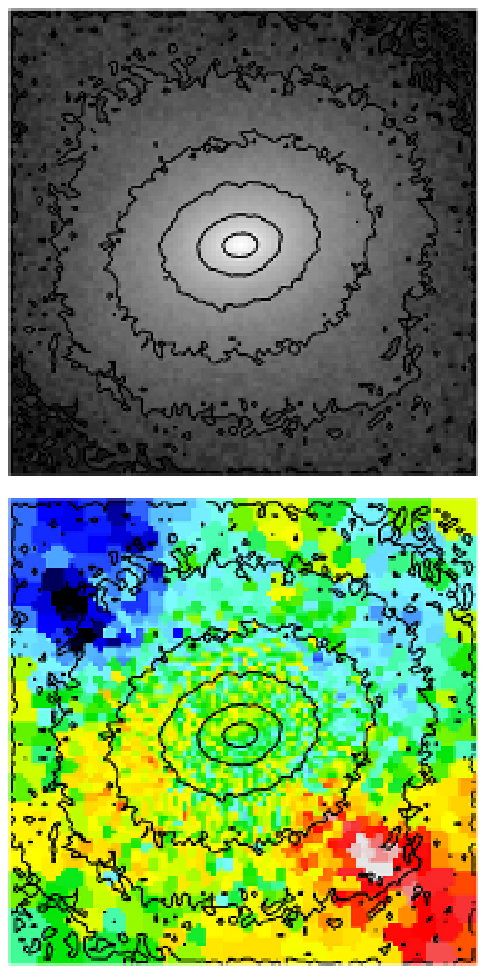, width = 0.3\columnwidth} & \vline &
\epsfig{file=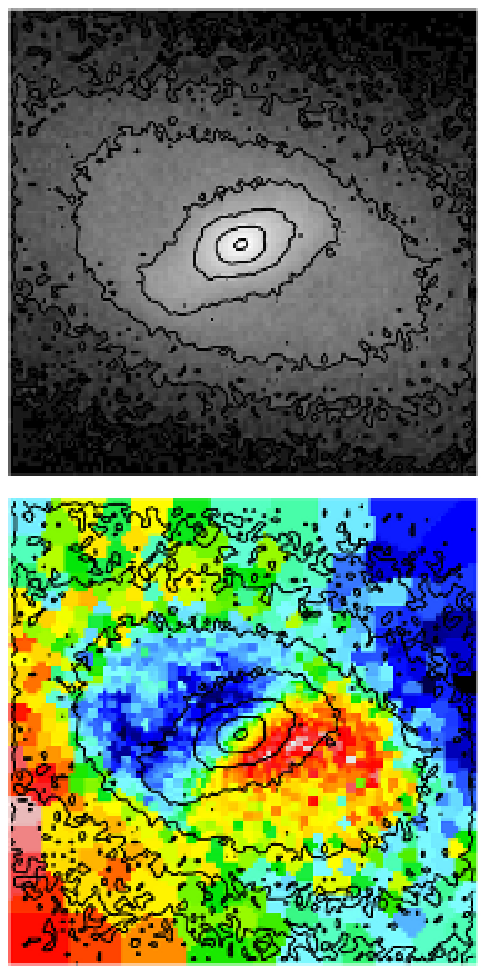, width = 0.3\columnwidth} & \epsfig{file=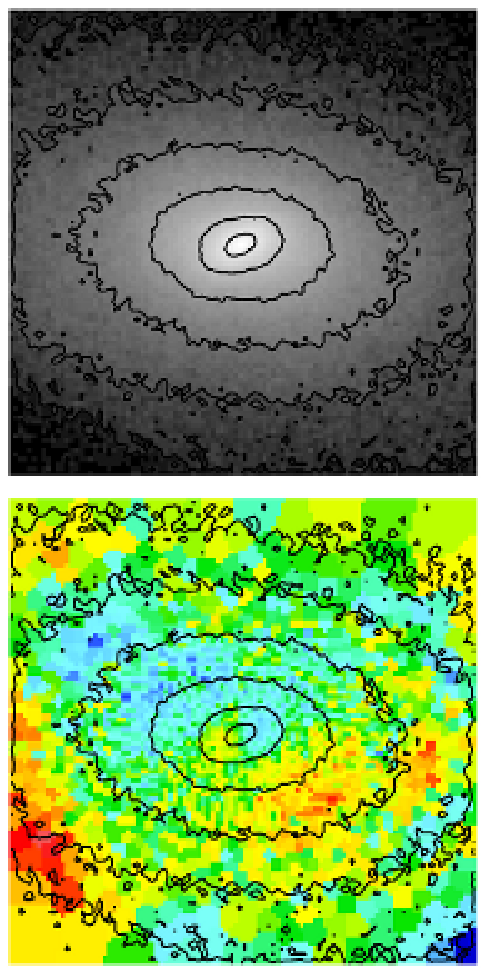, width = 0.3\columnwidth} \\
\multicolumn{2}{c}{\epsfig{file=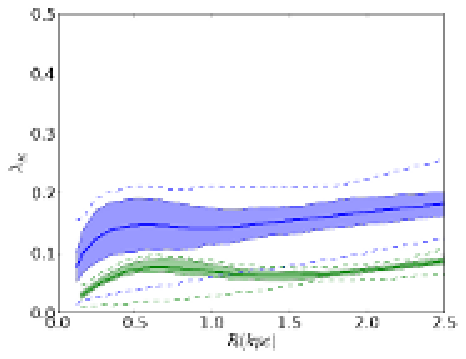, width = 0.66\columnwidth}} & \vline &
\multicolumn{2}{c}{\epsfig{file=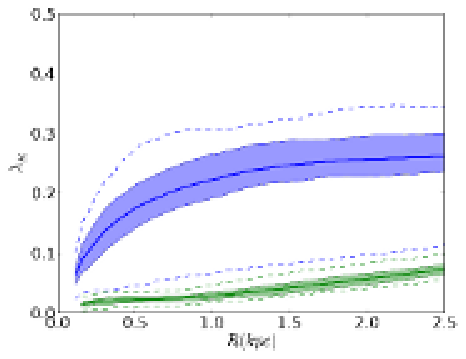, width = 0.66\columnwidth}} & \vline &
\multicolumn{2}{c}{\epsfig{file=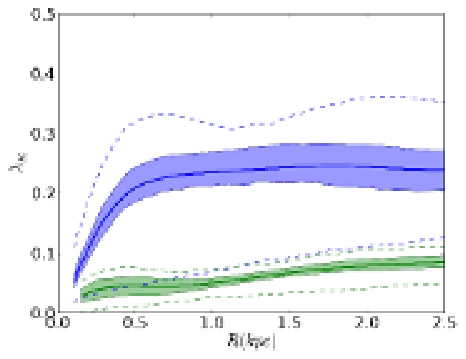, width = 0.66\columnwidth}} \\
  \end{tabular}
  \caption{Projected density and velocity fields for the \low{} and \med{} resolutions and their respective $\lambda_R$ profiles (the \low{} in blue, the \med{} in green) for the {\it slow-1, slow-2, slow-3} models (from left to right resp.).}
\label{fig:otherslow}
\end{center}
\end{figure*}

To ensure that the resolution effect in the formation of slow rotators is a systematic one, and not specific to one simulated orbit, we have selected other major mergers that form slow rotators at high-enough resolution in a larger simulation sample (Bois et al. in preparation), and re-simulated them at lower resolution. These three additional mergers were not simulated at the \high{} but at a resolution which is actually a bit higher than the \med{}, with a spatial resolution of 58~pc and $2\times 10^6$ particles per component par galaxy (i.e. a total of $1.2 \times 10^7$ particles). They were also resimulated at the same \low{} level as the previous models. The gas fraction in these models is 10\%, so as to check that a high gas fraction is not required for the resolution effects to arise. Compared to the original slow-rotator simulation, one orbital parameter is changed in each case. Model {\it slow-1} has a retrograde orbit for both galaxies, model {\em slow-2} has a pericenter distance of 25~kpc, model {\it slow-3} has an orbit inclination of 25 degrees for both discs.

The projected stellar density maps and line-of-sight velocity fields are shown for these models, at standard- and high-resolutions, under the flattest projections in each case, on Figure~\ref{fig:otherslow} (the $\lambda_R$ profiles being also shown in this figure). These three merger remnants are slow rotators (at least at one effective radius) at \med, but at \low{} they all have a much higher angular momentum $\lambda_R$, and a velocity field aligned with their morphological axis -- the high-resolution slow rotators have important kinematic misalignments and KDCs.

These additional cases confirm the effect found and analyzed in detail in the original slow-rotator model, with about the same degree of discrepancy between the standard-resolution and higher-resolution models. Because we have spanned the four slowest rotators among the 1:1 mergers from the Bois et al. (in preparation) sample, the resolution effect seems to strongly affect the modeling of a significant number of slow-rotating ellipticals, if not all, even with modest gas fractions (here 10\% of the baryonic mass).

\section{Discussion and Conclusion}
\label{sec:conclusions}
In this paper, we have studied the effect of numerical resolutions (spatial and mass resolution) on the global properties of merger remnants. Our simulations at ``standard'' resolution, are comparable to the majority of merger simulation samples published in the last few years: the spatial resolution (gravitational softening and typical hydrodynamical smoothing lengths) is 180~pc, and the number of particles $\sim$ 10$^{5}$ per galaxy and per component (gas, stars, and dark halo). These simulations have been compared to models of the same mergers with increased resolution, up to 32~pc and almost 10$^{7}$ particle per galaxy and per component.

We have analyzed the morphology and kinematics of the relaxed merger remnants. In particular, we have studied whether they are ``fast rotators'', with an apparent spin parameter $\lambda_R > 0.1$ and have small misalignments between the morphological and kinematic axes, i.e. in broad terms early-type galaxies with significant flattening and rotational support. At the opposite end, ``slow rotators'' are systems with a low $\lambda_R \leq 0.1$ (at one effective radius), large kinematic misalignments, i.e. early-type galaxies dominated by (anisotropic) pressure support and low residual rotation. Such slow rotators usually have central KDCs in our high-enough resolution simulations.

Our mains findings can be summarized as follows:
\begin{itemize}
\item The formation of \fasts{} is not significantly affected by numerical resolution. Models that produce fast rotators at the highest resolution also result in fast rotators at lower resolution, with some random fluctuation of their properties but no sign of systematic variation in the morphology or angular momentum profile against resolution.

\item The formation of \slows{} is greatly affected by numerical resolution. Models that produce slow rotators at the highest resolution  result in much faster rotators at lower, standard resolution. The effect is present, but relatively minor, in purely collisionless dry mergers. Discrepancies become major in wet mergers, even in cases with modest gas fractions like 10\% of the baryonic mass.

\item These effects cannot be attributed to our choice of initial conditions or interaction orbits, but actually relate to the physical treatment of the merging process itself. In particular, small-scale density fluctuations increase at high resolution, and they participate to scattering stellar orbits and largely influence the final degree of relaxation and orbital structure in the merger remnants.

\item The effect of gas on the properties of merger remnants is generally considered to consist in preserving a higher angular momentum, in particular through enhancing the survival/rebuilding of disc components in merger remnants. We find that this picture is incomplete: at high resolution, gas still reforms discy components, but also forms a large number of dense substructures (massive star clusters, dense nuclei, tails and filaments, etc) that trigger rapid variations of the gravitational potential and the degree of relaxation of the final system. This effect is missed with a too low resolution. At high-enough resolution, adding gas to a given merger does not necessarily increase the rotational support of the final merger remnant; we even find a case of a wet merger with 20\% of gas that has a final angular momentum parameter $\lambda_R$ slightly lower than the corresponding dry merger.
\end{itemize}

At the present stage, our results do not indicate how frequently real slowly-rotating ellipticals were formed by binary (wet) mergers of disc galaxies, but they show that this can be a robust pathway for their formation. In the course of the \atlas{}\ project, we are conducting, analyzing, and comparing a large set of numerical simulations for various formation mechanisms, in order to derive which is (are) the main formation mode(s) of real slow rotators in the nearby Universe. Our present results already indicate the limitations of existing samples of galaxy merger simulations, and will then serve to estimate the required resolution, the limitations of numerical models and their possible biases.

More generally, the immediate implications of these findings on our understanding of early-type galaxy formation are:
\begin{itemize}
\item High resolution in simulations of major mergers does not just allow to resolve small-scale structures like nuclear systems and star clusters, but impacts the whole global properties of the elliptical-like merger remnants, at least for the slow-rotating ones.
\item The formation of slow-rotating elliptical galaxies can be achieved through a major merger relatively more easily than previously believed. It can be frequent even in wet mergers with relatively high gas fraction, and with late-type, disc-dominated progenitor galaxies. 
\item Repeated mergers and/or dry mergers of galaxies that are already early-type systems are thus not the only theoretical path to produce slow-rotating galaxies. Major mergers of two dic galaxies, including wet mergers, can produce slow-rotating early-type galaxies. Further studies are needed to determine how common this formation mechanism is for slow-rotators. 
\item Quantitative comparisons of major merger simulation results with the observed properties of real early-type galaxies require high resolution models. A typical requirement, according to our study, would be a spatial resolution better than 100~pc for both the gravitational N-body aspects (i.e., softening length) and the hydrodynamical ones (for instance, the size of groups of particles other with quantities are smoothed in SPH models). The mass resolution should correspond to at least $\sim 10^6$ particle per galaxy per component, which typically corresponds to a mass resolution $\sim 10^4$~M$_{\sun}$ for the gas discs of bright spiral galaxies. We find reasonable convergence above this resolution, but cannot rule out that some systematic effects still exist; in any case simulations below this resolution level show clear and strong resolution effects. Unfortunately, most published samples of major merger simulations are below this typical resolution limit.
\item The small-scale properties of interstellar gas and clustered star formation are important for the global, large-scale properties of merger remnants. Simulations directly resolving gas cooling down to low temperatures, the formation of cold (molecular) gas clouds and star formation therein, are highly desirable to understand the whole process of early-type galaxy formation. Modern hydrodynamic codes are promising in this respect \citep[e.g.][]{B09,kim09}.
\end{itemize}

\section*{Acknowledgments}
We are most grateful to the referee, Joshua Barnes, for constructive comments that very significantly improved the presentation of our results. This work was supported by Agence Nationale de la Recherche under contract ANR-08-BLAN-0274-01. Simulations were performed at CEA-CCRT using HPC resources from GENCI, grant 2009-042192.

\appendix
\section{Additional morphological and kinematic parameters for the merger remnants}
\label{app:complete}

In this appendix, we provide further details on the morphology and kinematics of the four merger remnants analysed in the present paper.

   \subsection{Morphology \label{app:morph}}
\begin{figure*}
  \centering
  \epsfig{file=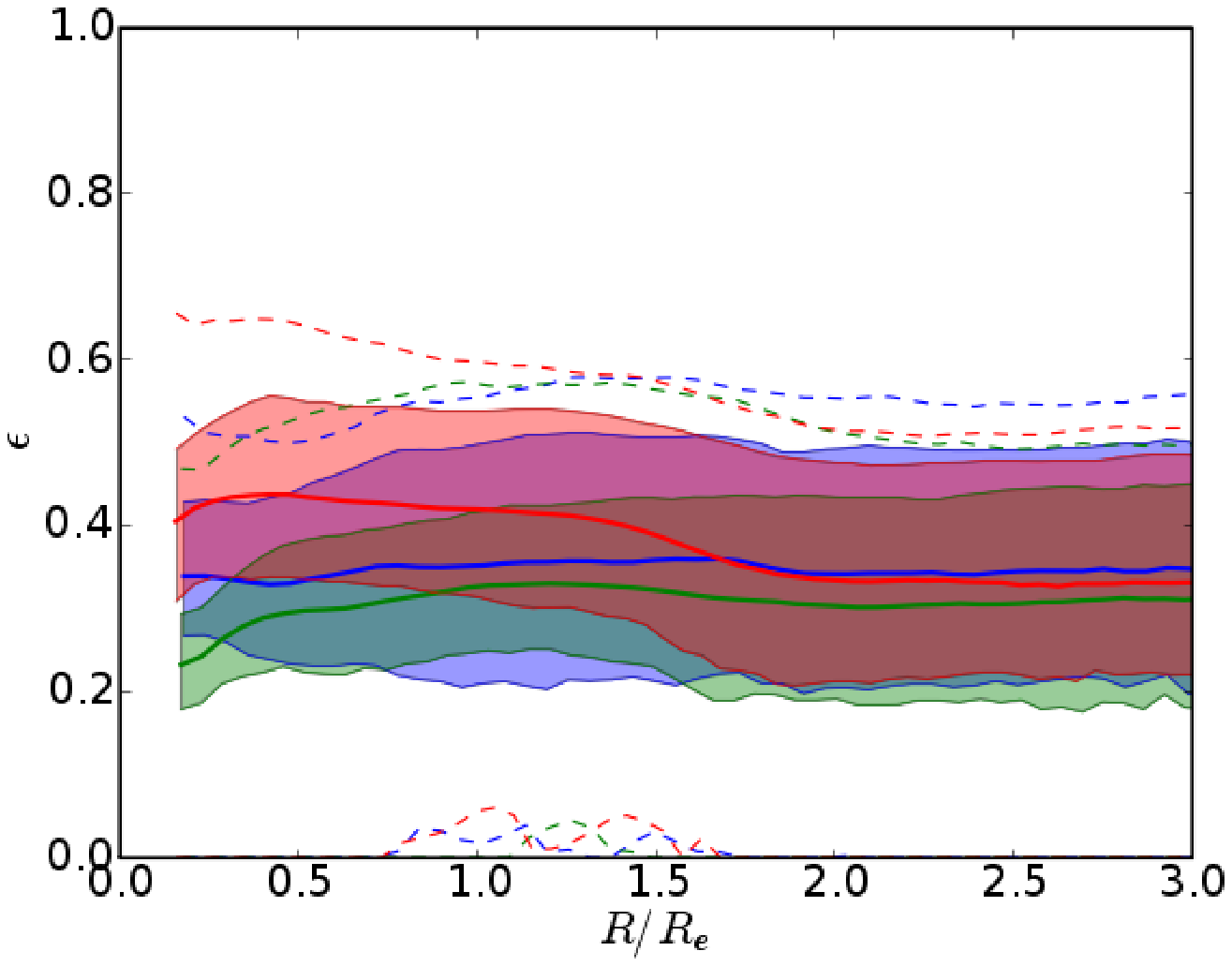, width=0.8\columnwidth}
  \epsfig{file=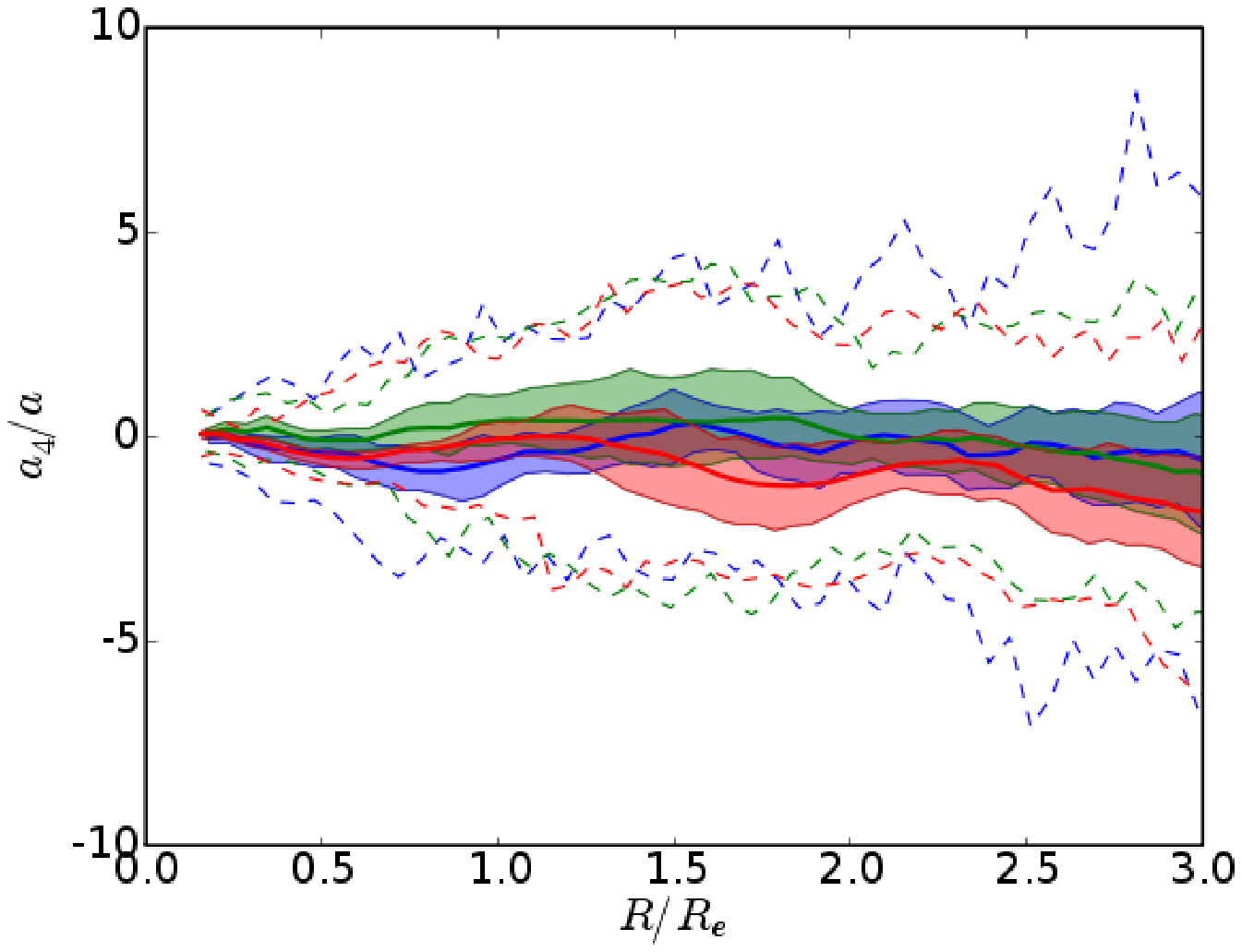, width=0.8\columnwidth}
  \epsfig{file=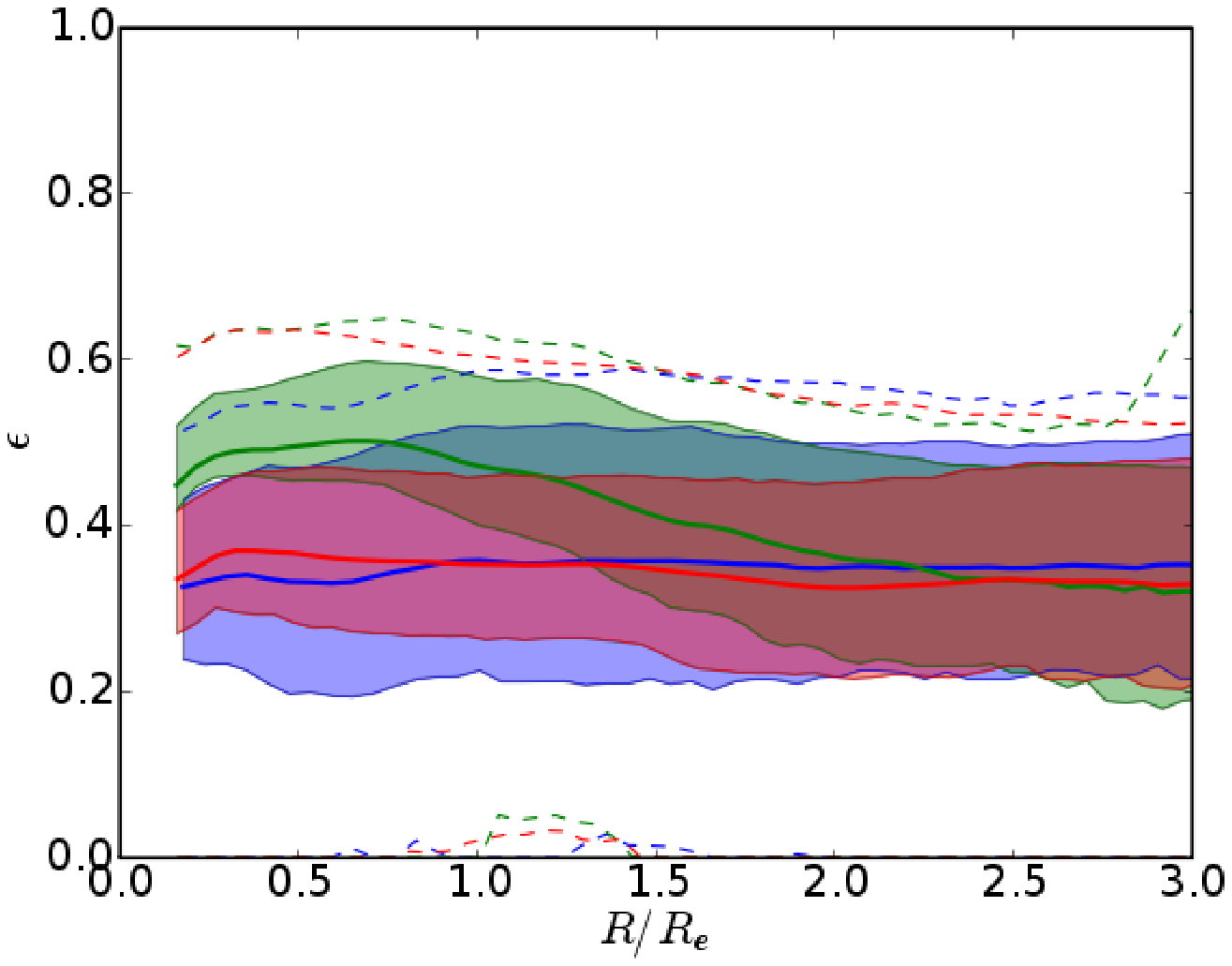, width=0.8\columnwidth}
  \epsfig{file=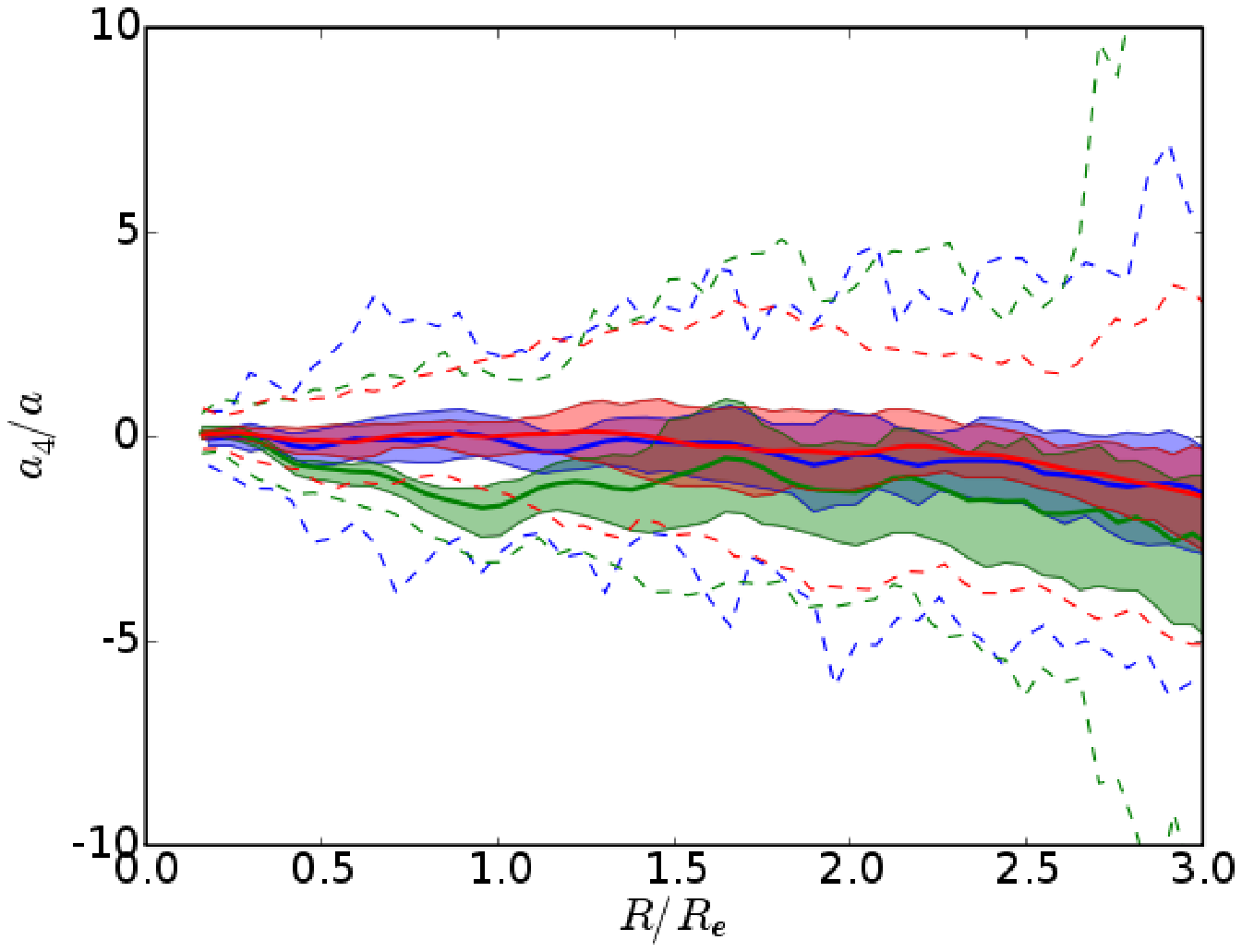, width=0.8\columnwidth}
  \epsfig{file=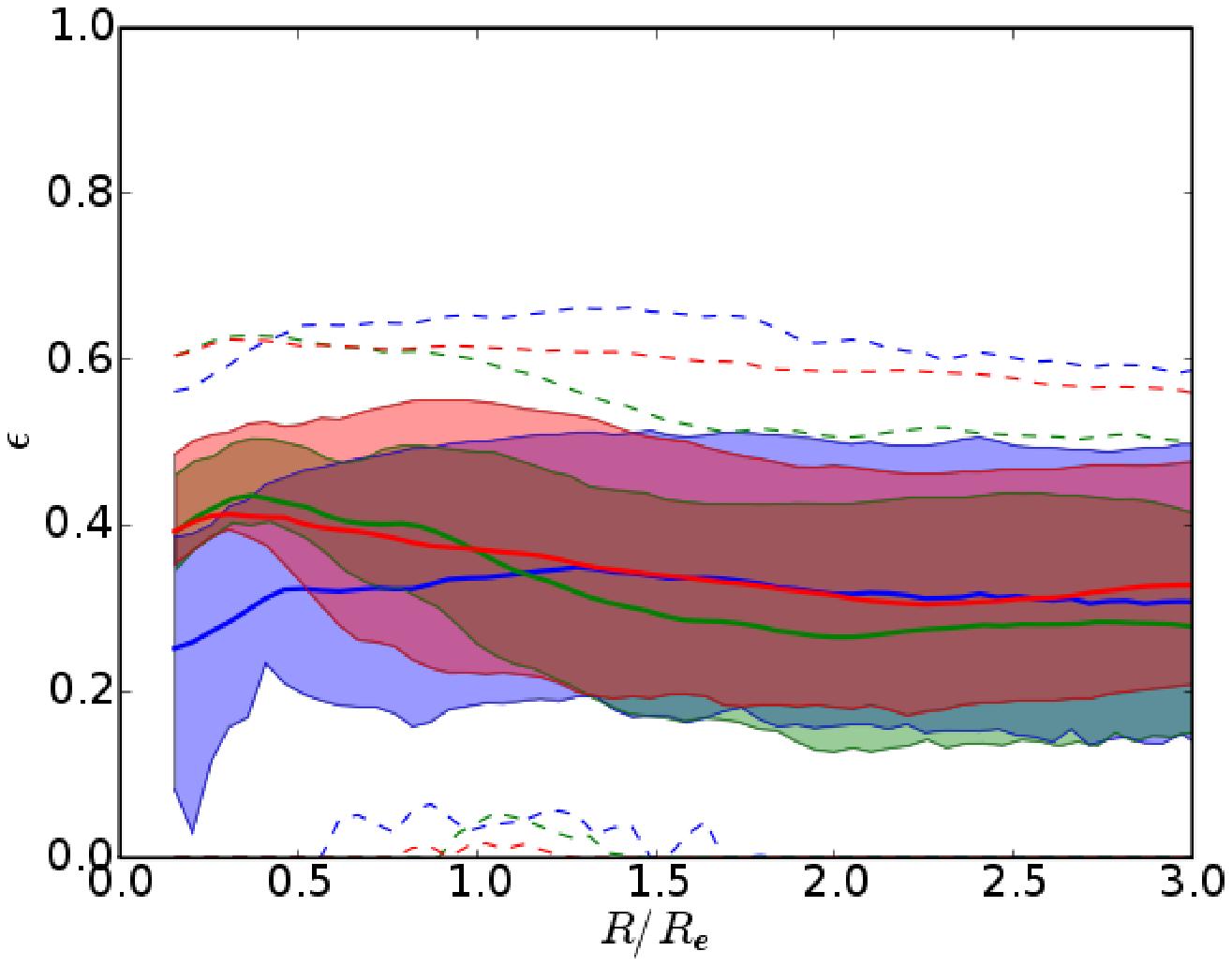, width=0.8\columnwidth}
  \epsfig{file=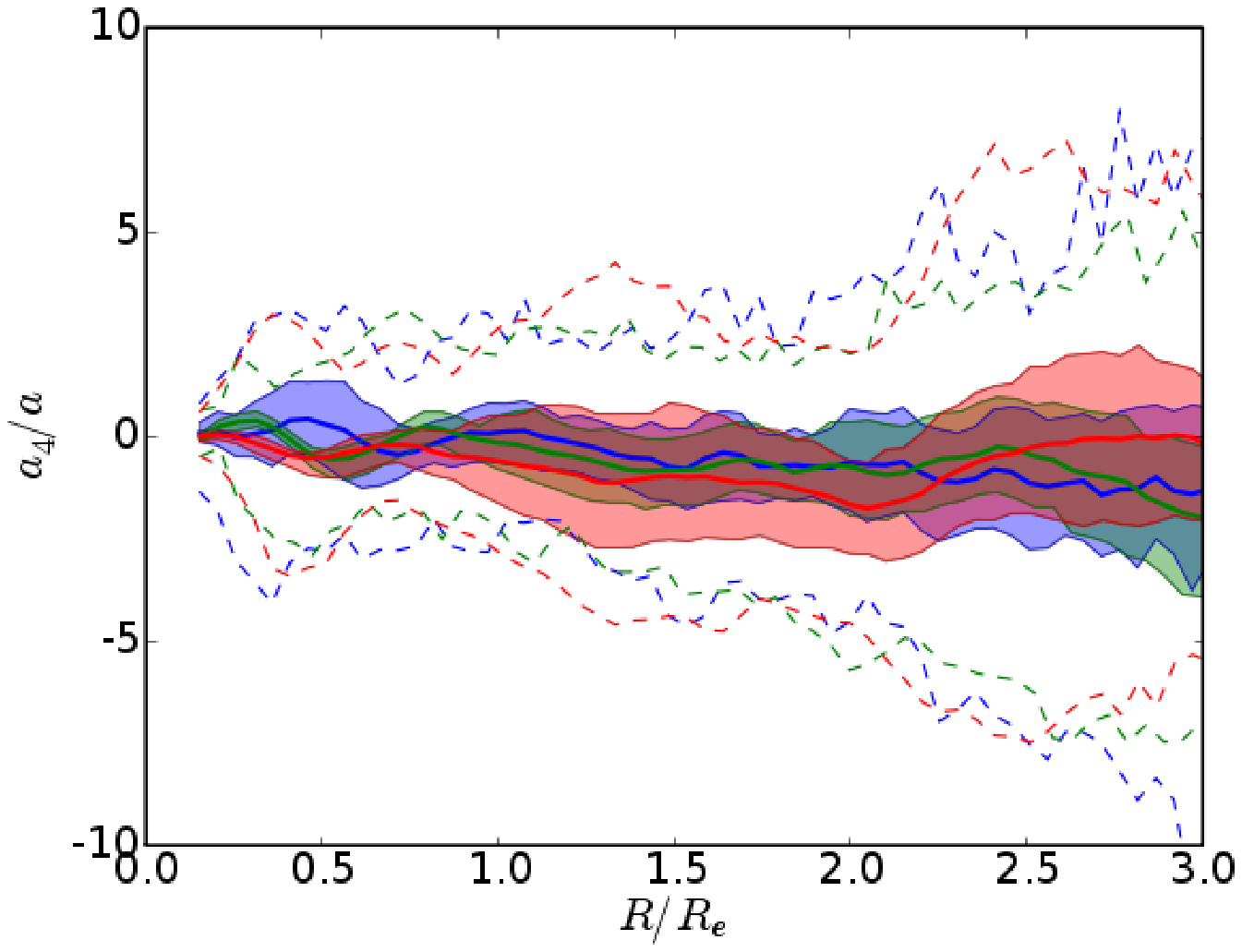, width=0.8\columnwidth}
  \epsfig{file=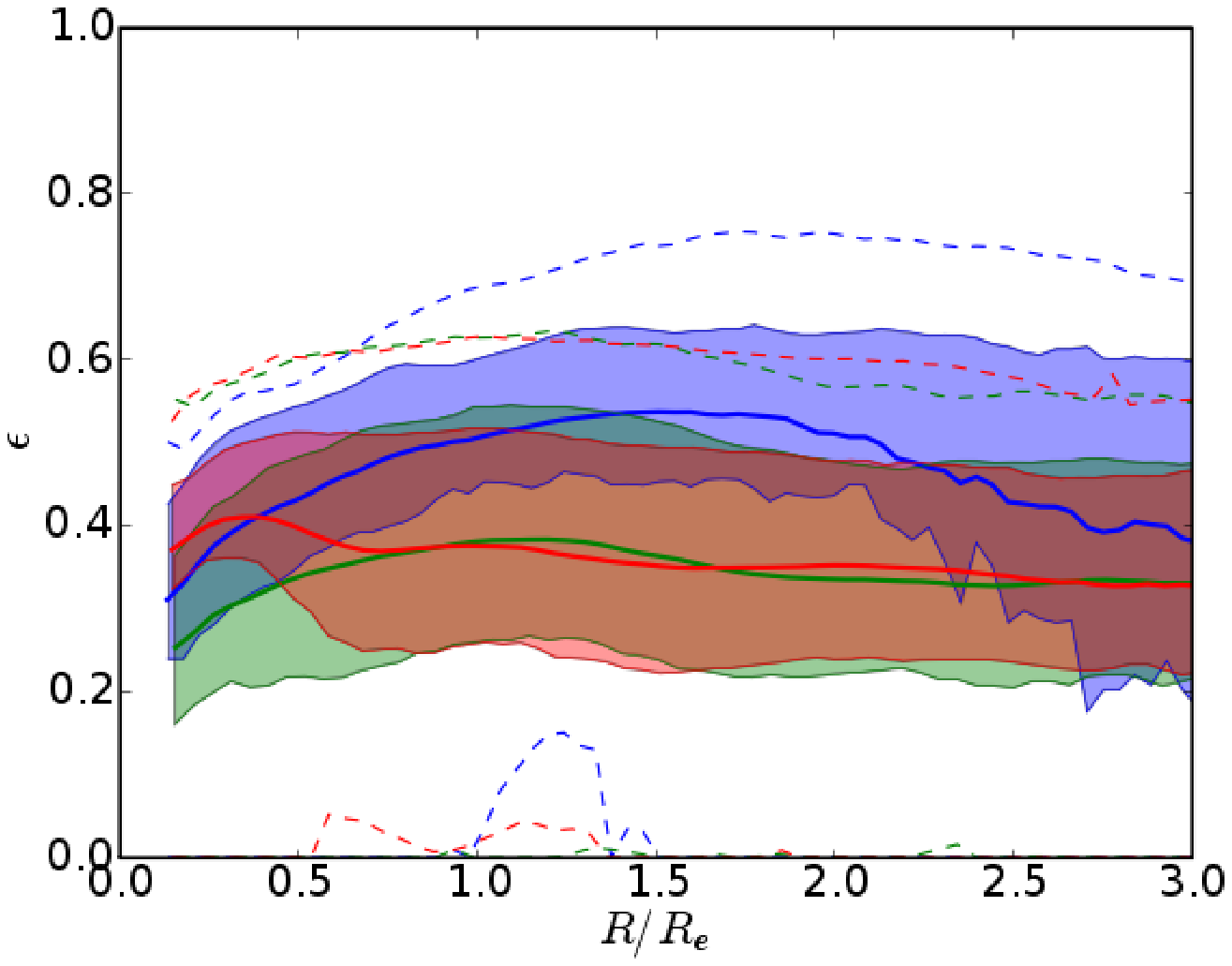, width=0.8\columnwidth}
  \epsfig{file=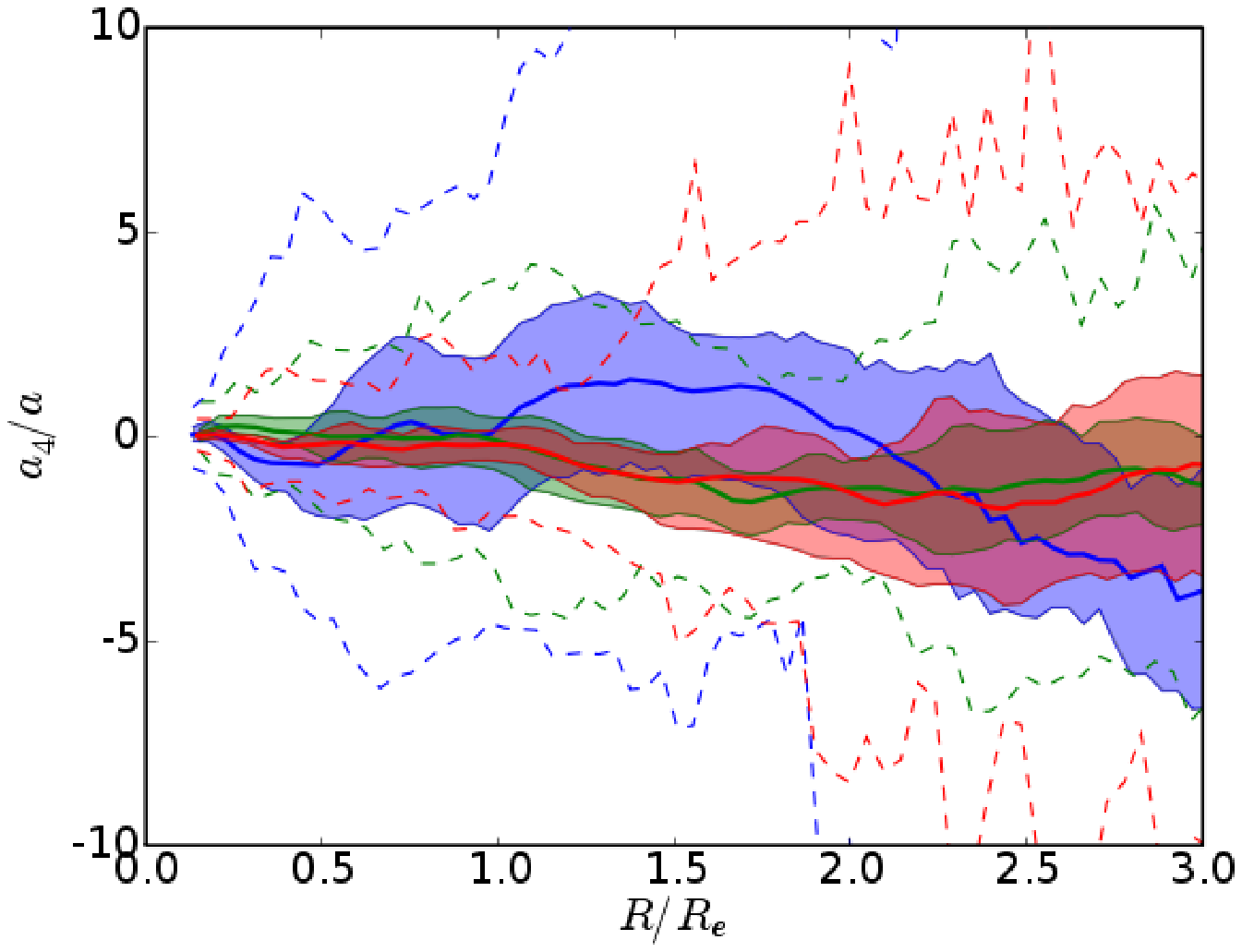, width=0.8\columnwidth}
  \caption{Ellipticity and $a_4/a$ profiles (left and right panels resp.) of the four simulations as a function of $R/R_e$. From top to bottom : Simulations Dry-Fast, Wet-Fast, Dry-Slow and Wet-Slow. The three resolutions are shown with different colours : the \low{} in blue, the \med{} in green and the \high{} in red. For each resolution, we plot five lines which correspond to the minimum and maximum at each radii (dashed lines), the mean value (thick solid lines) and the first and third quartiles (thin solid lines). The inter-quartiles space (which correspond to 50\% of all projections) is filled with the colour associated with the resolution.}
  \label{figA:ella4}
\end{figure*}

The ellipticity $\epsilon$ and $a_4/a$ profiles are shown in Figure~\ref{figA:ella4} as a function of $R/R_e$. The apparent differences sketched in Sect.~\ref{sec:IVmaps} are confirmed quantitatively in the radial ellipticity profiles. Within 1.5~$R_e$, there are small differences in the Dry/Wet-Fast and Dry-Slow simulations. The ellipticity outside 1.5~$R_e$ is however quite similar at all resolutions for these three simulations: the minimum ellipticity is basically 0, the mean is $0.33 \pm 0.03$ and the maximum is $0.55 \pm 0.05$ at 2~$R_e$ for all three \low{}, \med{}, \high{}. 

The Wet-Slow simulation shows much larger differences. Outside 0.5~$R_e$, the \med{} and \high{} are similar. The ellipticity profile of the \low{} has then a completely different appearance: between 0.6 and 2 $R_e$, 75\% of the projections have an ellipticity higher than 0.4, and the reached maximum in $\epsilon$ is 0.75 (versus $\sim 0.6$ for the other two resolutions). In the outer part ($R > 2 R_e$) the ellipticity of most of the projections is decreasing but its maximum is still larger than 0.7.

The same trends are observed in the $a_4/a$ profiles. In the Dry/Wet-Fast and Dry-Slow simulations at all three resolutions, the mean $a_4/a$ is around 0. Then 50\% of the projections are between -2 and 2\%. The \med{} Wet-Fast and the \high{} Dry-Fast simulations are only slightly more boxy. The $a_4/a$ profile of the Wet-Slow simulation dramatically confirm what we observe for the ellipticity. The \low{} clearly departs from the \med{} and \high{}, which are quite similar. Between 0.5 and 1~$R_e$, the projections of the \low{} span a very large range of $a_4/a$. Between 1 and 2~$R_e$, 75\% of the projections have a discy shape, and the isophotes of the merger remnant become then increasingly boxy going outwards.

   \subsection{Kinematics \label{app:kinem}}
The Wet-Slow simulation has been treated in the paper, we will thus focus on the three other simulations.

\begin{figure*}
  \centering
  \epsfig{file=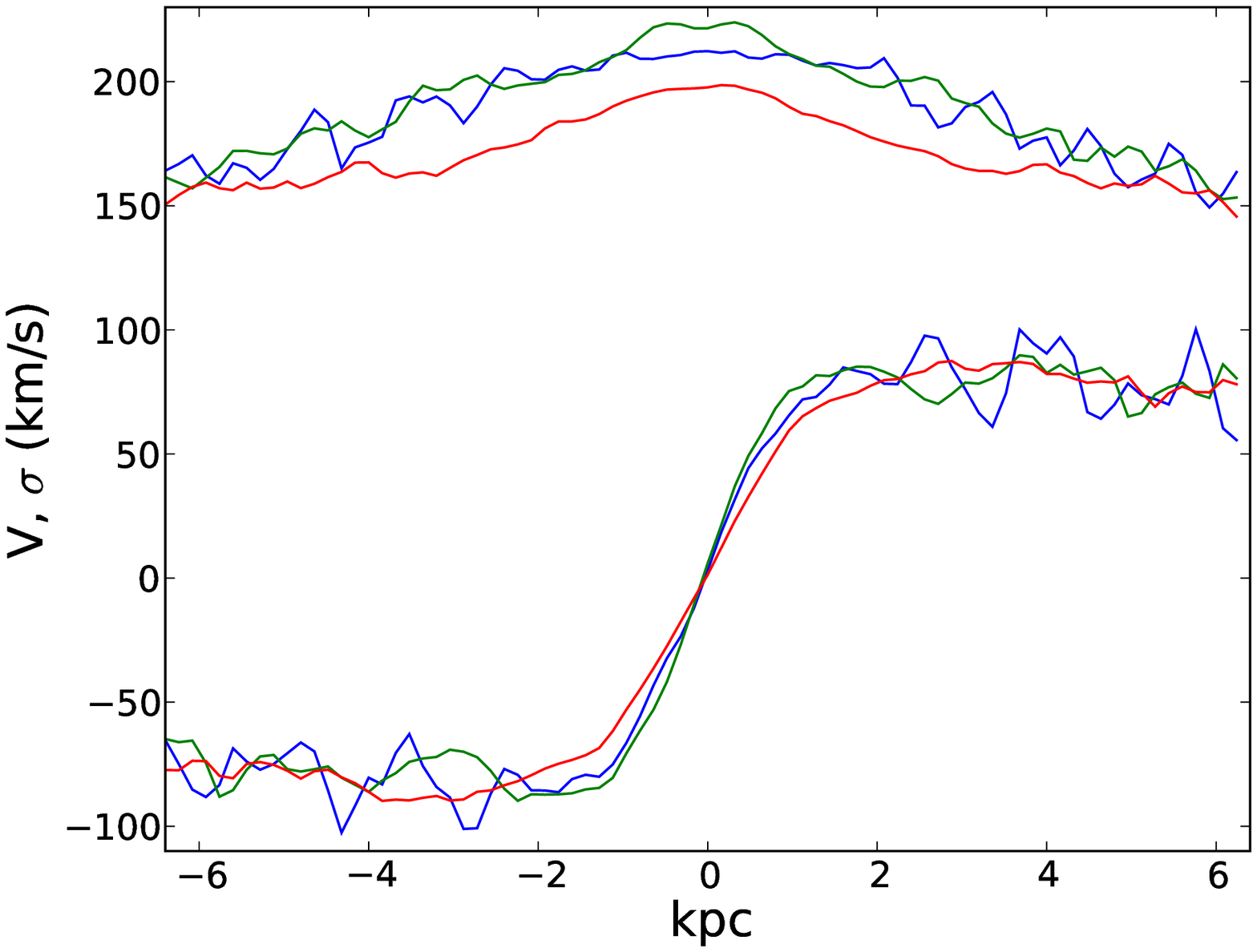, width=0.8\columnwidth}
  \epsfig{file=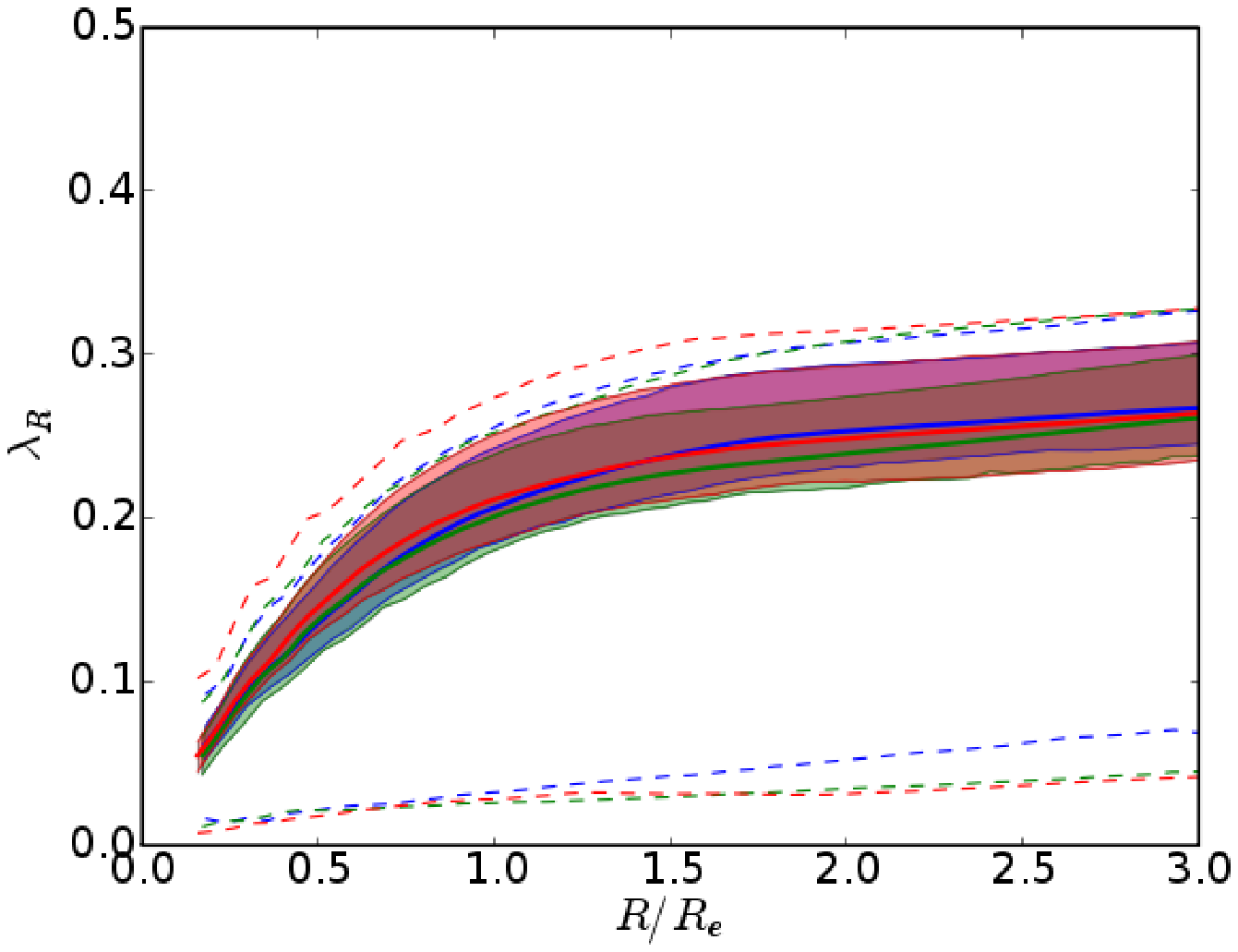, width=0.8\columnwidth}
  \epsfig{file=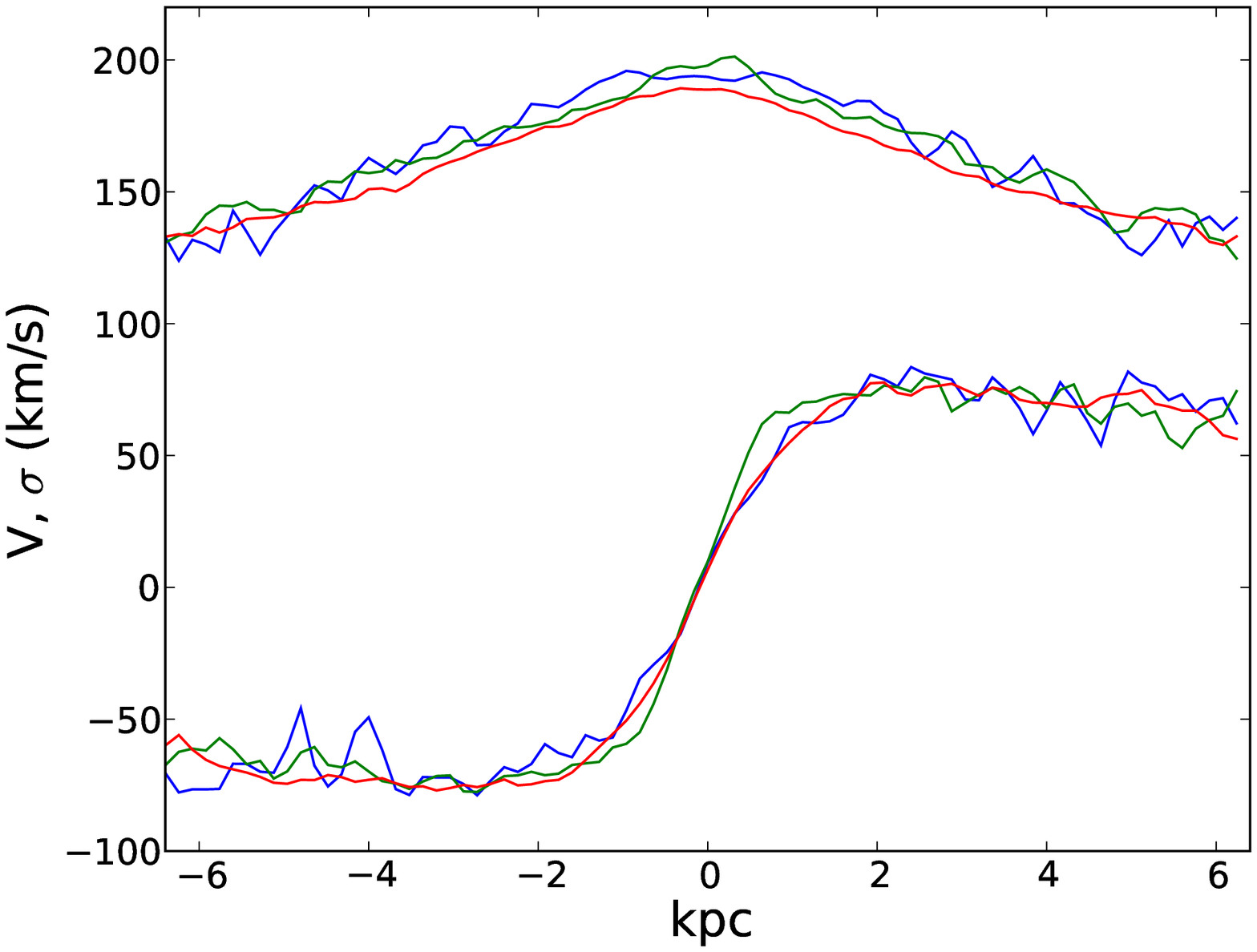, width=0.8\columnwidth}
  \epsfig{file=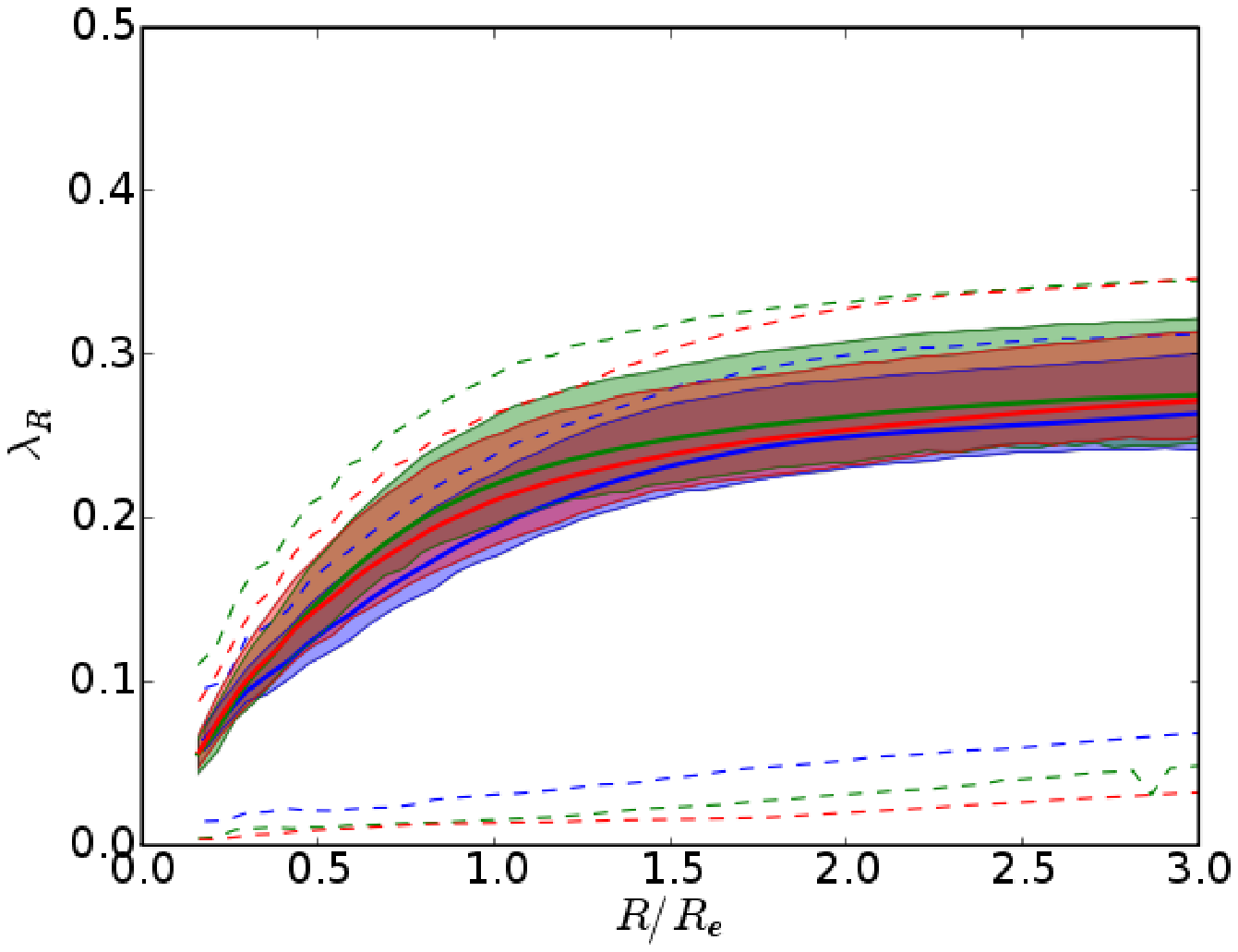, width=0.8\columnwidth}
  \epsfig{file=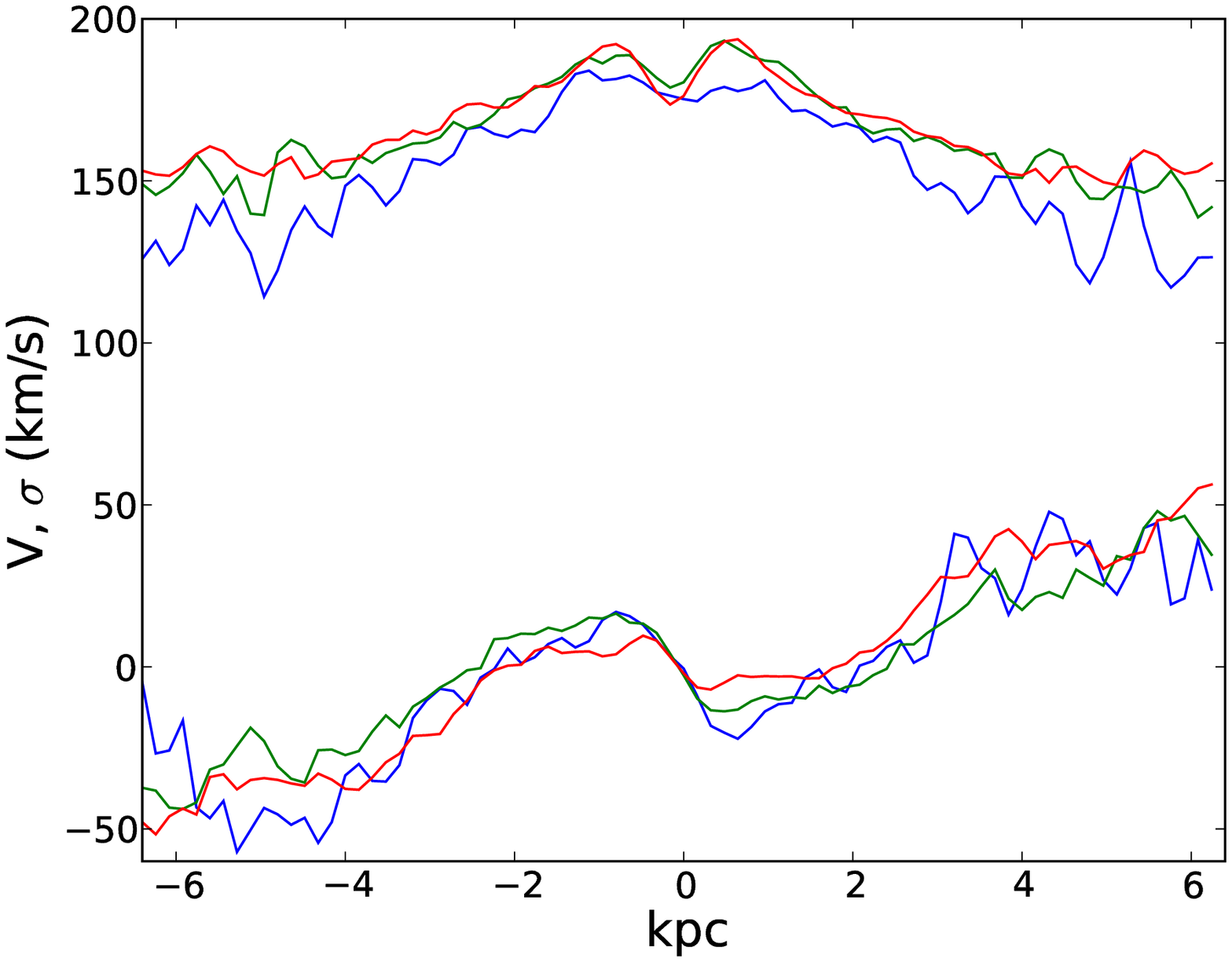, width=0.8\columnwidth}
  \epsfig{file=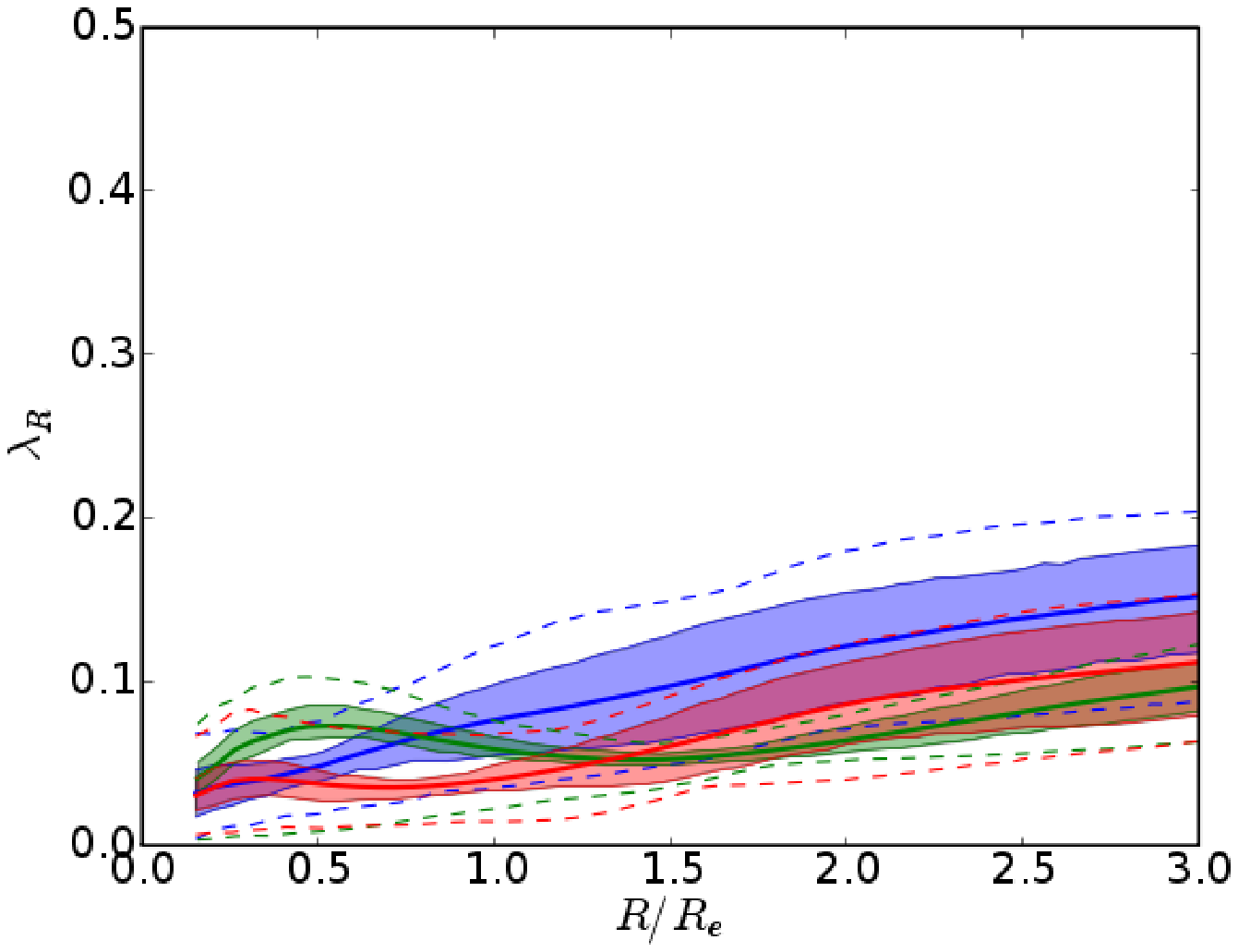, width=0.8\columnwidth}
  \epsfig{file=v_sigma_wet_dr.eps, width=0.8\columnwidth}
  \epsfig{file=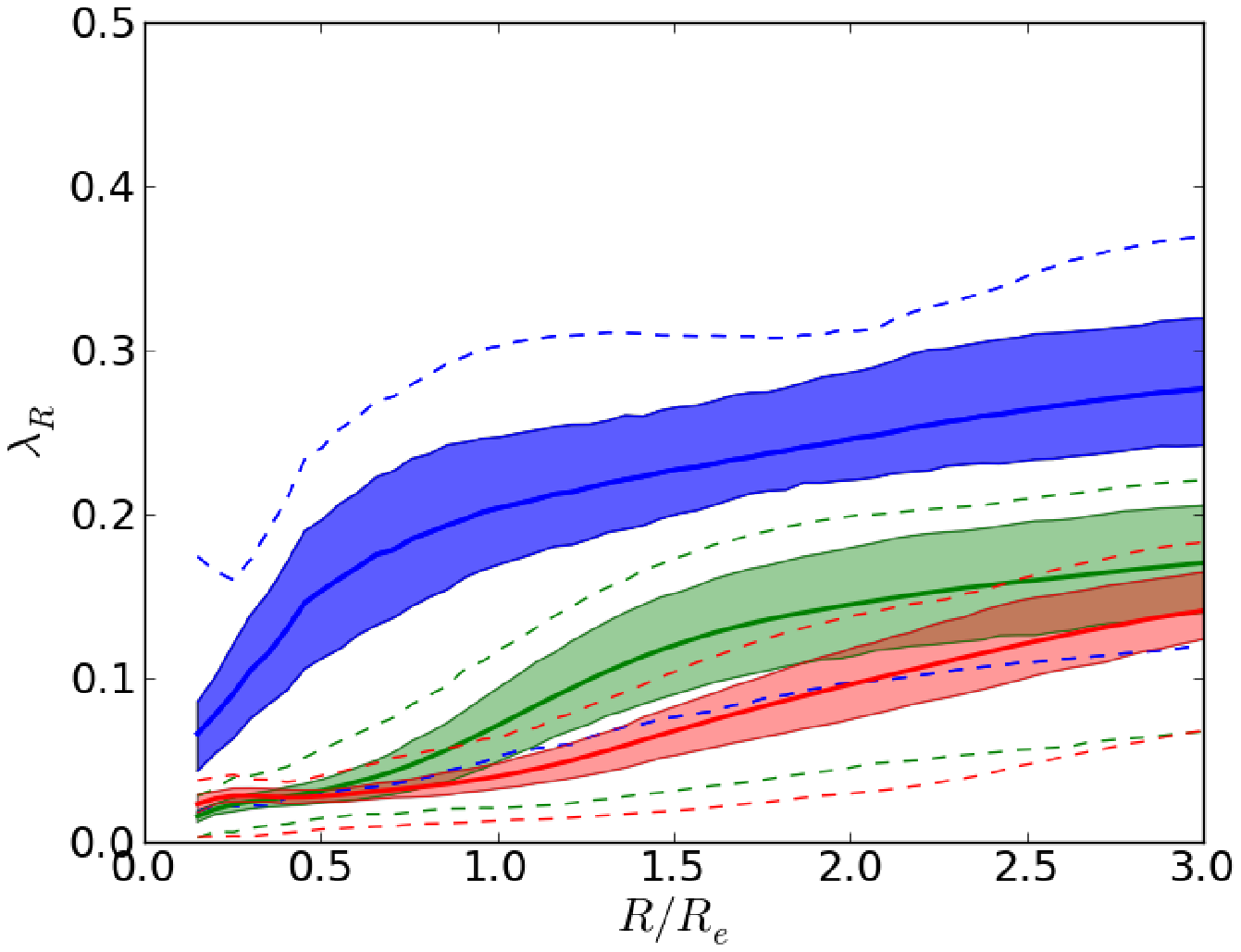, width=0.8\columnwidth}
  \caption{{\bf Left Panels} The radial velocity and velocity dispersion profiles (in km\ s$^{-1}$) for the mean ellipticity projection along the global kinematic position angle (radius in kpc). {\bf Right Panels} $\lambda_R$ profiles as a function of $R/R_e$. From top to bottom : Simulations Dry-Fast, Wet-Fast, Dry-Slow and Wet-Slow. The three resolutions are shown with different colours : the \low{} in blue, the \med{} in green and the \high{} in red.}
  \label{figA:vslr}
\end{figure*}

Left panels of Figure~\ref{figA:vslr} show the velocity and velocity dispersion curves for the mean-ellipticity projection along the global kinematic position angle. In the Dry-Fast simulation, the central slope of the rotation curve at \high{} is slightly shallower, and the dispersion about 15\% smaller, but in the outer part the velocity amplitude is similar at all three resolutions, with a velocity of about 60~km\ s$^{-1}$ at 6~kpc, and dispersions values going to about $150$~km\ s$^{-1}$. The Wet-Fast simulations show consistent velocity profiles at all three resolutions, with a velocity amplitude of 60~km\ s$^{-1}$ at 6~kpc, and dispersion decreasing outwards down to $\sim 125$~km\ s$^{-1}$. Again, the velocity curves for the Dry-Slow simulation are all very similar, but these profiles clearly reveal the previously observed KDC which appears here as a kpc-size core counter-rotating with respect to the outer part. Note the \low{} dispersion profile which is about 10\% smaller than for the other two higher resolutions.

In right panels of Figure~\ref{figA:vslr}, we now compare the simulations using the apparent angular momentum $\lambda_R$. These figures clearly show that the Dry/Wet-Fast simulations (top  and second from top) both result in \fasts{}, the mean values of $\lambda_R$ is 0.2 and the maximum about 0.25 at 1~$R_e$ for the three resolutions. This confirms our previously mentioned results that the spatial and mass resolutions do not seem to have a significant effect on these merger remnants.

The analysis of the morphology and kinematics of the Dry-Slow simulation did show mild differences in the remnants for varying resolutions, the $\lambda_R$ profiles exacerbate these small discrepancies. At \low{}, $\lambda_R$ is an increasing function of radius, with 75\% of all projections having values below 0.1 at 1 $R_e$, and 25\% above 0.1. However, if we are taking into account the projection which maximises $\lambda_R$, the \low{} remnant should be classified as a \fast. In the same context, both the \med{} and \high{} are classified as \slows. They have not the same profiles but have a similar overall behaviour: $\lambda_R$ first increases up to about $0.5$~$R_e$, and then decreases (up to $1.5$~$R_e$ for the \med{} and 1~$R_e$ for the \high). Outside 1.5~$R_e$, $\lambda_R$ increases again outwards. Such a $\lambda_R$ profile is the clear signature of large-scale KDCs as mentioned in \citealt{ems07} \citep[see also][]{mcdermid06}.


\label{lastpage}

\end{document}